\documentclass[review]{elsarticle}

\makeatletter
\def\ps@pprintTitle{%
   \let\@oddhead\@empty
   \let\@evenhead\@empty
   \let\@oddfoot\@empty
   \let\@evenfoot\@oddfoot
}
\makeatother

\newcommand{\n}{\noindent}
\newcommand{\mb}{\mathbf}

\newcommand{\x}{\times}
\newcommand{\Z}{\mathbb{Z}}
\newcommand{\N}{\mathbb{N}}
\newcommand{\R}{\mathbb{R}}
\newcommand\bigzero{\makebox(0,0){\text{\huge0}}}
\renewcommand{\epsilon}{\varepsilon}

\usepackage[dvipsnames]{xcolor}
\usepackage{amssymb,amsmath,mdframed,ytableau,tikz,extarrows,graphicx,float,enumerate}
\usepackage[mathscr]{eucal}
\usepackage[nottoc]{tocbibind}
\usetikzlibrary{calc,arrows,math,snakes}
\usepackage[raggedright]{titlesec}
\usepackage[linktocpage=true,colorlinks=true]{hyperref}

\numberwithin{equation}{section}

\usepackage{amsthm}
\newtheorem{thm}{Theorem}[section]
\newtheorem{lem}[thm]{Lemma}
\newtheorem{cor}[thm]{Corollary}
\newtheorem{con}[thm]{Conjecture}

\newtheorem{defn}[thm]{Definition}
\newtheorem{nota}[thm]{Notation}
\newtheorem{ex}[thm]{Example}
\newtheorem{rem}[thm]{Remark}

\title{The Ghost-Box-Ball System: A Unified Perspective on Soliton Cellular Automata, the RSK Algorithm and Phase Shifts}

\author{Nicholas M. Ercolani\fnref{fn1}}
\ead{ercolani@math.arizona.edu}
\author{Jonathan Ramalheira-Tsu\fnref{fn2}}
\ead{jramalheiratsu@math.arizona.edu}

\fntext[fn1]{Department of Mathematics, The University of Arizona, Tucson, AZ 85721-0089 (\href{mailto:ercolani@math.arizona.edu}{ercolani@math.arizona.edu}). Supported by NSF grant DMS-1615921.}
\fntext[fn2]{Department of Mathematics, The University of Arizona, Tucson, AZ 85721-0089 (\href{mailto:jramalheiratsu@math.arizona.edu}{jramalheiratsu@math.arizona.edu}). Supported by NSF grant DMS-1615921.}

\begin{document}
\begin{abstract}
In this paper, we introduce the ghost-box-ball system, which is an extended version of the classical soliton cellular automaton. It is initially motivated as a mechanism for making precise a connection between the Schensted insertion (of the Robinson-Schensted-Knuth correspondence) and the dynamical process of the box-ball system. In addition to this motivation, we explore generalisations of classical notions of the box-ball system, including the solitonic phenomenon, the asymptotic sorting property, and the invariant shape construction. 

We analyse the ghost-box-ball system beyond its initial relevance to the Robinson-Schensted-Knuth correspondence, unpacking its relationship to its underlying dynamical evolution on a coordinatisation and using a mechanism for augmenting a regular box-ball configuration to study the classical ultradiscrete phase shift phenomenon.
\end{abstract}


\maketitle

\tableofcontents

\section{Introduction}
\subsection{Background}
The analogy between computational algorithms and dynamical systems is a natural one that received a concrete realization in Conway's seminal work on {\it cellular automata} (CA) \cite{bib:g}, \cite{bib:w}. Within this setting it is perhaps not surprising that, in time, CA corresponding to integrable dynamical systems would be identified; and, indeed, this was realized in the pioneering work of Takahashi and Satsuma \cite{bib:ts} \cite{bib:tts} on so-called {\it box-ball systems} (BBS) identifying the striking {\it solitary wave} character of these systems. Over the past three decades there has been an explosion of interest in this system and its many variations. For more on this we refer the reader to the very good survey papers \cite{bib:tokihiro} and \cite{bib:ikt}. A most remarkable outgrowth of this circle of ideas has been the discovery of deep connections with combinatorics and representation theory. One of the most notable examples of this is the analogy between box-ball systems and the Robinson-Schensted-Knuth (RSK) algorithm, one of the fundamental combinatorial tools of modern representation theory, \cite{bib:kirillov} and \cite{bib:ny}. The focus of this paper is to make this analogy more precise through an extension to what we refer to as ghost box-ball systems (GBBS). A strong motivation for doing this comes from recent developments related to integrable stochastic processes (\cite{bib:o}, \cite{bib:bbo}) motivated, in turn, by integrable systems approaches to random matrix theory.  

\subsection{The Box-Ball System and the Robinson-Schensted-Knuth Correspondence}

In order to provide an overview of the results in this paper we first give, in this subsection, a brief summary of BBS and the RSK algorithm.

\subsubsection{Box-Ball Systems}
A cellular automaton is a special type of discrete dynamical system with both discrete time steps and a discrete (in fact finite) number of states. Of particular interest is the box-ball system (BBS) which was introduced in 1990 by Takahashi and Satsuma \cite{bib:ts}.\\[4pt]

\begin{figure}[H]
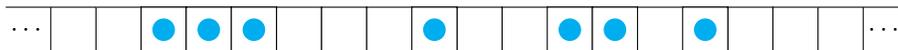

\centering
\tikz[scale=0.6]{
\foreach \x in {0,1,2,3,4,5,6,7,8,9,10,11,12,13,14,15}
{\draw[fill=white]  (\x,3) -- (\x+1,3) -- (\x+1,4) -- (\x,4) -- cycle;			
}
\foreach \x in {1,2,3,7,10,11,13}
{\draw[fill=white]  (\x,3) -- (\x+1,3) -- (\x+1,4) -- (\x,4) -- cycle;			
\fill[cyan] (\x+0.5,3.5) circle (0.25);
}
\foreach \x in {}
{\draw[fill=white]  (\x,3) -- (\x+1,3) -- (\x+1,4) -- (\x,4) -- cycle;			
\fill[red] (\x+0.5,3.5) circle (0.25);
}
\foreach \x in {16}
{\draw[fill=white,white]  (\x,3) -- (\x+2,3) -- (\x+2,4) -- (\x,4) -- cycle;
\draw[-] (\x,3) -- (\x,4);
\draw[-] (\x,3) -- (\x+2,3);
\draw[-] (\x,4) -- (\x+2,4);
\draw[-] (\x+1,3) -- (\x+1,4);
\node at (\x+1.5,3.5) {$\cdots$};
}
\foreach \x in {0}
{\draw[fill=white,white]  (\x,3) -- (\x-2,3) -- (\x-2,4) -- (\x,4) -- cycle;
\draw[-] (\x,3) -- (\x,4);
\draw[-] (\x,3) -- (\x-2,3);
\draw[-] (\x,4) -- (\x-2,4);
\draw[-] (\x-1,3) -- (\x-1,4);
\node at (\x-1.5,3.5) {$\cdots$};
}
}
\caption{A Box-Ball State}\label{firstbbsexfordef}
\end{figure}

\begin{defn}\label{bbsoneszeroesdefn}
The (basic) box-ball system consists of a one-dimensional infinite array of boxes with a finite number of the boxes filled with balls, and no more than one ball in each box (see, for example, Figure \ref{firstbbsexfordef}). \\[3pt]
More formally, the phase space of this system, which we denote by $\text{BBS}$, can be identified with the space of binary sequences
$$\{0,1\}^\Z,$$
with all but finitely many entries equal to zero, so that $1$'s correspond to filled boxes and $0$'s to empty boxes.
\end{defn}

\subsubsection{The Box-Ball Evolution}\label{bbesubsec}
A simple evolution rule is provided for the box-ball dynamics:\index{Basic Box-Ball Evolution}\index{Box-Ball System}
\begin{enumerate}[(1)]
\item Take the left-most ball that has not been moved and move it to the left-most empty box to its right.
\item Repeat (1) until all balls have been moved precisely once.
\end{enumerate}

\n Since the algorithm requires one to know which balls have been moved, we can, without technically changing the algorithm, introduce a colour-coding based on whether balls have moved or not. Balls will be blue until they have moved, after which they will become red. When all balls are red, the colours should be reset to blue, ready for the next time step. Or, equivalently, a $0$-th step of colouring all balls blue should be prescribed. We will use the latter for a minor benefit in brevity. Below is an example of the evolution with this colour-coding, with each ball move separated into a sub-step:

\begin{figure}[H]
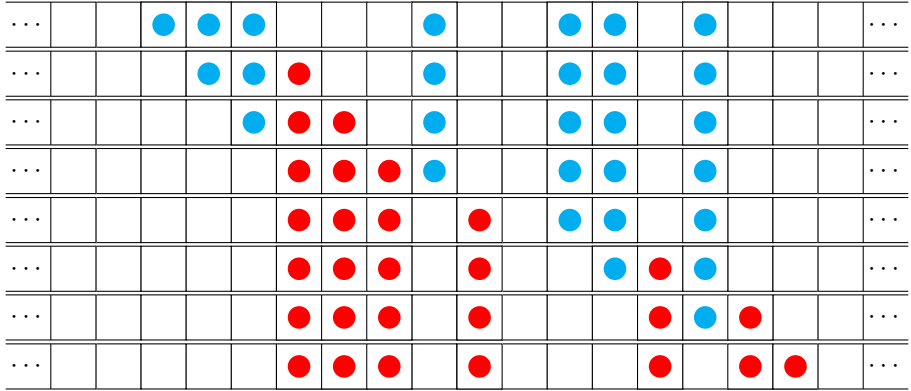

\centering
\tikz[scale=0.6]{
\foreach \x in {0,1,2,3,4,5,6,7,8,9,10,11,12,13,14,15}
{\draw[fill=white]  (\x,3) -- (\x+1,3) -- (\x+1,4) -- (\x,4) -- cycle;			
}
\foreach \x in {1,2,3,7,10,11,13}
{\draw[fill=white]  (\x,3) -- (\x+1,3) -- (\x+1,4) -- (\x,4) -- cycle;			
\fill[cyan] (\x+0.5,3.5) circle (0.25);
}
\foreach \x in {}
{\draw[fill=white]  (\x,3) -- (\x+1,3) -- (\x+1,4) -- (\x,4) -- cycle;			
\fill[red] (\x+0.5,3.5) circle (0.25);
}
\foreach \x in {16}
{\draw[fill=white,white]  (\x,3) -- (\x+2,3) -- (\x+2,4) -- (\x,4) -- cycle;
\draw[-] (\x,3) -- (\x,4);
\draw[-] (\x,3) -- (\x+2,3);
\draw[-] (\x,4) -- (\x+2,4);
\draw[-] (\x+1,3) -- (\x+1,4);
\node at (\x+1.5,3.5) {$\cdots$};
}
\foreach \x in {0}
{\draw[fill=white,white]  (\x,3) -- (\x-2,3) -- (\x-2,4) -- (\x,4) -- cycle;
\draw[-] (\x,3) -- (\x,4);
\draw[-] (\x,3) -- (\x-2,3);
\draw[-] (\x,4) -- (\x-2,4);
\draw[-] (\x-1,3) -- (\x-1,4);
\node at (\x-1.5,3.5) {$\cdots$};
}
}
\tikz[scale=0.6]{
\foreach \x in {0,1,2,3,4,5,6,7,8,9,10,11,12,13,14,15}
{\draw[fill=white]  (\x,3) -- (\x+1,3) -- (\x+1,4) -- (\x,4) -- cycle;			
}
\foreach \x in {2,3,7,10,11,13}
{\draw[fill=white]  (\x,3) -- (\x+1,3) -- (\x+1,4) -- (\x,4) -- cycle;			
\fill[cyan] (\x+0.5,3.5) circle (0.25);
}
\foreach \x in {4}
{\draw[fill=white]  (\x,3) -- (\x+1,3) -- (\x+1,4) -- (\x,4) -- cycle;			
\fill[red] (\x+0.5,3.5) circle (0.25);
}
\foreach \x in {16}
{\draw[fill=white,white]  (\x,3) -- (\x+2,3) -- (\x+2,4) -- (\x,4) -- cycle;
\draw[-] (\x,3) -- (\x,4);
\draw[-] (\x,3) -- (\x+2,3);
\draw[-] (\x,4) -- (\x+2,4);
\draw[-] (\x+1,3) -- (\x+1,4);
\node at (\x+1.5,3.5) {$\cdots$};
}
\foreach \x in {0}
{\draw[fill=white,white]  (\x,3) -- (\x-2,3) -- (\x-2,4) -- (\x,4) -- cycle;
\draw[-] (\x,3) -- (\x,4);
\draw[-] (\x,3) -- (\x-2,3);
\draw[-] (\x,4) -- (\x-2,4);
\draw[-] (\x-1,3) -- (\x-1,4);
\node at (\x-1.5,3.5) {$\cdots$};
}
}
\tikz[scale=0.6]{
\foreach \x in {0,1,2,3,4,5,6,7,8,9,10,11,12,13,14,15}
{\draw[fill=white]  (\x,3) -- (\x+1,3) -- (\x+1,4) -- (\x,4) -- cycle;			
}
\foreach \x in {3,7,10,11,13}
{\draw[fill=white]  (\x,3) -- (\x+1,3) -- (\x+1,4) -- (\x,4) -- cycle;			
\fill[cyan] (\x+0.5,3.5) circle (0.25);
}
\foreach \x in {4,5}
{\draw[fill=white]  (\x,3) -- (\x+1,3) -- (\x+1,4) -- (\x,4) -- cycle;			
\fill[red] (\x+0.5,3.5) circle (0.25);
}
\foreach \x in {16}
{\draw[fill=white,white]  (\x,3) -- (\x+2,3) -- (\x+2,4) -- (\x,4) -- cycle;
\draw[-] (\x,3) -- (\x,4);
\draw[-] (\x,3) -- (\x+2,3);
\draw[-] (\x,4) -- (\x+2,4);
\draw[-] (\x+1,3) -- (\x+1,4);
\node at (\x+1.5,3.5) {$\cdots$};
}
\foreach \x in {0}
{\draw[fill=white,white]  (\x,3) -- (\x-2,3) -- (\x-2,4) -- (\x,4) -- cycle;
\draw[-] (\x,3) -- (\x,4);
\draw[-] (\x,3) -- (\x-2,3);
\draw[-] (\x,4) -- (\x-2,4);
\draw[-] (\x-1,3) -- (\x-1,4);
\node at (\x-1.5,3.5) {$\cdots$};
}
}
\tikz[scale=0.6]{
\foreach \x in {0,1,2,3,4,5,6,7,8,9,10,11,12,13,14,15}
{\draw[fill=white]  (\x,3) -- (\x+1,3) -- (\x+1,4) -- (\x,4) -- cycle;			
}
\foreach \x in {7,10,11,13}
{\draw[fill=white]  (\x,3) -- (\x+1,3) -- (\x+1,4) -- (\x,4) -- cycle;			
\fill[cyan] (\x+0.5,3.5) circle (0.25);
}
\foreach \x in {4,5,6}
{\draw[fill=white]  (\x,3) -- (\x+1,3) -- (\x+1,4) -- (\x,4) -- cycle;			
\fill[red] (\x+0.5,3.5) circle (0.25);
}
\foreach \x in {16}
{\draw[fill=white,white]  (\x,3) -- (\x+2,3) -- (\x+2,4) -- (\x,4) -- cycle;
\draw[-] (\x,3) -- (\x,4);
\draw[-] (\x,3) -- (\x+2,3);
\draw[-] (\x,4) -- (\x+2,4);
\draw[-] (\x+1,3) -- (\x+1,4);
\node at (\x+1.5,3.5) {$\cdots$};
}
\foreach \x in {0}
{\draw[fill=white,white]  (\x,3) -- (\x-2,3) -- (\x-2,4) -- (\x,4) -- cycle;
\draw[-] (\x,3) -- (\x,4);
\draw[-] (\x,3) -- (\x-2,3);
\draw[-] (\x,4) -- (\x-2,4);
\draw[-] (\x-1,3) -- (\x-1,4);
\node at (\x-1.5,3.5) {$\cdots$};
}
}
\tikz[scale=0.6]{
\foreach \x in {0,1,2,3,4,5,6,7,8,9,10,11,12,13,14,15}
{\draw[fill=white]  (\x,3) -- (\x+1,3) -- (\x+1,4) -- (\x,4) -- cycle;			
}
\foreach \x in {10,11,13}
{\draw[fill=white]  (\x,3) -- (\x+1,3) -- (\x+1,4) -- (\x,4) -- cycle;			
\fill[cyan] (\x+0.5,3.5) circle (0.25);
}
\foreach \x in {4,5,6,8}
{\draw[fill=white]  (\x,3) -- (\x+1,3) -- (\x+1,4) -- (\x,4) -- cycle;			
\fill[red] (\x+0.5,3.5) circle (0.25);
}
\foreach \x in {16}
{\draw[fill=white,white]  (\x,3) -- (\x+2,3) -- (\x+2,4) -- (\x,4) -- cycle;
\draw[-] (\x,3) -- (\x,4);
\draw[-] (\x,3) -- (\x+2,3);
\draw[-] (\x,4) -- (\x+2,4);
\draw[-] (\x+1,3) -- (\x+1,4);
\node at (\x+1.5,3.5) {$\cdots$};
}
\foreach \x in {0}
{\draw[fill=white,white]  (\x,3) -- (\x-2,3) -- (\x-2,4) -- (\x,4) -- cycle;
\draw[-] (\x,3) -- (\x,4);
\draw[-] (\x,3) -- (\x-2,3);
\draw[-] (\x,4) -- (\x-2,4);
\draw[-] (\x-1,3) -- (\x-1,4);
\node at (\x-1.5,3.5) {$\cdots$};
}
}
\tikz[scale=0.6]{
\foreach \x in {0,1,2,3,4,5,6,7,8,9,10,11,12,13,14,15}
{\draw[fill=white]  (\x,3) -- (\x+1,3) -- (\x+1,4) -- (\x,4) -- cycle;			
}
\foreach \x in {11,13}
{\draw[fill=white]  (\x,3) -- (\x+1,3) -- (\x+1,4) -- (\x,4) -- cycle;			
\fill[cyan] (\x+0.5,3.5) circle (0.25);
}
\foreach \x in {4,5,6,8,12}
{\draw[fill=white]  (\x,3) -- (\x+1,3) -- (\x+1,4) -- (\x,4) -- cycle;			
\fill[red] (\x+0.5,3.5) circle (0.25);
}
\foreach \x in {16}
{\draw[fill=white,white]  (\x,3) -- (\x+2,3) -- (\x+2,4) -- (\x,4) -- cycle;
\draw[-] (\x,3) -- (\x,4);
\draw[-] (\x,3) -- (\x+2,3);
\draw[-] (\x,4) -- (\x+2,4);
\draw[-] (\x+1,3) -- (\x+1,4);
\node at (\x+1.5,3.5) {$\cdots$};
}
\foreach \x in {0}
{\draw[fill=white,white]  (\x,3) -- (\x-2,3) -- (\x-2,4) -- (\x,4) -- cycle;
\draw[-] (\x,3) -- (\x,4);
\draw[-] (\x,3) -- (\x-2,3);
\draw[-] (\x,4) -- (\x-2,4);
\draw[-] (\x-1,3) -- (\x-1,4);
\node at (\x-1.5,3.5) {$\cdots$};
}
}
\tikz[scale=0.6]{
\foreach \x in {0,1,2,3,4,5,6,7,8,9,10,11,12,13,14,15}
{\draw[fill=white]  (\x,3) -- (\x+1,3) -- (\x+1,4) -- (\x,4) -- cycle;			
}
\foreach \x in {13}
{\draw[fill=white]  (\x,3) -- (\x+1,3) -- (\x+1,4) -- (\x,4) -- cycle;			
\fill[cyan] (\x+0.5,3.5) circle (0.25);
}
\foreach \x in {4,5,6,8,12,14}
{\draw[fill=white]  (\x,3) -- (\x+1,3) -- (\x+1,4) -- (\x,4) -- cycle;			
\fill[red] (\x+0.5,3.5) circle (0.25);
}
\foreach \x in {16}
{\draw[fill=white,white]  (\x,3) -- (\x+2,3) -- (\x+2,4) -- (\x,4) -- cycle;
\draw[-] (\x,3) -- (\x,4);
\draw[-] (\x,3) -- (\x+2,3);
\draw[-] (\x,4) -- (\x+2,4);
\draw[-] (\x+1,3) -- (\x+1,4);
\node at (\x+1.5,3.5) {$\cdots$};
}
\foreach \x in {0}
{\draw[fill=white,white]  (\x,3) -- (\x-2,3) -- (\x-2,4) -- (\x,4) -- cycle;
\draw[-] (\x,3) -- (\x,4);
\draw[-] (\x,3) -- (\x-2,3);
\draw[-] (\x,4) -- (\x-2,4);
\draw[-] (\x-1,3) -- (\x-1,4);
\node at (\x-1.5,3.5) {$\cdots$};
}
}
\tikz[scale=0.6]{
\foreach \x in {0,1,2,3,4,5,6,7,8,9,10,11,12,13,14,15}
{\draw[fill=white]  (\x,3) -- (\x+1,3) -- (\x+1,4) -- (\x,4) -- cycle;			
}
\foreach \x in {}
{\draw[fill=white]  (\x,3) -- (\x+1,3) -- (\x+1,4) -- (\x,4) -- cycle;			
\fill[cyan] (\x+0.5,3.5) circle (0.25);
}
\foreach \x in {4,5,6,8,12,14,15}
{\draw[fill=white]  (\x,3) -- (\x+1,3) -- (\x+1,4) -- (\x,4) -- cycle;			
\fill[red] (\x+0.5,3.5) circle (0.25);
}
\foreach \x in {16}
{\draw[fill=white,white]  (\x,3) -- (\x+2,3) -- (\x+2,4) -- (\x,4) -- cycle;
\draw[-] (\x,3) -- (\x,4);
\draw[-] (\x,3) -- (\x+2,3);
\draw[-] (\x,4) -- (\x+2,4);
\draw[-] (\x+1,3) -- (\x+1,4);
\node at (\x+1.5,3.5) {$\cdots$};
}
\foreach \x in {0}
{\draw[fill=white,white]  (\x,3) -- (\x-2,3) -- (\x-2,4) -- (\x,4) -- cycle;
\draw[-] (\x,3) -- (\x,4);
\draw[-] (\x,3) -- (\x-2,3);
\draw[-] (\x,4) -- (\x-2,4);
\draw[-] (\x-1,3) -- (\x-1,4);
\node at (\x-1.5,3.5) {$\cdots$};
}
}
\caption{A box-ball system time evolution (one time step).}\label{firstbbsexample}
\end{figure}	

\subsubsection{The Robinson-Schensted-Knuth Correspondence}\label{introsecrsk}

RSK is an algorithm for the direct sum decomposition of tensor products of representations of the unitary groups $U(k)$. This general area is referred to as Schur-Weyl theory \cite{bib:kuperberg}. It has important applications for quantum-many-body theory and quantum field theory. \\

\n The fundamental building block of RSK is {\it Schensted insertion}. Its description is a bit involved (and will be more fully developed in Section \ref{chapterrskschensted}) but for our introductory purposes here it will suffice to point out that Schensted insertion can be reduced to a discrete evolution on
$$\mathcal{R} := \bigcup_{n\in\N} \mathcal{R}_n,$$
where $\mathcal{R}_n := \N_0^n\x \N_0^n$ and $\N_0:=\N\cup\{0\}$. The foundation for Schensted insertion is a prescription for taking an input pair of sequences $$(\mb{a},\mb{x})=((a_1,\ldots,a_n),(x_1,\ldots,x_n)),$$ 
which encode words (weakly increasing sequences of positive integers) from an alphabet $\{1,\ldots,n\}$,  and transforming the input sequences into an output pair of sequences $$(\mb{b},\mb{y})=((b_1,\ldots,b_n),(y_1,\ldots,y_n))$$
which encode the result of performing Schensted insertion. This encoding is introduced in full detail in Section \ref{schenstedinsertionsectionrsk}.\\

\n This dynamic evolution on $\mathcal{R}$ (pairs of $n$-tuples) is what, going forward, we will refer to as 
\begin{equation}
    \text{RSK}:\mathcal{R}\to\mathcal{R}. \label{rskdefndynamic}
\end{equation}

\subsection{Statement of Results}

There is a natural connection between BBS and RSK given by a process known as tropicalization (or Maslov dequantization) that will be described in Section \ref{introsectrop}. However, this connection is not a precise correspondence. The main point of this paper is to address that problem. We do this by means of introducing a (ghost) background, or environment, against which the basic box-ball system moves and interacts. Once this is developed, our main result may indeed be summarized as showing that the following diagram (from Figure \ref{commdiagrskgbbs}) commutes: 

\begin{figure}[H]
\centering
\tikz[scale=1.5]{
\node (A) at (0,0) {GBBS};
\node (B) at (2,0) {GBBS};
\node (C) at (0,-2) {$\mathcal{B}^0$};
\node (D) at (2,-2) {$\mathcal{B}^0$};
\node (E) at (0,-4) {$\mathcal{R}$};
\node (F) at (2,-4) {$\mathcal{R}$};
\draw[->] (A) -- (B) node[midway,above]  {$\hat{\varrho}$};
\draw[->] (E) -- (F) node[midway,above]  {RSK};
\draw[->] (C) -- (A) node[midway,left]  {$C^{-1}$};
\draw[->] (B) -- (D) node[midway,right]  {$C$};
\draw[->] (E) -- (C) node[midway,left]  {$\phi_{\text{RSK}\to\text{BBS}}$};
\draw[->] (D) -- (F) node[midway,right]  {$\phi_{\text{BBS}\to\text{RSK}}$};
}
\end{figure}

\n where the lower arrow denotes the RSK dynamics (\ref{rskdefndynamic}). The map $\hat{\varrho}$ is the dynamics for our extended \textit{ghost-box-ball system}, which is constructed in Section \ref{chaptergbbs}. Finally, $\mathcal{B}^0$ represents the \textit{precise} coordinatisation linking the dynamics of $\text{GBBS}$ to that of the algorithm of RSK. This is developed in
Section \ref{rskcoordbbs}.\\

\n In fact, we will see that much more is true: our construction yields a detailed stage-by-stage correspondence between the fine structure of the respective systems.\\

\n Prior to this work, a relationship between RSK and an advanced version of the box-ball system was developed by Fukuda \cite{bib:fukuda}. However, this advanced box-ball system requires various extra features not automatically possessed by the dequantisation mentioned above. As such, our system is simpler, and this simplicity may open the way to connections with continuous space - continuous time systems that have a deeper relation to Lie theory and classical solitonic structures. Some discussion of that will be presented in Sections \ref{chapterextensions} - \ref{chapterconclusions}. We will also discuss compare our work with that of \cite{bib:fukuda} in Section \ref{fukudasecfourrems}.\\

\n In the remainder of this paper, Sections \ref{secnamebbs} - \ref{chapterrskschensted} provide the detailed background for the main ingredients we have just surveyed. The precise statement of our results and their proofs is carried out in Section \ref{chaptergbbs}. Natural extensions of those results are presented in Section \ref{chapterextensions}, where we discuss the intrinsic dynamics of the ghost-box-ball system, as well as its phase shift phenomenon. Further motivation for what has been done here is presented in Section \ref{chapterconclusions}.

\section{The Box-Ball System}\label{secnamebbs}

\subsection{Soliton Behaviour and the Sorting Property}

\n The box-ball system is sometimes referred to as a soliton cellular automaton. To appreciate this reference, we think of an entire box-ball configuration as being the soliton. In the classical setting, a soliton is thought of as being composed of masses (or pulses) that are nonlinearly related. In the box-ball setting, a ``mass'' corresponds to a consecutive sequence of balls. One may observe (see below) that such a block travels with velocity equal to the number of balls in it, so that larger blocks have velocity greater than smaller blocks. As with classical soliton masses, during the course of its evolution, the blocks may collide, and temporarily change their sizes. However, asymptotically in both forward and backward (discrete) time ($t$), the sizes of blocks comprising the soliton are the same. We will therefore refer to such a configuration as an \textit{$N$-soliton}, if the total number of blocks asymptotically is $N$.\\

\n After blocks collide, they come out of the collision ordered with the longer blocks ahead of smaller blocks, but having a \textit{phase shift} due to the nonlinearity. By phase shift here, we mean the difference between where the block ends up after the collision and where the block would have been if it were not for the collision. For now, we take this for granted. A more detailed analysis will be given in Section \ref{bbsphsshftsecmn}, along with a conjectured, explicit formula for the phase shift.\\[4pt]

\n In the following figure, we illustrate how the blocks become ordered after sufficiently many time evolutions. Once sorted, they travel with their respective velocities, never to collide again.

\begin{figure}[H]
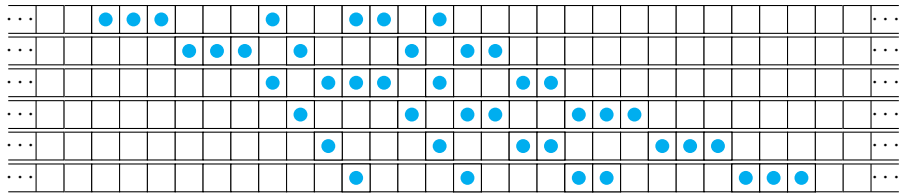

\centering
\tikz[scale=0.37]{
\foreach \x in {0,1,2,3,4,5,6,7,8,9,10,11,12,13,14,15,16,17,18,19,20,21,22,23,24,25,26,27}
{\draw[fill=white]  (\x,3) -- (\x+1,3) -- (\x+1,4) -- (\x,4) -- cycle;			
}
\foreach \x in {1,2,3,7,10,11,13}
{\draw[fill=white]  (\x,3) -- (\x+1,3) -- (\x+1,4) -- (\x,4) -- cycle;			
\fill[cyan] (\x+0.5,3.5) circle (0.25);
}
\foreach \x in {}
{\draw[fill=white]  (\x,3) -- (\x+1,3) -- (\x+1,4) -- (\x,4) -- cycle;			
\fill[red] (\x+0.5,3.5) circle (0.25);
}
\foreach \x in {28}
{\draw[fill=white,white]  (\x,3) -- (\x+2,3) -- (\x+2,4) -- (\x,4) -- cycle;
\draw[-] (\x,3) -- (\x,4);
\draw[-] (\x,3) -- (\x+2,3);
\draw[-] (\x,4) -- (\x+2,4);
\draw[-] (\x+1,3) -- (\x+1,4);
\node at (\x+1.6,3.5) {\small$\cdots$};
}
\foreach \x in {0}
{\draw[fill=white,white]  (\x,3) -- (\x-2,3) -- (\x-2,4) -- (\x,4) -- cycle;
\draw[-] (\x,3) -- (\x,4);
\draw[-] (\x,3) -- (\x-2,3);
\draw[-] (\x,4) -- (\x-2,4);
\draw[-] (\x-1,3) -- (\x-1,4);
\node at (\x-1.5,3.5) {\small$\cdots$};
}
}
\tikz[scale=0.37]{
\foreach \x in {0,1,2,3,4,5,6,7,8,9,10,11,12,13,14,15,16,17,18,19,20,21,22,23,24,25,26,27}
{\draw[fill=white]  (\x,3) -- (\x+1,3) -- (\x+1,4) -- (\x,4) -- cycle;			
}
\foreach \x in {4,5,6,8,12,14,15}
{\draw[fill=white]  (\x,3) -- (\x+1,3) -- (\x+1,4) -- (\x,4) -- cycle;			
\fill[cyan] (\x+0.5,3.5) circle (0.25);
}
\foreach \x in {}
{\draw[fill=white]  (\x,3) -- (\x+1,3) -- (\x+1,4) -- (\x,4) -- cycle;			
\fill[red] (\x+0.5,3.5) circle (0.25);
}
\foreach \x in {28}
{\draw[fill=white,white]  (\x,3) -- (\x+2,3) -- (\x+2,4) -- (\x,4) -- cycle;
\draw[-] (\x,3) -- (\x,4);
\draw[-] (\x,3) -- (\x+2,3);
\draw[-] (\x,4) -- (\x+2,4);
\draw[-] (\x+1,3) -- (\x+1,4);
\node at (\x+1.6,3.5) {\small$\cdots$};
}
\foreach \x in {0}
{\draw[fill=white,white]  (\x,3) -- (\x-2,3) -- (\x-2,4) -- (\x,4) -- cycle;
\draw[-] (\x,3) -- (\x,4);
\draw[-] (\x,3) -- (\x-2,3);
\draw[-] (\x,4) -- (\x-2,4);
\draw[-] (\x-1,3) -- (\x-1,4);
\node at (\x-1.5,3.5) {\small$\cdots$};
}
}
\tikz[scale=0.37]{
\foreach \x in {0,1,2,3,4,5,6,7,8,9,10,11,12,13,14,15,16,17,18,19,20,21,22,23,24,25,26,27}
{\draw[fill=white]  (\x,3) -- (\x+1,3) -- (\x+1,4) -- (\x,4) -- cycle;			
}
\foreach \x in {7,9,10,11,13,16,17}
{\draw[fill=white]  (\x,3) -- (\x+1,3) -- (\x+1,4) -- (\x,4) -- cycle;			
\fill[cyan] (\x+0.5,3.5) circle (0.25);
}
\foreach \x in {}
{\draw[fill=white]  (\x,3) -- (\x+1,3) -- (\x+1,4) -- (\x,4) -- cycle;			
\fill[red] (\x+0.5,3.5) circle (0.25);
}
\foreach \x in {28}
{\draw[fill=white,white]  (\x,3) -- (\x+2,3) -- (\x+2,4) -- (\x,4) -- cycle;
\draw[-] (\x,3) -- (\x,4);
\draw[-] (\x,3) -- (\x+2,3);
\draw[-] (\x,4) -- (\x+2,4);
\draw[-] (\x+1,3) -- (\x+1,4);
\node at (\x+1.6,3.5) {\small$\cdots$};
}
\foreach \x in {0}
{\draw[fill=white,white]  (\x,3) -- (\x-2,3) -- (\x-2,4) -- (\x,4) -- cycle;
\draw[-] (\x,3) -- (\x,4);
\draw[-] (\x,3) -- (\x-2,3);
\draw[-] (\x,4) -- (\x-2,4);
\draw[-] (\x-1,3) -- (\x-1,4);
\node at (\x-1.5,3.5) {\small$\cdots$};
}
}
\tikz[scale=0.37]{
\foreach \x in {0,1,2,3,4,5,6,7,8,9,10,11,12,13,14,15,16,17,18,19,20,21,22,23,24,25,26,27}
{\draw[fill=white]  (\x,3) -- (\x+1,3) -- (\x+1,4) -- (\x,4) -- cycle;			
}
\foreach \x in {8,12,14,15,18,19,20}
{\draw[fill=white]  (\x,3) -- (\x+1,3) -- (\x+1,4) -- (\x,4) -- cycle;			
\fill[cyan] (\x+0.5,3.5) circle (0.25);
}
\foreach \x in {}
{\draw[fill=white]  (\x,3) -- (\x+1,3) -- (\x+1,4) -- (\x,4) -- cycle;			
\fill[red] (\x+0.5,3.5) circle (0.25);
}
\foreach \x in {28}
{\draw[fill=white,white]  (\x,3) -- (\x+2,3) -- (\x+2,4) -- (\x,4) -- cycle;
\draw[-] (\x,3) -- (\x,4);
\draw[-] (\x,3) -- (\x+2,3);
\draw[-] (\x,4) -- (\x+2,4);
\draw[-] (\x+1,3) -- (\x+1,4);
\node at (\x+1.6,3.5) {\small$\cdots$};
}
\foreach \x in {0}
{\draw[fill=white,white]  (\x,3) -- (\x-2,3) -- (\x-2,4) -- (\x,4) -- cycle;
\draw[-] (\x,3) -- (\x,4);
\draw[-] (\x,3) -- (\x-2,3);
\draw[-] (\x,4) -- (\x-2,4);
\draw[-] (\x-1,3) -- (\x-1,4);
\node at (\x-1.5,3.5) {\small$\cdots$};
}
}
\tikz[scale=0.37]{
\foreach \x in {0,1,2,3,4,5,6,7,8,9,10,11,12,13,14,15,16,17,18,19,20,21,22,23,24,25,26,27}
{\draw[fill=white]  (\x,3) -- (\x+1,3) -- (\x+1,4) -- (\x,4) -- cycle;			
}
\foreach \x in {9,13,16,17,21,22,23}
{\draw[fill=white]  (\x,3) -- (\x+1,3) -- (\x+1,4) -- (\x,4) -- cycle;			
\fill[cyan] (\x+0.5,3.5) circle (0.25);
}
\foreach \x in {}
{\draw[fill=white]  (\x,3) -- (\x+1,3) -- (\x+1,4) -- (\x,4) -- cycle;			
\fill[red] (\x+0.5,3.5) circle (0.25);
}
\foreach \x in {28}
{\draw[fill=white,white]  (\x,3) -- (\x+2,3) -- (\x+2,4) -- (\x,4) -- cycle;
\draw[-] (\x,3) -- (\x,4);
\draw[-] (\x,3) -- (\x+2,3);
\draw[-] (\x,4) -- (\x+2,4);
\draw[-] (\x+1,3) -- (\x+1,4);
\node at (\x+1.6,3.5) {\small$\cdots$};
}
\foreach \x in {0}
{\draw[fill=white,white]  (\x,3) -- (\x-2,3) -- (\x-2,4) -- (\x,4) -- cycle;
\draw[-] (\x,3) -- (\x,4);
\draw[-] (\x,3) -- (\x-2,3);
\draw[-] (\x,4) -- (\x-2,4);
\draw[-] (\x-1,3) -- (\x-1,4);
\node at (\x-1.5,3.5) {\small$\cdots$};
}
}
\tikz[scale=0.37]{
\foreach \x in {0,1,2,3,4,5,6,7,8,9,10,11,12,13,14,15,16,17,18,19,20,21,22,23,24,25,26,27}
{\draw[fill=white]  (\x,3) -- (\x+1,3) -- (\x+1,4) -- (\x,4) -- cycle;			
}
\foreach \x in {10,14,18,19,24,25,26}
{\draw[fill=white]  (\x,3) -- (\x+1,3) -- (\x+1,4) -- (\x,4) -- cycle;			
\fill[cyan] (\x+0.5,3.5) circle (0.25);
}
\foreach \x in {}
{\draw[fill=white]  (\x,3) -- (\x+1,3) -- (\x+1,4) -- (\x,4) -- cycle;			
\fill[red] (\x+0.5,3.5) circle (0.25);
}
\foreach \x in {28}
{\draw[fill=white,white]  (\x,3) -- (\x+2,3) -- (\x+2,4) -- (\x,4) -- cycle;
\draw[-] (\x,3) -- (\x,4);
\draw[-] (\x,3) -- (\x+2,3);
\draw[-] (\x,4) -- (\x+2,4);
\draw[-] (\x+1,3) -- (\x+1,4);
\node at (\x+1.6,3.5) {\small$\cdots$};
}
\foreach \x in {0}
{\draw[fill=white,white]  (\x,3) -- (\x-2,3) -- (\x-2,4) -- (\x,4) -- cycle;
\draw[-] (\x,3) -- (\x,4);
\draw[-] (\x,3) -- (\x-2,3);
\draw[-] (\x,4) -- (\x-2,4);
\draw[-] (\x-1,3) -- (\x-1,4);
\node at (\x-1.5,3.5) {\small$\cdots$};
}
}
\caption{The sorting property of the box-ball system.}\label{sortingpropbbs}
\end{figure}

\subsubsection{Invariants of the Box-Ball System}\label{invsofbbs}
Invariants of the box-ball system may be expressed in terms of combinatorial structures known as \textit{Young diagrams}.

\begin{defn}\index{Young Diagram} \index{Ferrers Diagram}
Let $n\in\N$ and let $\lambda=(\lambda_1,\lambda_2,\ldots,\lambda_k)$ be a partition of $n$, \textit{i.e.} $\lambda_i\in\N$ for each $i$, and $\lambda_1\geq\lambda_2\geq\cdots\geq\lambda_k$ and $\sum \lambda_i = n$. Associated to $\lambda$ is the Young diagram (or Ferrers diagram) of shape $\lambda$ which is composed of $\lambda_1$ boxes in the first row, $\lambda_2$ boxes in the second row, $\ldots$, and $\lambda_k$ boxes in the $k$-th row. The boxes are of equal size and aligned in a grid, justified to the left.
\end{defn}

\begin{ex}
The partition $\lambda=(5,3,3,2,1)$, which partitions $n=14$, has Young diagram: \vspace{0.2cm}$$\ydiagram{5,3, 3,2, 1}$$
\end{ex}

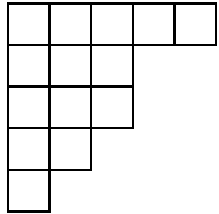
\n A Young diagram for a box-ball state can be described in terms of balls and boxes, but it is slightly more convenient to represent the box-ball system as a sequence of 0's (for empty boxes) and 1's (for boxes with balls), which is made explicit in \cite{bib:tts} and \cite{bib:tokihiro}. The construction goes as follows:\index{Conserved Shape of the BBS}
\begin{itemize}
\item Let $p_1$ be the number of $10$'s in the sequence.
\item Eliminate all of these $10$'s, and let $p_2$ be the number of $10$'s in the resulting sequence.
\item Repeat this process until no $10$'s remain.
\item The sequence $(p_1,p_2,\ldots)$ is weakly decreasing, hence a partition of the number of balls. It can be represented by a Young diagram by taking the $j$\textsuperscript{th} column to have $p_j$ boxes.
\end{itemize}

\n For example, representing the first box-ball system in Figure \ref{sortingpropbbs} as a sequence of $1$'s and $0$'s, as described in Definition \ref{bbsoneszeroesdefn}, we have
$$\cdots ~~0 ~~0 ~~1 ~~1 ~\fbox{1 ~0} ~0 ~~0 ~\fbox{1~ 0} ~0~~ 1 ~\fbox{1 0}  ~\fbox{1 0}~ 0~~0 ~~ \cdots$$
In the above, there are four instances of $10$'s in the sequence, so $p_1=4$. Removing those instances of $10$'s, we obtain
$$\cdots ~~0 ~~0 ~~1 ~\fbox{1 0}~0 ~~0~\fbox{1 0}~ 0 ~~ \cdots$$
to see that $p_2=2$. Next, we have
$$\cdots ~~0 ~~0 ~\fbox{1 0}~0~~ 0 ~~ \cdots$$
there is only one $10$ here, so $p_3=1$. The last removal of the $10$ above yields a sequence with no $1$'s:
$$\cdots ~~0 ~~0 ~~0~~ 0 ~~ \cdots$$
Thus, the process terminates and we have the sequence $(4,2,1)$, which we represent in the following Young diagram
\begin{figure}[H]
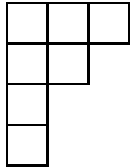

\centering
\ydiagram{3,2,1,1}
\caption{The invariant shape of the box-ball system(s) in Figure \ref{sortingpropbbs}}
\end{figure}

\begin{thm}\label{ttstheoreminvariantshape}
\cite{bib:tts} The sequence of $p_i$'s are time invariants of the box-ball evolution. Equivalently, the Young diagram is invariant under the box-ball evolution.
\end{thm}

A proof of Theorem \ref{ttstheoreminvariantshape} is given in Section 5 of \cite{bib:tts}, using combinatorial techniques.

\begin{rem}
The row lengths then act as a \textit{signature} for the system: if the column lengths of the Young diagram are the $p_i$'s, then the row lengths give the asymptotic lengths of the blocks. One can see this heuristically by noting that as time goes to $\pm \infty$, the blocks will be sufficiently separated by empty boxes so that each block provides precisely one ``$10$'' for each particle comprising it.
\end{rem}

\subsubsection{Coordinates on the Box-Ball System}
Suppose at time $t$, one has $N$ blocks in the soliton. Let $Q_1^t$, $Q_2^t$, $\ldots$, $Q_N^t$ denote the lengths of these blocks, taken from left to right. Let $W_1^t$, $W_2^t$, $\ldots$, $W_{N-1}^t$ denote the lengths of the sets of empty boxes between the $N$ sets of filled boxes, again taken from left to right. Lastly, let $W_0^t$ and $W_N^t$ be formally defined to be $\infty$, reflecting the fact that the empty boxes continue infinitely in both directions.\\[4pt]

\n The following theorem gives evolution equations for these coordinates. They can be found, for example, in \cite{bib:tokihiro}.

\begin{thm}\label{thmbbscoords}\index{Box-Ball Coordinate Dynamics}(\cite{bib:tokihiro})
The coordinates $(W_0^t,Q_1^t,W_1^t,\ldots,Q_N^t,W_N^t)$ evolve under the box ball dynamics according to
\begin{align}
W_0^{t+1}&=W_N^{t+1}=\infty\\
W_i^{t+1}&=Q_{i+1}^t+W_i^t-Q_i^{t+1},~~~~~~~~~~~~~~~~~~~i=1,\ldots,N-1\label{Witplusoneeqnbbs}\\
Q_i^{t+1}&=\min\left(W_i^{t},\sum_{j=1}^i Q_j^t-\sum_{j=1}^{i-1}
Q_j^{t+1}\right),~~~~~i=1,\ldots,N,\label{Qitplusoneeqnbbs}
\end{align}
\end{thm}

\begin{rem}
By Theorem \ref{ttstheoreminvariantshape}, $p_1$ is invariant under the box-ball evolution. Since $p_1$ is the number of blocks of a box-ball state, it follows that the number of blocks is invariant under this evolution. Pairing this with Theorem \ref{thmbbscoords}, we see each $W_i^t>0$ for each $i$ and for all time. Furthermore, since there are always $N$ blocks, each $Q_i^t>0$ (by definition of a block). 
\end{rem}

\begin{ex}
Take the initial state in Figure \ref{firstbbsexample}:
\begin{figure}[H]
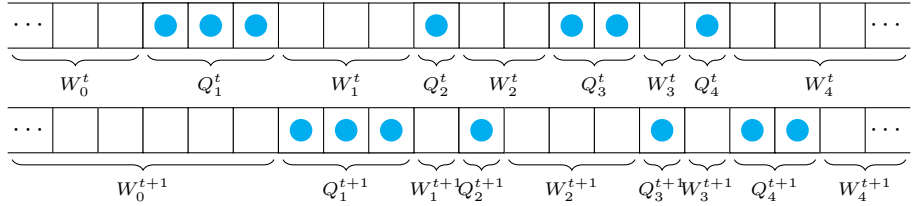

\centering
\tikz[scale=0.6]{
\foreach \x in {0,1,2,3,4,5,6,7,8,9,10,11,12,13,14,15}
{\draw[fill=white]  (\x,3) -- (\x+1,3) -- (\x+1,4) -- (\x,4) -- cycle;			
}
\foreach \x in {1,2,3,7,10,11,13}
{\draw[fill=white]  (\x,3) -- (\x+1,3) -- (\x+1,4) -- (\x,4) -- cycle;			
\fill[cyan] (\x+0.5,3.5) circle (0.25);
}
\foreach \x in {}
{\draw[fill=white]  (\x,3) -- (\x+1,3) -- (\x+1,4) -- (\x,4) -- cycle;			
\fill[red] (\x+0.5,3.5) circle (0.25);
}
\foreach \x in {16}
{\draw[fill=white,white]  (\x,3) -- (\x+2,3) -- (\x+2,4) -- (\x,4) -- cycle;
\draw[-] (\x,3) -- (\x,4);
\draw[-] (\x,3) -- (\x+2,3);
\draw[-] (\x,4) -- (\x+2,4);
\draw[-] (\x+1,3) -- (\x+1,4);
\node at (\x+1.5,3.5) {$\cdots$};
}
\foreach \x in {0}
{\draw[fill=white,white]  (\x,3) -- (\x-2,3) -- (\x-2,4) -- (\x,4) -- cycle;
\draw[-] (\x,3) -- (\x,4);
\draw[-] (\x,3) -- (\x-2,3);
\draw[-] (\x,4) -- (\x-2,4);
\draw[-] (\x-1,3) -- (\x-1,4);
\node at (\x-1.5,3.5) {$\cdots$};
}
\draw [decorate,decoration={brace,amplitude=4pt}] (0.9,2.85) -- (-1.9,2.85) node [black,midway,yshift=-0.4cm] {\scriptsize{$W_0^t$}};
\draw [decorate,decoration={brace,amplitude=4pt}] (3.9,2.85) -- (1.1,2.85) node [black,midway,yshift=-0.4cm] {\scriptsize{$Q_1^t$}};
\draw [decorate,decoration={brace,amplitude=4pt}] (6.9,2.85) -- (4.1,2.85) node [black,midway,yshift=-0.4cm] {\scriptsize{$W_1^t$}};
\draw [decorate,decoration={brace,amplitude=4pt}] (7.9,2.85) -- (7.1,2.85) node [black,midway,yshift=-0.4cm] {\scriptsize{$Q_2^t$}};
\draw [decorate,decoration={brace,amplitude=4pt}] (9.9,2.85) -- (8.1,2.85) node [black,midway,yshift=-0.4cm] {\scriptsize{$W_2^t$}};
\draw [decorate,decoration={brace,amplitude=4pt}] (11.9,2.85) -- (10.1,2.85) node [black,midway,yshift=-0.4cm] {\scriptsize{$Q_3^t$}};
\draw [decorate,decoration={brace,amplitude=4pt}] (12.9,2.85) -- (12.1,2.85) node [black,midway,yshift=-0.4cm] {\scriptsize{$W_3^t$}};
\draw [decorate,decoration={brace,amplitude=4pt}] (13.9,2.85) -- (13.1,2.85) node [black,midway,yshift=-0.4cm] {\scriptsize{$Q_4^t$}};
\draw [decorate,decoration={brace,amplitude=4pt}] (17.9,2.85) -- (14.1,2.85) node [black,midway,yshift=-0.4cm] {\scriptsize{$W_4^t$}};
}
\tikz[scale=0.6]{
\foreach \x in {0,1,2,3,4,5,6,7,8,9,10,11,12,13,14,15}
{\draw[fill=white]  (\x,3) -- (\x+1,3) -- (\x+1,4) -- (\x,4) -- cycle;			
}
\foreach \x in {4,5,6,8,12,14,15}
{\draw[fill=white]  (\x,3) -- (\x+1,3) -- (\x+1,4) -- (\x,4) -- cycle;			
\fill[cyan] (\x+0.5,3.5) circle (0.25);
}
\foreach \x in {}
{\draw[fill=white]  (\x,3) -- (\x+1,3) -- (\x+1,4) -- (\x,4) -- cycle;			
\fill[red] (\x+0.5,3.5) circle (0.25);
}
\foreach \x in {16}
{\draw[fill=white,white]  (\x,3) -- (\x+2,3) -- (\x+2,4) -- (\x,4) -- cycle;
\draw[-] (\x,3) -- (\x,4);
\draw[-] (\x,3) -- (\x+2,3);
\draw[-] (\x,4) -- (\x+2,4);
\draw[-] (\x+1,3) -- (\x+1,4);
\node at (\x+1.5,3.5) {$\cdots$};
}
\foreach \x in {0}
{\draw[fill=white,white]  (\x,3) -- (\x-2,3) -- (\x-2,4) -- (\x,4) -- cycle;
\draw[-] (\x,3) -- (\x,4);
\draw[-] (\x,3) -- (\x-2,3);
\draw[-] (\x,4) -- (\x-2,4);
\draw[-] (\x-1,3) -- (\x-1,4);
\node at (\x-1.5,3.5) {$\cdots$};
}
\draw [decorate,decoration={brace,amplitude=4pt}] (3.9,2.85) -- (-1.9,2.85) node [black,midway,yshift=-0.4cm] {\scriptsize{$W_0^{t+1}$}};
\draw [decorate,decoration={brace,amplitude=4pt}] (6.9,2.85) -- (4.1,2.85) node [black,midway,yshift=-0.4cm] {\scriptsize{$Q_1^{t+1}$}};
\draw [decorate,decoration={brace,amplitude=4pt}] (7.9,2.85) -- (7.1,2.85) node [black,midway,yshift=-0.4cm] {\scriptsize{$W_1^{t+1}$}};
\draw [decorate,decoration={brace,amplitude=4pt}] (8.9,2.85) -- (8.1,2.85) node [black,midway,yshift=-0.4cm] {\scriptsize{$Q_2^{t+1}$}};
\draw [decorate,decoration={brace,amplitude=4pt}] (11.9,2.85) -- (9.1,2.85) node [black,midway,yshift=-0.4cm] {\scriptsize{$W_2^{t+1}$}};
\draw [decorate,decoration={brace,amplitude=4pt}] (12.9,2.85) -- (12.1,2.85) node [black,midway,yshift=-0.4cm] {\scriptsize{$Q_3^{t+1}$}};
\draw [decorate,decoration={brace,amplitude=4pt}] (13.9,2.85) -- (13.1,2.85) node [black,midway,yshift=-0.4cm] {\scriptsize{$W_3^{t+1}$}};
\draw [decorate,decoration={brace,amplitude=4pt}] (15.9,2.85) -- (14.1,2.85) node [black,midway,yshift=-0.4cm] {\scriptsize{$Q_4^{t+1}$}};
\draw [decorate,decoration={brace,amplitude=4pt}] (17.9,2.85) -- (16.1,2.85) node [black,midway,yshift=-0.4cm] {\scriptsize{$W_4^{t+1}$}};
}
\caption{The box-ball coordinates on a box-ball system and its time evolution.}
\end{figure}
\n Under the time evolution, the coordinates evolve as
\begin{equation}
(\infty,3,3,1,2,2,1,1,\infty)\mapsto (\infty,3,1,1,3,1,1,2,\infty)
\end{equation}
\end{ex}

\n We now introduce some fundamental definitions, which are key to the way we will later extend the classical box-ball system, that serve to distinguish between the box-ball evolution and the induced coordinate evolution.

\begin{defn}\label{defnofboxballandcoordspaces}
The $\text{BBS}$ phase space of $n$-soliton states is coordinatised by sequences of the form
$$(\infty, Q_1,W_1,\ldots,W_{n-1},Q_n,\infty)\in \{\infty\}\x \N^{2n-1}\x\{\infty\}=:\mathcal{B}_n.$$
The full BBS phase space is coordinatised by $$\mathcal{B}:=\bigcup\limits_{n\in\N}\mathcal{B}_n.$$
Let $C:\text{BBS}\to \mathcal{B}$ denote the map taking a box-ball system state to its coordinates. We further define $\varrho:\text{BBS}\to \text{BBS}$ to be the box-ball evolution and $\chi:\mathcal{B}\to \mathcal{B}$ to be the corresponding evolution on coordinates given in Theorem \ref{thmbbscoords}.
\end{defn}

\n With these definitions, we have an immediate corollary of Theorem \ref{thmbbscoords}.

\begin{cor}\label{commdiagbbscoords}
The following diagram commutes
\begin{figure}[H]
\centering
\tikz[scale=0.7]{
\node (1) at (0,0) {BBS};
\node (2) at (3,0) {BBS};
\node (3) at (0,-3) {$\mathcal{B}$};
\node (4) at (3,-3) {$\mathcal{B}$};
\draw[->] (1) -- (2) node[above,midway] {$\varrho$};
\draw[->] (1) -- (3) node[left,midway] {$C$};
\draw[->] (3) -- (4) node[above,midway] {$\chi$};
\draw[->] (2) -- (4) node[right,midway] {$C$};
}.
\end{figure}
\end{cor}

\section{RSK and gRSK}\label{chapterrskschensted}

\subsection{The Robinson-Schensted-Knuth Correspondence}\label{sectionRSK}
\n In this section, we provide the background and some basic motivation behind the Robinson-Schensted-Knuth correspondence and Schensted insertion. We begin with a review of some of the combinatorial objects of interest, the RSK equations describing Schensted word insertion, and Kirillov's geometric lifting of the (tropical) RSK equations to the geometric RSK (gRSK) equations. We will be following the papers \cite{bib:ad} by Aldous and Diaconis and \cite{bib:ny} by Noumi and Yamada, and the book \cite{bib:aigner} by Aigner.\\

\n The coverage of this background is fairly in-depth, with examples provided to aid in following the rather algorithmic constructions presented here. However, what is most pertinent to this paper are the equations for Schensted insertion which are described in Corollary \ref{corrskeqnssummary}.\\

\subsubsection{Schensted Insertion}\label{schenstedinsertionsectionrsk}
We describe in this section, following the treatment and notation given by \cite{bib:ny}, the process known as Schensted insertion, but only to the extent needed for understanding the RSK equations: Schensted insertion into a \textit{word}.\vspace{0.3cm}

\begin{defn}
A (weakly increasing) \textbf{word} in an \textbf{alphabet} $\{1,2,\ldots,n\}$ is a sequence $l_1,l_2,\ldots,l_k$ with $l_i\in \{1,2,\ldots,n\}$ for each $i$ and for which $l_i\leq l_{i+1}$ for each $1\leq i< k$.
\end{defn}\vspace{0.1cm}

\begin{rem}
For convenience of expression, when it is appropriate to do so and when it creates no confusion, we will drop the commas when writing words. So, for example, the word $1,1,1,2,2,4,4,4$ will simply be written as $11122444$.\\
Additionally, all words we work with will be weakly increasing, so that is what ``word'' will be taken to mean from now on.
\end{rem}

\n The process of Schensted word insertion will be an evolution on pairs of words: one will start with an \textit{initial word}, into which another word (the \textit{insertion word}) will be inserted. This process will result in a \textit{new word} and a byproduct of the insertion, in the form of a \textit{bumped word}. The following diagram is typical for representing this structure:
\begin{figure}[H]
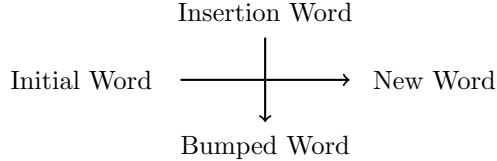

\centering
\tikz[scale=0.45]{
\node at (-5.3,0) {Initial Word~~};
\node at (0,2) {Insertion Word};
\node at (5,0) {New Word};
\node at (0,-2) {Bumped Word};
\draw[->,thick] (-2.5,0) -- (2.5,0);
\draw[->,thick] (0,1.25) -- (0,-1.25);
}
\caption{Pictorial representation of Schensted insertion of a word into a word.}\label{pictorialrepnofschensted}
\end{figure}	

\begin{nota}
Let $n\in\N$ be fixed (giving the bound on the alphabet for the words). Each word in this alphabet can be represented by (and identified with) an $n$-tuple in $\N_0^n$, where $\N_0=\N\cup\{0\}$, by letting the $i$-th entry be the number of instances of $i$ in the word. Since words are weakly increasing, there is no loss of information from doing this. For example, with $n=5$, the 5-tuple $(3,2,0,3,0)$ represents the word $w=11122444=1^32^23^04^35^0$.\\
\n For the words involved in Schensted insertion, we adopt this notation as follows:
\begin{itemize}
\item $\mb{x}=(x_1,\ldots,x_n)$: $n$-tuple for the \textbf{initial} word,
\item $\mb{a}=(a_1,\ldots,a_n)$: $n$-tuple for the \textbf{insertion} word, 
\item $\mb{y}=(y_1,\ldots,y_n)$: $n$-tuple for the \textbf{new} word, 
\item $\mb{b}=(b_1,\ldots,b_n)$: $n$-tuple for the \textbf{bumped} word.
\end{itemize}
\end{nota}\vspace{0.2cm}

\n We begin by prescribing the rules for inserting a letter (a number) into a word. Word insertion is then the iterative application of letter insertion.

\begin{defn}\label{schenstednumintoword}
The process of \textbf{Schensted insertion of a letter (number) into a word} is given as follows. To insert a number $a\in\{1,\ldots,n\}$ into a word $w=1^{w_1}2^{w_2}\cdots n^{w_n}$:
\begin{enumerate}
\item Look for the left-most number in $w$ that is strictly greater than $a$, if it exists, then replace that number by $a$. The replaced number, which has been removed from the word, is now bumped.
\item If no such number exists in $w$, then append $a$ to the end of $w$.
\end{enumerate}
\end{defn}

\begin{rem}
In performing this procedure, the inserted number, $a$, will always find a place in the word, whilst keeping the word weakly increasing. It will either do so by bumping something out of the word or it will do so by growing the word.
\end{rem}
 \vspace{0.2cm}
 
\begin{ex}
Here are some examples with the initial word given by $\mb{x}=1112234$:
\begin{enumerate}
\item Inserting $1$ creates a new word of $1111234$, bumping out $2$.
\item Inserting $3$ creates a new word of $1112233$, bumping out $4$
\item Inserting $4$ creates a new word of $11122344$, bumping nothing.
\item Inserting $7$ creates a new word of $11122347$, bumping nothing.
\end{enumerate}
\end{ex} \vspace{0.1cm}

\begin{defn}\label{schenstedwordintoword}
The process of \textbf{Schensted insertion of a word into another word} is defined as follows. Perform letter insertion (Definition \ref{schenstednumintoword}) iteratively, reading in the insertion word from left to right.
\end{defn}

\begin{rem}This process does indeed produce a new pair of weakly increasing words:
\begin{enumerate}
\item Since the word being inserted is weakly increasing, so too must the sequence of bumped letters when concatenated. This concatenation forms the bumped word.
\item By the nature of letter insertion, the new word is always weakly increasing.
\end{enumerate}
\end{rem}\vspace{0.1cm}

\begin{ex}\label{schenstedexamplewordword}
We insert the word $111334$ into $1144555$:
\begin{enumerate}
    \item We insert $1$ into $1144555$, resulting in $1114555$ (bumping $4$),
    \item we insert $1$ into $1114555$, resulting in $1111555$ (bumping $4$),
    \item we insert $1$ into $1111555$, resulting in $1111155$ (bumping $5$),
    \item we insert $3$ into $1111155$, resulting in $1111135$ (bumping $5$),
    \item we insert $3$ into $1111135$, resulting in $1111133$ (bumping $5$),
    \item we insert $4$ into $1111133$, resulting in $11111334$ (nothing is bumped).
\end{enumerate}
The new word is then $11111334$ and the bumped word is $44555$. Or, using the pictorial representation in Figure \ref{pictorialrepnofschensted}, we summarise this insertion as
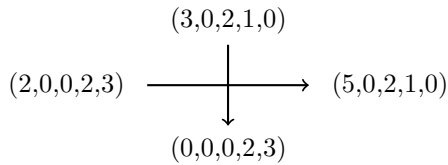
\begin{figure}[H]
\centering
\tikz[scale=0.43]{
\node at (-5,0) {(2,0,0,2,3)};
\node at (0,2) {(3,0,2,1,0)};
\node at (5,0) {(5,0,2,1,0)};
\node at (0,-2) {(0,0,0,2,3)};
\draw[->,thick] (-2.5,0) -- (2.5,0);
\draw[->,thick] (0,1.25) -- (0,-1.25);
}
\end{figure}
\end{ex}

\subsubsection{Explicit Formul{\ae} for the RSK Dynamics}\label{secexplformulae}

\n Write $\mb{x}=1^{x_1}2^{x_2}\cdots n^{x_n}$ for the word into which we wish to insert the word $\mb{a}=1^{a_1}2^{a_2}\cdots n^{a_n}$, writing $\mb{y}=1^{y_1}2^{y_2}\cdots n^{y_n}$ for the result, and write $\mb{b}=1^{b_1}2^{b_2}\cdots n^{b_n}$. Thus, we want to know how $(y_1,\ldots,y_n)$ and $(b_1,\ldots,b_n)$ arise from $(x_1,\ldots,x_n)$ and $(a_1,\ldots,a_n)$.\\[5pt]

\n In order to simplify the calculations, we introduce new variables by taking partial sums:
\begin{equation}\xi_j=x_1+\cdots+x_j,~~\eta_j=y_1+\cdots+y_j\end{equation}
for $j=1,\ldots,n$. These will serve to reformulate the information in a ``max-plus'' form (Lemma \ref{wordins}), which will be crucial for detropicalisation, discussed in the subsequent section.

\n The $y$'s can then be recovered from the $\eta$'s as $y_1=\eta_1$ and $y_j=\eta_j-\eta_{j-1}$ for $j>1$. The $b$'s can be obtained once the $y$'s are known. For example, $b_j$ represents the number of $j$'s bumped from $w$. Since $w$ started with $x_j$ consecutive $j$'s, and we introduced $a_j$ consecutive $j$'s, and ended up with $y_j$ consecutive $j$'s, the number of bumped $j$'s must equal $b_j=x_j+a_j-y_j=a_j+\xi_j-\xi_{j-1}-\eta_j+\eta_{j-1}$ for $j>1$. For $j=1$, note that $1$ cannot be bumped, so $y_1=x_1+a_1$, so that $b_1=0$ always.

\begin{lem}(\cite{bib:ny})\label{wordins}
When $v=k^a$ and $w=1^{x_1}2^{x_2}\cdots n^{x_n}$, we get\vspace{0.25cm}
\begin{equation}\eta_j=\left\{\begin{array}{cl}
\xi_j & \text{if }j<k\\
\xi_k+a & \text{if }j=k\\
\max\{\eta_k,\xi_j\} & \text{if }j>k
\end{array}\right..\end{equation}
\end{lem}
\vspace{0.1cm}
\begin{proof}
If $\xi_j>\eta_k$, then some of the $j$'s `survived' the bumping. This means the bumping did not get past the last $j$, and so the number of boxes with numbers $\leq j$ is still given by $\xi_j$, hence $\eta_j=\xi_j$. If $\xi_j=\eta_k$, then the bumping got to the very last $j$, which still gives the same conclusion of $\eta_j=\xi_j$. However, if $\xi_j<\eta_k$, then this means that the bumping eradicated all instances of $j$ (hence, by the nature of the insertion algorithm, no numbers $k<l\leq j$ remain) and so $\eta_j=\eta_k$, hence $\eta_j=\max\{\eta_k,\xi_j\}$ for all $j>k$.
\end{proof}

\begin{cor}(\cite{bib:ny})
Inserting $v=1^{a_1}2^{a_2}\cdots n^{a_n}$ into $w=1^{x_1}2^{x_2}\cdots n^{x_n}$, one has
\begin{equation}\eta_j=\max_{1\leq k\leq j}\{x_1+\cdots+x_k+a_k+\cdots +a_j\}\end{equation}
for all $j$.
\end{cor}
\begin{proof}
Applying Lemma \ref{wordins} recursively, one obtains:
\begin{align*}
\text{\footnotesize{$(\xi_1,\xi_2,\xi_3,\ldots)$}} & \text{\footnotesize{$\xrightarrow{~1^{a_1}~} (\xi_1+a_1=\eta_1,\max\{\eta_1,\xi_2\},\max\{\eta_1,\xi_3\},\ldots,\max\{\eta_1,\xi_j\},\ldots)$}}\\
& \text{\footnotesize{$\xrightarrow{~2^{a_2}~} (\eta_1,\max\{\eta_1,\xi_2\}+a_2=\eta_2,\max\{\eta_1,\eta_2,\xi_3\},\ldots,\max\{\eta_1,\eta_2,\xi_j\},\ldots)$}}\\
& \text{\footnotesize{$\xrightarrow{~3^{a_3}~} (\eta_1,\eta_2,\max\{\eta_1,\eta_2,\xi_3\}+a_3,\ldots,\max\{\eta_1,\eta_2,\eta_3,\xi_j\},\ldots).$}}
\end{align*}
Thus, $\eta_j=\max\{\eta_1,\eta_2,\ldots,\eta_{j-1},\xi_j\}+a_j$ for all $j>1$ and $\eta_1=\xi_1+a_1$.\\[5pt]
Since the $\eta$'s are weakly-increasing, we have 

\begin{equation}\eta_j=\max\{\eta_{j-1},\xi_j\}+a_j=\max\{\eta_{j-1}+a_j,\xi_j+a_j\}\label{eqnetamax}\end{equation} for all $j>1$.\\[5pt]
Unpacking this (to remove the $\eta_l$'s, for $l<j$), we get
\begin{align}
\eta_j&=\max\{\max\{\eta_{j-2}+a_{j-1},\xi_{j-1}+a_{j-1}\}+a_j,\xi_j+a_j\}\\
&=\max\{\eta_{j-2}+a_{j-1}+a_j,\xi_{j-1}+a_{j-1}+a_j,\xi_j+a_j\}\\
&=\cdots=\max_{1\leq k\leq j}\{\xi_k+a_k+a_{k+1}+\ldots+a_j\}\\
&=\max_{1\leq k\leq j}\{x_1+\cdots+x_k+a_k+\ldots+a_j\}\label{etamaxsum}
\end{align}
which also covers $j=1$.
\end{proof}

\n To summarise, one now has
$$\scalebox{0.8}{$y_j=\left\{\begin{array}{cl}
x_1+a_1 & \text{if }j=1\\
\max\limits_{1\leq k\leq j}\{x_1+\cdots+x_k+a_k+\cdots +a_j\}-\max\limits_{1\leq k\leq j-1}\{x_1+\cdots+x_k+a_k+\cdots +a_{j-1}\} & \text{if }j>1
\end{array}\right.$}$$
and $b_j=x_j+a_j-y_j$ for all $j$.\\[5pt]

\n However, for the purpose of convenient calculation, dividing the computation into phases, using Equation \ref{etamaxsum}, one has

\begin{cor}(\cite{bib:ny})\label{corrskeqnssummary}
Given input coordinates $(a_1,\ldots,a_n)$ and $(x_1,\ldots,x_n)$, one obtains the output coordinates $(b_1,\ldots,b_n)$ and $(y_1,\ldots,y_n)$ as follows:
\begin{enumerate}
\item compute $\xi_j=x_1+\cdots+x_j$ for $j=1,\ldots,n$,
\item compute $\eta_j=\max\{\eta_{j-1},\xi_j\}+a_j$ recursively for $j=1,\ldots,n$, initialising with $\eta_1=\xi_1+a_1$,
\item the $y$-coordinates are obtained by taking $y_1=\eta_1$ and $y_j=\eta_j-\eta_{j-1}$ for $j=2,\ldots,n$,
\item the $b$-coordinates are obtained by taking $b_1=0$ and $b_j=a_j+x_j-y_j$ for $j=2,\ldots,n$.
\end{enumerate}
\end{cor}

\n The above description of Schensted insertion as a dynamical evolution on pairs of $n$-tuples motivates the following definition that will be key in relating RSK and the box-ball system.

\begin{defn}\index{$\mathcal{R}_n$}\index{$\mathcal{R}$}\label{rskeqnsdefn}
Define
$$\mathcal{R}_n=(\N_0^n)^2$$ 
which is the set of pairs of $n$-tuples. Schensted insertion defines a map $\text{RSK}_n:\mathcal{R}_n\to \mathcal{R}_n$ by taking the first (respectively, last) $n$ coordinates of $(\mb{a},\mb{x})\in \mathcal{R}_n$ to be the insertion (respectively, initial) word in the RSK algorithm, and letting $\text{RSK}_n(\mb{a},\mb{x})=(\mb{b},\mb{y})$. Define $\mathcal{R}=\bigcup\limits_{n\in \N}\mathcal{R}_n$, and let $\text{RSK}:\mathcal{R}\to\mathcal{R}$ be defined naturally.
\end{defn}

\begin{ex}
Take $w=1^22^33^14^05^2=1^22^33^15^2$ and $v=1^12^24^15^1$, i.e. ${\bf{x}}=(2,3,1,0,2)$ and ${\bf{a}}=(1,2,0,1,1)$, where emboldened letters denote the vectors of the corresponding variables. According to Equation \ref{etamaxsum}, we should have 
$$\eta_1=3,~\eta_2=\max\{5,7\}=7,~\eta_3=\max\{5,7,6\}=7,$$
$$\eta_4=\max\{6,8,7,7\}=8,~\eta_5=\max\{7,9,8,8,9\}=9.$$
Thus,
$$y_1=3,~~y_2=7-3=4,~~y_3=7-7=0,~~y_4=8-7=1,~~y_5=9-8=1.$$
So,
$$b_1=1+2-3=0,~~b_2=2+3-4=1 ,~~b_3=0+1-0=1 ,$$
$$b_4=1+0-1=0 ,~~b_5=1+2-1=2.$$
To check this, let us perform the word insertion, using the notation $v\leftarrow w$ for the result of Schensted inserting a word $w$ into a word $v$:
\begin{align*}
11222355 \leftarrow 12245 & = 11122355 \leftarrow 2245~~~~&\text{(bumped 2)}\\
& = 11122255 \leftarrow 245~~~~&\text{(bumped 3)}\\
& = 11122225 \leftarrow 45~~~~&\text{(bumped 5)}\\
& = 11122224 \leftarrow 5~~~~&\text{(bumped 5)}\\
& = 111222245  ~~~~&\text{(no bumps)}
\end{align*}

\n So, the word $w'=1^32^44^15^1$ is left, and $v'=2^13^15^2$ is bumped. This gives ${\bf{y}}=(3,4,0,1,1)$ and ${\bf{b}}=(0,1,1,0,2)$ agreeing with the results of the formul{\ae}.\\

\n Using the notation of Figure \ref{pictorialrepnofschensted}, we can represent this evolution as
\begin{figure}[H]
\centering
\tikz[scale=0.6]{
\node at (-5,0) {$\mathbf{x}=(2,3,1,0,2)$};
\node at (0,2) {$\mathbf{a}=(1,2,0,1,1)$};
\node at (5,0) {$\mathbf{y}=(3,4,0,1,1)$};
\node at (0,-2) {$\mathbf{b}=(0,1,1,0,2)$};
\draw[->,thick] (-2.5,0) -- (2.5,0);
\draw[->,thick] (0,1.25) -- (0,-1.25);
}
\caption{Schensted word insertion represented as in Figure \ref{pictorialrepnofschensted}}\label{nycrossdiagram}
\end{figure}

\end{ex}

\subsection{Ultradiscretisation and the Tropical Semiring}\label{introsectrop}

\n Tropical mathematics (\cite{bib:lmrs}, \cite{bib:l}, \cite{bib:v}) is the study of the max-plus semiring, which we will now define.  In this section, we follow the presentation given by Maslov \cite{bib:lmrs}. The structure of the semiring $(\R_{\geq 0},+,\x)$ is carried over to the set $S=\R\cup\{-\infty\}$ by a family of bijections $D_\hbar$, for $\hbar>0$, given by 

\begin{equation}
D_\hbar(x)=\left\{\begin{array}{ccl}
\hbar\ln x &~~~& \text{if }x\neq 0\\
-\infty && \text{if }x= 0
\end{array}\right..
\end{equation}

\n This induces a family of semirings, parametrised by $\hbar>0$, $(S,\oplus_\hbar,\otimes_\hbar)$ with operations given by
\begin{align}
a\oplus_\hbar b&=D_\hbar(D_\hbar^{-1}(a)+D_\hbar^{-1}(b))
=\left\{\begin{array}{cl}
\hbar\ln(e^{a/\hbar}+e^{b/\hbar}) & \text{if }a,b\neq -\infty\\
\max(a,b) & \text{otherwise}
\end{array}\right.\\
a\otimes_\hbar b&=D_\hbar(D_\hbar^{-1}(a)D_\hbar^{-1}(b))
=a+b.
\end{align}

\n In the limit, $\hbar\to 0$, Maslov `dequantises' $(\R_{\geq 0},+,\x)$ to obtain the tropical semiring $(\R\cup\{-\infty\},\max,+)$, where its addition is the usual $\max$ operation and its multiplication operation is usual addition, hence the name ``max-plus semiring''.\\

\n Maslov views this construction as an analogue of the correspondence principle from quantum mechanics, with $(\R_{\geq 0},+,\x)$ as the quantum object and $(\R\cup\{-\infty\},\max,+)$ as its classical counterpart.

\subsubsection{Ultradiscretisation and Subtraction-Free Rational Functions}\label{sectdetrop}
In general, one can apply the above process to rational maps. However, when subtraction is present, one may encounter the so-called \textit{minus-sign problem} (see, for example, \cite{bib:kknt}). If a rational map is subtraction-free, then we may safely apply the above construction of Maslov, to perform what is often referred to as \textit{ultradiscretisation}. We will see this in action in Section \ref{ultradiscofgRSK}.

\subsection{Kirillov's Geometric Lifting: gRSK}\label{kirillovgrsksection}
The formul{\ae} in the previous section involve only the operations max and addition, hence the formul{\ae} live in the tropical max-plus algebra.

\n We will make the change of operations:
$$(\max,+) ~~~\rightarrow ~~~ (+, \cdot)$$
to the formul{\ae} in the Corollary \ref{corrskeqnssummary}, making the necessary algebraic analogue for the `additive' identities ($0 \to 1$) to go from
\begin{align*}
\xi_j&= x_1+\cdots + x_j~~~~~\forall \, j=1,\ldots,n\\
\eta_1&=\xi_1+a_1\\
\eta_j&=\max\{\eta_{j-1},\xi_j\}+a_j~~~~~\forall \, j=2,\ldots,n
\end{align*}
and
\begin{align*}
y_1&=\eta_1\\
y_j&=\eta_j-\eta_{j-1}~~~~~\forall \, j=2,\ldots,n\\
b_1&=0\\
b_j&=a_j+x_j-y_j=a_j+\xi_j-\xi_{j-1}-\eta_j+\eta_{j-1}~~~~~\forall \, j=2,\ldots,n.
\end{align*}
to the (de)tropicalised analogue:
\begin{align*}
\xi_j&= x_1\cdots x_j~~~~~\forall \, j=1,\ldots,n\\
\eta_1&=\xi_1a_1\\
\eta_j&=(\eta_{j-1}+\xi_j)a_j~~~~~\forall \, j=2,\ldots,n
\end{align*}
and
\begin{align*}
y_1&=\eta_1\\
y_j&=\dfrac{\eta_j}{\eta_{j-1}}~~~~~\forall \, j=2,\ldots,n\\
b_1&=1\\
b_j&=a_j\dfrac{x_j}{y_j}=a_j\dfrac{\xi_j\eta_{j-1}}{\xi_{j-1}\eta_j}~~~~~\forall \, j=2,\ldots,n.
\end{align*}

\begin{lem}(\cite{bib:ny})\label{GRSK}
Returning to the original variables of $$x_1,\ldots, x_n,a_1,\ldots,a_n,y_1,\ldots,y_n,b_1,\ldots,b_n,$$ the above formul{\ae} reduce to the following system $({\bf{x}},{\bf{a}})\mapsto ({\bf{y}},{\bf{b}})$:\vspace{0.2cm}
\begin{equation}\label{gRSKequations}
\def\arraystretch{2.0}
\left\{\begin{array}{l}
b_1 = 1\\
a_1x_1=y_1\\
a_jx_j=y_jb_j ~~~~~\forall \, j=2,\ldots,n\\
\dfrac{1}{a_1}+\dfrac{1}{x_2}=\dfrac{1}{b_2}\\
\dfrac{1}{a_j}+\dfrac{1}{x_{j+1}}  = \dfrac{1}{y_j}+\dfrac{1}{b_{j+1}}~~~~~\forall \, j=2,\ldots,n.
\end{array}\right.
\end{equation}

\end{lem}
\begin{proof} The first two formul{\ae} are by virtue of $\eta_1=y_1$ and $\xi_1=x_1$. For the other formul{\ae}, take $\eta_1=\xi_1a_1$ and $\eta_j=(\eta_{j-1}+\xi_j)a_j$ and rearrange to get \vspace{0.3cm}
$$\left\{\begin{array}{lll}
\dfrac{\eta_1}{\xi_1a_1}=1&&(*)\\\\
\dfrac{\eta_j-\eta_{j-1}a_j}{\xi_ja_j}=1&~~\forall~j=2,\ldots,n~~~~&(*_j)
\end{array}\right.$$
\n Equating $(*)$ and $(*_2)$ yields\vspace{0.1cm}
\begin{align*}
&\dfrac{\eta_2-\eta_1a_2}{\xi_2a_2}=\dfrac{\eta_1}{\xi_1a_1}\\
\Rightarrow ~~& \dfrac{y_2}{x_2a_2}-\dfrac{1}{x_2} =\dfrac{1}{a_1}\\
\Rightarrow ~~& \dfrac{1}{b_2}-\dfrac{1}{x_2} =\dfrac{1}{a_1}
\end{align*}

\n resulting in the third formula.\\[5pt]
Equating $(*_{j+1})$ and $(*_j)$ for $j=2,\ldots,n$ yields\vspace{0.1cm}
\begin{align*}
&\dfrac{\eta_{j+1}-\eta_ja_{j+1}}{\xi_{j+1}a_{j+1}} = \dfrac{\eta_j-\eta_{j-1}a_j}{\xi_ja_j}\\
\Rightarrow ~~&  \dfrac{\eta_j}{\xi_j}\left(\dfrac{y_{j+1}}{x_{j+1}a_{j+1}}-\dfrac{1}{x_{j+1}}\right)=\dfrac{\eta_j}{\xi_j}\left(\dfrac{1}{a_j}-\dfrac{1}{y_j}\right)\\
\Rightarrow ~~&  \dfrac{y_{j+1}}{x_{j+1}a_{j+1}}+\dfrac{1}{y_j} = \dfrac{1}{a_j}+\dfrac{1}{x_{j+1}}\\
\Rightarrow ~~& \dfrac{1}{b_{j+1}}+\dfrac{1}{y_j} = \dfrac{1}{a_j}+\dfrac{1}{x_{j+1}}.
\end{align*}
\end{proof}

\subsubsection{A Matrix Representation of the Geometric RSK}\label{sectmatrixformgrsk}
Returning to the system of equations presented in Lemma \ref{GRSK}, and letting bars denote reciprocals (i.e. $\bar{x}:=\frac{1}{x}$), one obtains the following equations:
\begin{align*}
\bar{a}_1\bar{x}_1&=\bar{y}_1\\
\bar{a}_j\bar{x}_j&=\bar{y}_j\bar{b}_j ~~~~~~~~~~\,\forall \, j=2,\ldots,n\\
\bar{a}_1+\bar{x}_2&=\bar{b}_2\\
\bar{a}_j+\bar{x}_{j+1} & = \bar{y}_j+\bar{b}_{j+1}~~~~~\forall \, j=2,\ldots,n.
\end{align*}
This can be represented in the following form:\\[5pt]
\begin{equation}\label{upperuppergrskfactor}
\scalebox{0.83}{$
\arraycolsep=1.88pt\def\arraystretch{1.1}
\left(\begin{array}{ccccc}
\bar{a}_1 & 1 &  & & \bigzero\\
 & \bar{a}_2 & 1 &  & \\
& & \ddots & \ddots & \\
&&&\bar{a}_{n-1}&1\\
\bigzero&&&&\bar{a}_n
\end{array}\right)
\left(\begin{array}{ccccc}
\bar{x}_1 & 1 &  & &\bigzero \\
 & \bar{x}_2 & 1 &  & \\
& & \ddots & \ddots & \\
&&&\bar{x}_{n-1}&1\\
\bigzero&&&&\bar{x}_n
\end{array}\right)
=
\left(\begin{array}{ccccc}
\bar{y}_1 & 1 &  & &\bigzero \\
 & \bar{y}_2 & 1 &  & \\
& & \ddots & \ddots & \\
&&&\bar{y}_{n-1}&1\\
\bigzero&&&&\bar{y}_n
\end{array}\right)
\left(\begin{array}{ccccc}
1 & 0 &  & & \bigzero\\
 & \bar{b}_2 & 1 &  & \\
& & \ddots & \ddots & \\
&&&\bar{b}_{n-1}&1\\
\bigzero&&&&\bar{b}_n
\end{array}\right).$}
\end{equation}

\section{The Ghost-Box-Ball System} \label{chaptergbbs} 
In Section \ref{kirillovgrsksection}, we demonstrated how one can pass from the RSK equations (\textit{cf.} Corollary \ref{corrskeqnssummary}) to gRSK (as summarised in Lemma \ref{GRSK}). We begin by showing that, as one would hope, the ultradiscretisation of the gRSK equations indeed results in the original RSK equations. The purpose of this exercise, however, goes far beyond simply recovering the RSK equations: by performing the ultradiscretisation method on gRSK, we obtain a representation of the RSK equations in a form that lends itself to comparison with the box-ball coordinate evolution (\textit{cf.} Theorem \ref{thmbbscoords}).\\

\n We find that our comparison results in the need to be able to interpret the box-ball system when some of its coordinates are zero. This leads us to a new cellular automaton, extending the box-ball system by two new types of object, and we call this cellular automaton the ghost-box-ball system (GBBS). We prove some key results about our ghost-box-ball system, including its reduction to the original box-ball system under an operation we call \textit{exorcism}. The operation of exorcism, along with the properties of the GBBS, allows us to extend the classical Young diagram conserved quantity (\textit{cf.} Section \ref{invsofbbs}) of the box-ball system to the ghost-box-ball system. \\

\subsection{Ultradiscretisation of Geometric RSK}\label{ultradiscofgRSK}

\n We begin with an application of the ultradiscretisation process to the geometric RSK equations. Since we are just (re)tropicalising the detropicalised RSK equations, one should not be surprised to recover RSK. However, the process rewrites the RSK equations in a way that is essential for seeing the connection between RSK and the box-ball system.

\begin{lem}\label{gRSKtoRSKviaUD}
The ultradiscretisation of the geometric RSK equations (Equations \ref{gRSKequations}) results in the (tropical) RSK equations for Schensted insertion.
\end{lem}
\begin{proof}
We begin with the geometric RSK equations:
\begin{align}
y_1&=a_1x_1\label{grskeq1}\\
y_ib_i&=a_ix_i && i=2,\ldots,n\label{grskeq2}\\
b_2&=a_1+x_2 \label{grskeq3}\\
y_i+b_{i+1}&=a_i+x_{i+1} &&i=2,\ldots,n-1.\label{grskeq4}
\end{align}
Using Equations \ref{grskeq2} and \ref{grskeq4}, one obtains
\begin{align}
b_{i+1}&=x_{i+1}+a_i-y_i\\
&=x_{i+1}+\frac{a_i}{b_i}\left(b_i-x_i\right)\\
&=x_{i+1}+\frac{a_i}{b_i}\left(a_{i-1}-y_{i-1}\right)\\
=\cdots&=x_{i+1}+\dfrac{\prod_{j=1}^ia_j}{\prod_{j=2}^ib_j}
\end{align}
for $i=2,\ldots,n-1$.\\

\n Taking $b_1=1$, as it should be, the geometric RSK can be expressed as
\begin{align}
b_1&=1\label{bbscoordeq1}\\
y_i&=\frac{a_ix_i}{b_i}&&i=1,\ldots,n\\
b_{i+1}&=x_{i+1}+\frac{\prod_{j=1}^ia_j}{\prod_{j=2}^ib_j}&&i=1,\ldots,n-1\label{bbscoordeq3}
\end{align}
by following the convention of taking the empty product to be $1$.\\

\n With the evolution now expressed in a subtraction-free form (\textit{cf.} Section \ref{sectdetrop}), we begin ultradiscretisation, first by changing variables
$$a_i\to e^{-a_i(\epsilon)/\epsilon},~~~~~~x_i\to e^{-x_i(\epsilon)/\epsilon},~~~~~~y_i\to e^{-y_i(\epsilon)/\epsilon},~~~~~~b_i\to e^{-b_i(\epsilon)/\epsilon}.$$

\n The change of variables, applied to Equations \ref{bbscoordeq1} - \ref{bbscoordeq3}, yields
\begin{align}
e^{-\frac{1}{\epsilon}b_1(\epsilon)}&=1\\
e^{-\frac{1}{\epsilon}y_i(\epsilon)}&=e^{-\frac{1}{\epsilon}(a_i(\epsilon)+x_i(\epsilon)-b_i(\epsilon))}&&i=1,\ldots,n\\
e^{-\frac{1}{\epsilon}b_{i+1}(\epsilon)}&=e^{-\frac{1}{\epsilon}x_{i+1}(\epsilon)}+e^{-\frac{1}{\epsilon}\left(\sum_{j=1}^ia_j(\epsilon)-\sum_{j=2}^ib_j(\epsilon)\right)}&&i=1,\ldots,n-1,
\end{align}
\n where the empty sum is taken to be zero.\\

\n Solving for the exponentiated variables on the left-hand side, we get the following
\begin{align}
b_1(\epsilon)&=0\\
y_i(\epsilon)&=a_i(\epsilon)+x_i(\epsilon)-b_i(\epsilon)&&i=1,\ldots,n\\
b_{i+1}(\epsilon)&=-\epsilon\log\left(e^{-\frac{1}{\epsilon}x_{i+1}(\epsilon)}+e^{-\frac{1}{\epsilon}\left(\sum_{j=1}^ia_j(\epsilon)-\sum_{j=2}^ib_j(\epsilon)\right)}\right)&&i=1,\ldots,n-1.
\end{align}

\n The final step in ultradiscretisation is taking the limit as $\epsilon\to0^+$. If we abuse notation by recycling the original RSK variables in the ultradiscrete equations by letting, for example, $b_i=\lim\limits_{\epsilon\to0^+}b_i(\epsilon)$, we obtain the following ultradiscretisation of the geometric RSK equations
\begin{align}
b_1&=0\label{udgRSKeq1}\\
y_i&=a_i+x_i-b_i &&i=1,\ldots,n\label{udgRSKeq2}\\
b_{i+1}&=\min\left(x_{i+1},\left(\sum_{j=1}^ia_j-\sum_{j=2}^ib_j\right)\right)&&i=1,\ldots,n-1.\label{udgRSKeq3}
\end{align}

\n It remains to show that the solution to this system of equations solves the RSK equations (Corollary \ref{corrskeqnssummary}). Equations \ref{udgRSKeq1} and \ref{udgRSKeq2} are already part of the RSK equations. What is left of the RSK equations is for the following to hold
\begin{align}
\eta_1&=\xi_1+a_1\label{eqeta1}\\
\eta_j&=\max\{\eta_{j-1},\xi_j\}+a_j,~~~~~j=2,\ldots,n\label{eqetaj}
\end{align}
where $\eta_i=y_1+\cdots+y_i$ and $\xi_i=x_1+\cdots+x_i$ for $i=1,\ldots,n$. Since both the RSK equations and the ultradiscrete geometric RSK equations have a unique solution, this will complete the proof. We proceed directly:\\

\n Equation \ref{eqeta1} is equivalent to $y_1=x_1+a_1$, which clearly holds.\\

\n For Equation \ref{eqetaj}, we use the defining equations for the ultradiscrete geometric RSK equations in the following computation for $i\geq 1$:
\begin{align*}
y_{i+1}
&=a_{i+1}+x_{i+1}-b_{i+1}\\
&=a_{i+1}+x_{i+1}-\min\left(x_{i+1},\left(\sum_{j=1}^ia_j-\sum_{j=2}^ib_j\right)\right)\\
&=a_{i+1}+x_{i+1}-\min\left(x_{i+1},\left(\sum_{j=1}^ia_j-\sum_{j=2}^i(a_j+x_j-y_j)\right)\right)\\
&=a_{i+1}+x_{i+1}-\min\left(x_{i+1},\left(a_1+\sum_{j=2}^i(y_j-x_j)\right)\right)\\
&=a_{i+1}+x_{i+1}-\min\left(x_{i+1},\left(a_1-y_1+x_1+\sum_{j=1}^i(y_j-x_j)\right)\right)\\
&=a_{i+1}+x_{i+1}-\min\left(x_{i+1},\left(b_1+\eta_i-\xi_i\right)\right)\\
&=a_{i+1}+x_{i+1}+\max\left(-x_{i+1},\xi_i-\eta_i\right)\\
&=a_{i+1}+\max\left(0,x_{i+1}+\xi_i-\eta_i\right)\\
&=a_{i+1}+\max\left(0,\xi_{i+1}-\eta_i\right)\\
&=a_{i+1}+\max\left(\eta_i,\xi_{i+1}\right)-\eta_i.
\end{align*}
Thus,
\begin{equation}\eta_{i+1}=y_{i+1}+\eta_i=a_{i+1}+\max\left(\eta_i,\xi_{i+1}\right),\end{equation}
which completes the proof.
\end{proof}

\subsection{RSK Insertion and the Box-Ball Coordinates}\label{rskcoordbbs}
\n The key point of the derivation in Section \ref{ultradiscofgRSK}  is that we have now obtained the RSK equations in the following form:
\begin{align}
b_1&=0\\
y_i&=a_i+x_i-b_i &&i=1,\ldots,n\\
b_{i+1}&=\min\left(x_{i+1},\left(\sum_{j=1}^ia_j-\sum_{j=2}^ib_j\right)\right)&&i=1,\ldots,n-1.
\end{align}
which lends itself to comparison with the box-ball system equations for an $n+1$-soliton:
\begin{align}
W_i^{t+1}&=Q_{i+1}^t+W_i^t-Q_i^{t+1} &&i=1,\ldots,n\label{eqonemeowmoo}\\
Q_i^{t+1}&=\min\left(W_i^t,\sum_{j=1}^i Q_j^t-\sum_{j=1}^{i-1}Q_j^{t+1}\right)  &&i=1,\ldots,n+1.\label{eqtwomeowmoo}
\end{align}

\n In the box-ball system equations (Equations \ref{eqonemeowmoo} and \ref{eqtwomeowmoo}), we perform the following change of variables:
\begin{equation}Q_{i+1}^t=a_i, ~~~~W_i^t=x_i,~~~~W_i^{t+1}=y_i,~~~~Q_i^{t+1}=b_i,\label{changevarsbbstorsk}\end{equation}
producing the following
\begin{align}
y_i&=a_i+x_i-b_i&&i=1,\ldots,n\\
b_i&=\min\left(x_i,\sum_{j=1}^i a_{j-1}-\sum_{j=1}^{i-1}b_j\right) &&i=1,\ldots,n+1.
\end{align}

\n Since $b_1=\min(x_1,a_0)=\min(W_1^t,Q_1^t)$, to obtain the RSK condition $b_1=0$, we take $a_0=Q_1^t=0$. Under this condition, the box-ball equations now take the form
\begin{align}
b_1&=0\\
y_i&=a_i+x_i-b_i&&i=1,\ldots,n\\
b_{i+1}&=\min\left(x_{i+1},\sum_{j=0}^{i} a_j-\sum_{j=1}^i b_j\right) &&i=1,\ldots,n.
\end{align}

\n Finally, we make sense of $b_{n+1}$ in the above system:

\begin{equation}
b_{n+1}=\min\left(x_{n+1},\sum_{j=0}^n a_j-\sum_{j=1}^{n}b_j\right)=a_1+\cdots+a_n-b_2-\cdots-b_n
\end{equation}
since $x_{n+1}=W_{n+1}^t=\infty$.

\n Although the box-ball coordinate evolution was defined on $\mathcal{B}_n$ (as defined in Definition \ref{defnofboxballandcoordspaces}), in which all coordinates are positive integers, these equations naturally extend to coordinates which may contain zeroes and for which the dynamics satisfies $Q_1^{t+1}=0$ if $Q_1^t=0$. \\

\n With this extension in mind, we introduce a corresponding modification to the $n$-soliton phase space $\mathcal{B}_n$ with the following definition (where, recall, $\N_0$ denotes the natural numbers augmented by $0$):

\begin{defn}\label{defnextendedbbscoord}\index{$\mathcal{B}_n^0$}\index{$\mathcal{B}^0$}\index{$\mathcal{R}_n$}\index{$\mathcal{R}$}
Let 
\begin{align*}
    \mathcal{G}_n^0&=\{\infty\} \x \{0\}\x \N_0^{2(n-1)}\x \{\infty\}\\
    \mathcal{G}^0&=\bigcup\limits_{n\in\N}\mathcal{G}_n^0.
    \end{align*}
The dynamics $\chi: \mathcal{B}\to \mathcal{B}$ defined in Definition \ref{defnofboxballandcoordspaces} naturally extends to a dynamics $\chi^0:\mathcal{G}^0 \to \mathcal{G}^0$.
\end{defn} 

\begin{rem}
Note that $\mathcal{G}_n^0$ and $\mathcal{B}_n$ each have $2n-1$ finite coordinates, the difference is that the first is zero for $\mathcal{G}_n^0$ with the remaining allowed to be any non-negative integers, whereas all must be positive integers for $\mathcal{B}_n$.
\end{rem}

\n Recall the RSK phase space, $\mathcal{R}^n$, introduced in Definition \ref{rskeqnsdefn}. This phase space for RSK is seen to correspond naturally to the phase space $\mathcal{G}_{n+1}^0$ in that the former is a pair of $n$-tuples and the latter is a corresponding $n$-tuple of pairs.

\begin{defn}\index{$\phi_{\text{RSK}\to\text{BBS}}$}\index{$\phi_{\text{BBS}\to\text{RSK}}$}\label{phidefnsforref}
For a pair of sequences $\mb{a}=(a_1,\ldots,a_n),~\mb{x}=(x_1,\ldots,x_n)\in\mathbb{N}_0^n$, define a map $\phi_{\text{RSK}\to \text{BBS}}:\mathcal{R}^n\to \mathcal{G}_{n+1}^0$ by \vspace{0.2cm}
\begin{equation}\phi^n_{\text{RSK}\to \text{BBS}}(\mb{a},\mb{x})=(\infty,0,x_1,a_1,x_2,a_2,\ldots,x_n,a_n,\infty).\end{equation}

\n Conversely, for a sequence $\mb{z}\in \mathcal{G}_{n+1}^0$, define a map $\phi_{\text{BBS}\to \text{RSK}}: \mathcal{G}_{n+1}^0\to \mathcal{R}^n$ by\vspace{0.2cm}
\begin{equation}\phi^n_{\text{BBS}\to \text{RSK}}(\infty,0,z_1,\ldots,z_{2n},\infty)=((0,z_2,\ldots,z_{2(n-1)}),(z_1,z_3,\ldots,z_{2n-1})).\end{equation}

\n Let $\phi_{\text{RSK}\to \text{BBS}}:\mathcal{R}\to\mathcal{G}^0$ and $\phi_{\text{BBS}\to \text{RSK}}:\mathcal{G}^0\to\mathcal{R}$ be their natural extensions.
\end{defn}

\begin{rem}
Note that $\phi_{\text{RSK}\to\text{BBS}}$ is a bijective mapping from $\mathcal{R}^n$ to $\mathcal{G}_{n+1}^0$. However $\phi_{\text{BBS}\to\text{RSK}}$ is not its inverse, in fact $\phi_{\text{BBS}\to\text{RSK}}$ is neither injective nor surjective. This property of $\phi_{\text{BBS}\to\text{RSK}}$ is a consequence of the shift in the first equation of \ref{changevarsbbstorsk}.
\end{rem}

\n We summarise the calculations of this section in the following theorem:
\begin{thm}\label{rskISbBScoordevOL}
Under the box-ball evolution $\chi^0:\mathcal{G}_{n+1}^0\to \mathcal{G}_{n+1}^0$, RSK insertion is captured as the following
\begin{equation}
\chi^0(\infty,0,x_1,a_1,x_2,\ldots,a_{n-1},x_n,a_n,\infty)=\left(\infty,b_1,y_1,b_2,y_2,\ldots,b_n,y_n,b_{n+1},\infty\right),
\end{equation} 
noting that $b_1=0$.
\end{thm}

\begin{cor}\label{rskisbbscoords}
One has
\begin{equation}\text{RSK}=\phi_{\text{BBS}\to \text{RSK}}\circ \chi^0\circ \phi_{\text{RSK}\to \text{BBS}}.
\end{equation}
\end{cor}

\subsection{The Ghost-Box-Ball System}
We now introduce the ghost-box-ball system which is designed to be the cellular automaton realisation of $\chi^0$, given in Definition \ref{defnextendedbbscoord}. This amounts to modifying the original box-ball system to reflect the zeroes that we are allowing into the box-ball coordinates. Ultimately, what we want is a modified box-ball system into which one can encode an RSK pair, and from whose evolution one can read off the RSK output.

\begin{defn}\label{defnofgbbsstate}\index{Ghost-Box-Ball System (GBBS)}
A \textbf{ghost-box-ball system} consists of a one-dimensional infinite array of boxes with a finite number of boxes designated precisely one of the following three states (the rest of the boxes are \textit{empty}):
\begin{enumerate}
\item filled (with a ball),
\item filled ghost,
\item empty ghost,
\end{enumerate}
and subject to the following constraints:
\begin{enumerate}
\item a filled ghost may not be adjacent to another filled ghost, nor to a filled box, and
\item an empty ghost may not be adjacent to another empty ghost, nor to an empty box.
\end{enumerate}
We let $\text{GBBS}$ denote the set of all ghost-box-ball states.
\end{defn}

\begin{rem}
For the purpose of this section, we will only be interested in a particular class of ghost-box-ball states: the set of ghost-box-ball states for which the left-most box that isn't an empty box is a filled ghost. For the rest of this section, until our conclusions, we will use the term ``ghost-box-ball'' and the notation $\text{GBBS}$ to refer to this particular subset of interest. We will discuss extensions in Section \ref{chapterextensions}, where the full set of ghost-box-ball states will be studied.
\end{rem}

\n For a graphical representation of the ghost-box-ball states, we employ the following key:
\begin{enumerate}
\item An empty box shall be represented by \tikz[scale=0.5]{\draw[fill=white]  (0,0) -- (1,0) -- (1,1) -- (0,1) -- cycle;}
\item A filled box shall be represented by \tikz[scale=0.5]{\draw[fill=white]  (0,0) -- (1,0) -- (1,1) -- (0,1) -- cycle;\fill[cyan] (0.5,0.5) circle (0.25);}
\item A filled ghost shall be represented by \tikz[scale=0.5]{\draw[fill=black]  (0,0) -- (0.5,0) -- (0.5,1) -- (0,1) -- cycle;}
\item An empty ghost shall be represented by \tikz[scale=0.5]{\draw[fill=gray!40]  (0,0) -- (0.5,0) -- (0.5,1) -- (0,1) -- cycle;}
\end{enumerate}

\begin{ex} As an example, the following is a ghost-box-ball state:
\begin{figure}[H]
\centering
\tikz[scale=0.54]{
\foreach \y in {0}
{
\foreach \x in {1.0,7.5}
{\draw[fill=black]  (\x,\y-1) -- (\x+0.5,\y-1) -- (\x+0.5,\y) -- (\x,\y) -- cycle;			
}
\foreach \x in {5.5,7.0,14.0}
{\draw[fill=gray!40]  (\x,\y-1) -- (\x+0.5,\y-1) -- (\x+0.5,\y) -- (\x,\y) -- cycle;			
}
\foreach \x in {1.5,2.5,8.0,9.0,11.0,12.0,17.5}
{\draw[fill=white]  (\x,\y-1) -- (\x+1,\y-1) -- (\x+1,\y) -- (\x,\y) -- cycle;				
}
\foreach \x in {3.5,4.5,6.0,10.0,13.0,14.5,15.5,16.5}
{\draw[fill=white]  (\x,\y-1) -- (\x+1,\y-1) -- (\x+1,\y) -- (\x,\y) -- cycle;				
\fill[cyan] (\x+0.5,\y-0.5) circle (0.25);									
}
\foreach \x in {}
{\draw[fill=white]  (\x,\y-1) -- (\x+1,\y-1) -- (\x+1,\y) -- (\x,\y) -- cycle;				
\fill[red] (\x+0.5,\y-0.5) circle (0.25);									
}
\foreach \x in {18.5}
{\draw[fill=white,white]  (\x,\y-1) -- (\x+2,\y-1) -- (\x+2,\y) -- (\x,\y) -- cycle;
\draw[-] (\x,\y-1) -- (\x,\y);
\draw[-] (\x,\y-1) -- (\x+2,\y-1);
\draw[-] (\x,\y) -- (\x+2,\y);
\draw[-] (\x+1,\y-1) -- (\x+1,\y);
\node at (\x+1.5,\y-0.5) {$\cdots$};
}
\foreach \x in {1}
{\draw[fill=white,white]  (\x,\y-1) -- (\x-2,\y-1) -- (\x-2,\y) -- (\x,\y) -- cycle;
\draw[-] (\x,\y-1) -- (\x,\y);
\draw[-] (\x,\y-1) -- (\x-2,\y-1);
\draw[-] (\x,\y) -- (\x-2,\y);
\draw[-] (\x-1,\y-1) -- (\x-1,\y);
\node at (\x-1.5,\y-0.5) {$\cdots$};
}
}
}
\end{figure}
\n Note that no filled ghost is neighbours either a ball or another filled ghost, nor does any empty ghost neighbour an empty box or another empty ghost.
\end{ex}

\begin{rem}
Although ghost-box-ball states extend infinitely right by empty boxes, we will often truncate them in our depiction with the understanding that they still extend infinitely to the right. Similarly, since all ghost-box-ball states appearing in this section have as their left-most non-empty box a filled ghost, we will always truncate at the filled ghost in our graphical representations, knowing that there are infinitely many empty boxes to the left of the first filled ghost.\\

\n For example, the above ghost-box-ball state might be depicted as follows:
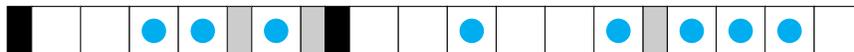
\begin{figure}[H]
\centering
\tikz[scale=0.65]{
\foreach \y in {0}
{
\foreach \x in {1.0,7.5}
{\draw[fill=black]  (\x,\y-1) -- (\x+0.5,\y-1) -- (\x+0.5,\y) -- (\x,\y) -- cycle;			
}
\foreach \x in {5.5,7.0,14.0}
{\draw[fill=gray!40]  (\x,\y-1) -- (\x+0.5,\y-1) -- (\x+0.5,\y) -- (\x,\y) -- cycle;			
}
\foreach \x in {1.5,2.5,8.0,9.0,11.0,12.0,17.5}
{\draw[fill=white]  (\x,\y-1) -- (\x+1,\y-1) -- (\x+1,\y) -- (\x,\y) -- cycle;				
}
\foreach \x in {3.5,4.5,6.0,10.0,13.0,14.5,15.5,16.5}
{\draw[fill=white]  (\x,\y-1) -- (\x+1,\y-1) -- (\x+1,\y) -- (\x,\y) -- cycle;				
\fill[cyan] (\x+0.5,\y-0.5) circle (0.25);									
}
\foreach \x in {}
{\draw[fill=white]  (\x,\y-1) -- (\x+1,\y-1) -- (\x+1,\y) -- (\x,\y) -- cycle;				
\fill[red] (\x+0.5,\y-0.5) circle (0.25);									
}
\foreach \x in {}
{\draw[fill=white,white]  (\x,\y-1) -- (\x+2,\y-1) -- (\x+2,\y) -- (\x,\y) -- cycle;
\draw[-] (\x,\y-1) -- (\x,\y);
\draw[-] (\x,\y-1) -- (\x+2,\y-1);
\draw[-] (\x,\y) -- (\x+2,\y);
\draw[-] (\x+1,\y-1) -- (\x+1,\y);
\node at (\x+1.5,\y-0.5) {$\cdots$};
}
}
}
\end{figure}
\end{rem}

\n We now define the following evolution rule on GBBS.

\begin{mdframed}[linewidth=1.2pt]
\begin{defn}\index{Ghost-Box-Ball Algorithm}\label{defnofgbbsevolalg}
\textbf{(The Ghost-Box-Ball Algorithm)}\index{Ghost-Box-Ball Algorithm}
\begin{enumerate}
\item Move each ball exactly once.
\item Move the leftmost unmoved ball to its nearest right empty box.
\item If a ball's new position has a filled ghost to its immediate right, materialise (create) an empty ghost between them.
\item If a ball's new position has a filled ghost to its immediate left, exorcise (delete) the ghost.
\item If a ball is moved from a position with an empty ghost right-adjacent of it, insert a filled ghost between the box vacated by the ball and the empty ghost.
\item If a ball is moved from a position with an empty ghost left-adjacent of it, exorcise that ghost.
\item Repeat (2)-(6) until all balls have been moved.
\end{enumerate}
\end{defn}
\end{mdframed}

~

\begin{rem}
Performing the whole ghost-box-ball algorithm constitutes one \textbf{time-step} of the evolution of a ghost-box-ball system, with each ball movement (together with any resulting materialisations and exorcisms) constituting a \textbf{stage} of the overall time-step.
\end{rem}

~\\[-25pt]

\begin{ex}\label{bbsexamplesteps}
We demonstrate the ghost-box-ball evolution, stage-by-stage below:
\begin{figure}[H]
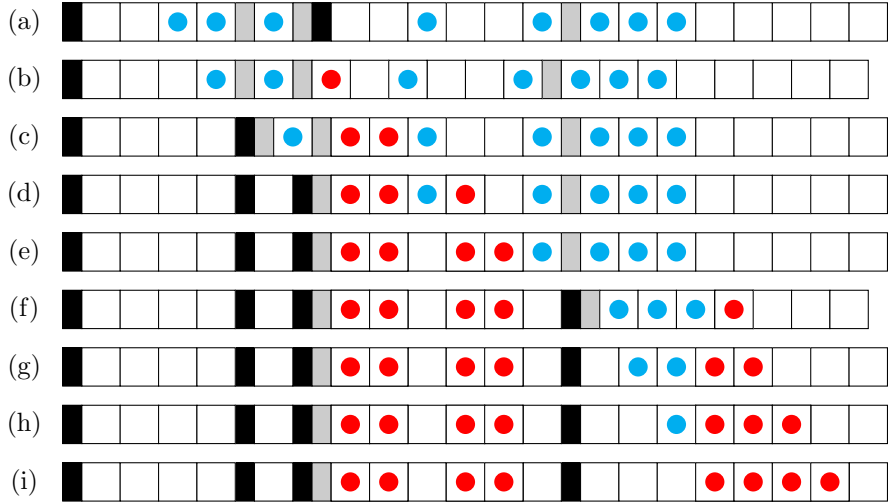

\centering
\tikz[scale=0.51]{
\foreach \y in {0}
{
\node at (0,\y-0.5) {(a)};
\foreach \x in {1.0,7.5}
{\draw[fill=black]  (\x,\y-1) -- (\x+0.5,\y-1) -- (\x+0.5,\y) -- (\x,\y) -- cycle;			
}
\foreach \x in {5.5,7.0,14.0}
{\draw[fill=gray!40]  (\x,\y-1) -- (\x+0.5,\y-1) -- (\x+0.5,\y) -- (\x,\y) -- cycle;			
}
\foreach \x in {1.5,2.5,8.0,9.0,11.0,12.0,17.5,18.5,19.5,20.5,21.5}
{\draw[fill=white]  (\x,\y-1) -- (\x+1,\y-1) -- (\x+1,\y) -- (\x,\y) -- cycle;				
}
\foreach \x in {3.5,4.5,6.0,10.0,13.0,14.5,15.5,16.5}
{\draw[fill=white]  (\x,\y-1) -- (\x+1,\y-1) -- (\x+1,\y) -- (\x,\y) -- cycle;				
\fill[cyan] (\x+0.5,\y-0.5) circle (0.25);									
}
\foreach \x in {}
{\draw[fill=white]  (\x,\y-1) -- (\x+1,\y-1) -- (\x+1,\y) -- (\x,\y) -- cycle;				
\fill[red] (\x+0.5,\y-0.5) circle (0.25);									
}
\foreach \x in {}
{\draw[fill=white,white]  (\x,\y-1) -- (\x+2,\y-1) -- (\x+2,\y) -- (\x,\y) -- cycle;
\draw[-] (\x,\y-1) -- (\x,\y);
\draw[-] (\x,\y-1) -- (\x+2,\y-1);
\draw[-] (\x,\y) -- (\x+2,\y);
\draw[-] (\x+1,\y-1) -- (\x+1,\y);
\node at (\x+1.5,\y-0.5) {$\cdots$};
}
}
\foreach \y in {-1.5}
{
\node at (0,\y-0.5) {(b)};
\foreach \x in {1.0}
{\draw[fill=black]  (\x,\y-1) -- (\x+0.5,\y-1) -- (\x+0.5,\y) -- (\x,\y) -- cycle;			
}
\foreach \x in {5.5,7.0,13.5}
{\draw[fill=gray!40]  (\x,\y-1) -- (\x+0.5,\y-1) -- (\x+0.5,\y) -- (\x,\y) -- cycle;			
}
\foreach \x in {1.5,2.5,3.5,8.5,10.5,11.5,17.0,18,19,20,21}
{\draw[fill=white]  (\x,\y-1) -- (\x+1,\y-1) -- (\x+1,\y) -- (\x,\y) -- cycle;				
}
\foreach \x in {4.5,6.0,9.5,12.5,14.0,15.0,16.0}
{\draw[fill=white]  (\x,\y-1) -- (\x+1,\y-1) -- (\x+1,\y) -- (\x,\y) -- cycle;				
\fill[cyan] (\x+0.5,\y-0.5) circle (0.25);									
}
\foreach \x in {7.5}
{\draw[fill=white]  (\x,\y-1) -- (\x+1,\y-1) -- (\x+1,\y) -- (\x,\y) -- cycle;				
\fill[red] (\x+0.5,\y-0.5) circle (0.25);									
}
\foreach \x in {}
{\draw[fill=white,white]  (\x,\y-1) -- (\x+2,\y-1) -- (\x+2,\y) -- (\x,\y) -- cycle;
\draw[-] (\x,\y-1) -- (\x,\y);
\draw[-] (\x,\y-1) -- (\x+2,\y-1);
\draw[-] (\x,\y) -- (\x+2,\y);
\draw[-] (\x+1,\y-1) -- (\x+1,\y);
\node at (\x+1.5,\y-0.5) {$\cdots$};
}
}
\foreach \y in {-3}
{
\node at (0,\y-0.5) {(c)};
\foreach \x in {1.0,5.5}
{\draw[fill=black]  (\x,\y-1) -- (\x+0.5,\y-1) -- (\x+0.5,\y) -- (\x,\y) -- cycle;			
}
\foreach \x in {6.0,7.5,14.0}
{\draw[fill=gray!40]  (\x,\y-1) -- (\x+0.5,\y-1) -- (\x+0.5,\y) -- (\x,\y) -- cycle;			
}
\foreach \x in {1.5,2.5,3.5,4.5,11.0,12.0,17.5,18.5,19.5,20.5,21.5}
{\draw[fill=white]  (\x,\y-1) -- (\x+1,\y-1) -- (\x+1,\y) -- (\x,\y) -- cycle;				
}
\foreach \x in {6.5,8.0,9.0,10.0,13.0,14.5,15.5,16.5}
{\draw[fill=white]  (\x,\y-1) -- (\x+1,\y-1) -- (\x+1,\y) -- (\x,\y) -- cycle;				
\fill[cyan] (\x+0.5,\y-0.5) circle (0.25);									
}
\foreach \x in {8,9}
{\draw[fill=white]  (\x,\y-1) -- (\x+1,\y-1) -- (\x+1,\y) -- (\x,\y) -- cycle;				
\fill[red] (\x+0.5,\y-0.5) circle (0.25);									
}
\foreach \x in {}
{\draw[fill=white,white]  (\x,\y-1) -- (\x+2,\y-1) -- (\x+2,\y) -- (\x,\y) -- cycle;
\draw[-] (\x,\y-1) -- (\x,\y);
\draw[-] (\x,\y-1) -- (\x+2,\y-1);
\draw[-] (\x,\y) -- (\x+2,\y);
\draw[-] (\x+1,\y-1) -- (\x+1,\y);
\node at (\x+1.5,\y-0.5) {$\cdots$};
}
}
\foreach \y in {-4.5}
{
\node at (0,\y-0.5) {(d)};
\foreach \x in {1.0,5.5,7.0}
{\draw[fill=black]  (\x,\y-1) -- (\x+0.5,\y-1) -- (\x+0.5,\y) -- (\x,\y) -- cycle;			
}
\foreach \x in {7.5,14.0}
{\draw[fill=gray!40]  (\x,\y-1) -- (\x+0.5,\y-1) -- (\x+0.5,\y) -- (\x,\y) -- cycle;			
}
\foreach \x in {1.5,2.5,3.5,4.5,6.0,12.0,17.5,18.5,19.5,20.5,21.5}
{\draw[fill=white]  (\x,\y-1) -- (\x+1,\y-1) -- (\x+1,\y) -- (\x,\y) -- cycle;				
}
\foreach \x in {8.0,9.0,10.0,11.0,13.0,14.5,15.5,16.5}
{\draw[fill=white]  (\x,\y-1) -- (\x+1,\y-1) -- (\x+1,\y) -- (\x,\y) -- cycle;				
\fill[cyan] (\x+0.5,\y-0.5) circle (0.25);									
}
\foreach \x in {8,9,11}
{\draw[fill=white]  (\x,\y-1) -- (\x+1,\y-1) -- (\x+1,\y) -- (\x,\y) -- cycle;				
\fill[red] (\x+0.5,\y-0.5) circle (0.25);									
}
\foreach \x in {}
{\draw[fill=white,white]  (\x,\y-1) -- (\x+2,\y-1) -- (\x+2,\y) -- (\x,\y) -- cycle;
\draw[-] (\x,\y-1) -- (\x,\y);
\draw[-] (\x,\y-1) -- (\x+2,\y-1);
\draw[-] (\x,\y) -- (\x+2,\y);
\draw[-] (\x+1,\y-1) -- (\x+1,\y);
\node at (\x+1.5,\y-0.5) {$\cdots$};
}
}
\foreach \y in {-6}
{
\node at (0,\y-0.5) {(e)};
\foreach \x in {1.0,5.5,7.0}
{\draw[fill=black]  (\x,\y-1) -- (\x+0.5,\y-1) -- (\x+0.5,\y) -- (\x,\y) -- cycle;			
}
\foreach \x in {7.5,14.0}
{\draw[fill=gray!40]  (\x,\y-1) -- (\x+0.5,\y-1) -- (\x+0.5,\y) -- (\x,\y) -- cycle;			
}
\foreach \x in {1.5,2.5,3.5,4.5,6.0,10.0,17.5,18.5,19.5,20.5,21.5}
{\draw[fill=white]  (\x,\y-1) -- (\x+1,\y-1) -- (\x+1,\y) -- (\x,\y) -- cycle;				
}
\foreach \x in {8.0,9.0,11.0,12.0,13.0,14.5,15.5,16.5}
{\draw[fill=white]  (\x,\y-1) -- (\x+1,\y-1) -- (\x+1,\y) -- (\x,\y) -- cycle;				
\fill[cyan] (\x+0.5,\y-0.5) circle (0.25);									
}
\foreach \x in {8,9,11,12}
{\draw[fill=white]  (\x,\y-1) -- (\x+1,\y-1) -- (\x+1,\y) -- (\x,\y) -- cycle;				
\fill[red] (\x+0.5,\y-0.5) circle (0.25);									
}
\foreach \x in {}
{\draw[fill=white,white]  (\x,\y-1) -- (\x+2,\y-1) -- (\x+2,\y) -- (\x,\y) -- cycle;
\draw[-] (\x,\y-1) -- (\x,\y);
\draw[-] (\x,\y-1) -- (\x+2,\y-1);
\draw[-] (\x,\y) -- (\x+2,\y);
\draw[-] (\x+1,\y-1) -- (\x+1,\y);
\node at (\x+1.5,\y-0.5) {$\cdots$};
}
}
\foreach \y in {-7.5}
{
\node at (0,\y-0.5) {(f)};
\foreach \x in {1.0,5.5,7.0,14.0}
{\draw[fill=black]  (\x,\y-1) -- (\x+0.5,\y-1) -- (\x+0.5,\y) -- (\x,\y) -- cycle;			
}
\foreach \x in {7.5,14.5}
{\draw[fill=gray!40]  (\x,\y-1) -- (\x+0.5,\y-1) -- (\x+0.5,\y) -- (\x,\y) -- cycle;			
}
\foreach \x in {1.5,2.5,3.5,4.5,6.0,10.0,13.0,19.0,20,21}
{\draw[fill=white]  (\x,\y-1) -- (\x+1,\y-1) -- (\x+1,\y) -- (\x,\y) -- cycle;				
}
\foreach \x in {8.0,9.0,11.0,12.0,15.0,16.0,17.0,18.0}
{\draw[fill=white]  (\x,\y-1) -- (\x+1,\y-1) -- (\x+1,\y) -- (\x,\y) -- cycle;				
\fill[cyan] (\x+0.5,\y-0.5) circle (0.25);									
}
\foreach \x in {8,9,11,12,18}
{\draw[fill=white]  (\x,\y-1) -- (\x+1,\y-1) -- (\x+1,\y) -- (\x,\y) -- cycle;				
\fill[red] (\x+0.5,\y-0.5) circle (0.25);									
}
\foreach \x in {}
{\draw[fill=white,white]  (\x,\y-1) -- (\x+2,\y-1) -- (\x+2,\y) -- (\x,\y) -- cycle;
\draw[-] (\x,\y-1) -- (\x,\y);
\draw[-] (\x,\y-1) -- (\x+2,\y-1);
\draw[-] (\x,\y) -- (\x+2,\y);
\draw[-] (\x+1,\y-1) -- (\x+1,\y);
\node at (\x+1.5,\y-0.5) {$\cdots$};
}
}
\foreach \y in {-9}
{
\node at (0,\y-0.5) {(g)};
\foreach \x in {1.0,5.5,7.0,14.0}
{\draw[fill=black]  (\x,\y-1) -- (\x+0.5,\y-1) -- (\x+0.5,\y) -- (\x,\y) -- cycle;			
}
\foreach \x in {7.5}
{\draw[fill=gray!40]  (\x,\y-1) -- (\x+0.5,\y-1) -- (\x+0.5,\y) -- (\x,\y) -- cycle;			
}
\foreach \x in {1.5,2.5,3.5,4.5,6.0,10.0,13.0,14.5,19.5,20.5,21.5}
{\draw[fill=white]  (\x,\y-1) -- (\x+1,\y-1) -- (\x+1,\y) -- (\x,\y) -- cycle;				
}
\foreach \x in {8.0,9.0,11.0,12.0,15.5,16.5,17.5,18.5}
{\draw[fill=white]  (\x,\y-1) -- (\x+1,\y-1) -- (\x+1,\y) -- (\x,\y) -- cycle;				
\fill[cyan] (\x+0.5,\y-0.5) circle (0.25);									
}
\foreach \x in {8,9,11,12,17.5,18.5}
{\draw[fill=white]  (\x,\y-1) -- (\x+1,\y-1) -- (\x+1,\y) -- (\x,\y) -- cycle;				
\fill[red] (\x+0.5,\y-0.5) circle (0.25);									
}
\foreach \x in {}
{\draw[fill=white,white]  (\x,\y-1) -- (\x+2,\y-1) -- (\x+2,\y) -- (\x,\y) -- cycle;
\draw[-] (\x,\y-1) -- (\x,\y);
\draw[-] (\x,\y-1) -- (\x+2,\y-1);
\draw[-] (\x,\y) -- (\x+2,\y);
\draw[-] (\x+1,\y-1) -- (\x+1,\y);
\node at (\x+1.5,\y-0.5) {$\cdots$};
}
}
\foreach \y in {-10.5}
{
\node at (0,\y-0.5) {(h)};
\foreach \x in {1.0,5.5,7.0,14.0}
{\draw[fill=black]  (\x,\y-1) -- (\x+0.5,\y-1) -- (\x+0.5,\y) -- (\x,\y) -- cycle;			
}
\foreach \x in {7.5}
{\draw[fill=gray!40]  (\x,\y-1) -- (\x+0.5,\y-1) -- (\x+0.5,\y) -- (\x,\y) -- cycle;			
}
\foreach \x in {1.5,2.5,3.5,4.5,6.0,10.0,13.0,14.5,15.5,20.5,21.5}
{\draw[fill=white]  (\x,\y-1) -- (\x+1,\y-1) -- (\x+1,\y) -- (\x,\y) -- cycle;				
}
\foreach \x in {8.0,9.0,11.0,12.0,16.5,17.5,18.5,19.5}
{\draw[fill=white]  (\x,\y-1) -- (\x+1,\y-1) -- (\x+1,\y) -- (\x,\y) -- cycle;				
\fill[cyan] (\x+0.5,\y-0.5) circle (0.25);									
}
\foreach \x in {8,9,11,12,17.5,18.5,19.5}
{\draw[fill=white]  (\x,\y-1) -- (\x+1,\y-1) -- (\x+1,\y) -- (\x,\y) -- cycle;				
\fill[red] (\x+0.5,\y-0.5) circle (0.25);									
}
\foreach \x in {}
{\draw[fill=white,white]  (\x,\y-1) -- (\x+2,\y-1) -- (\x+2,\y) -- (\x,\y) -- cycle;
\draw[-] (\x,\y-1) -- (\x,\y);
\draw[-] (\x,\y-1) -- (\x+2,\y-1);
\draw[-] (\x,\y) -- (\x+2,\y);
\draw[-] (\x+1,\y-1) -- (\x+1,\y);
\node at (\x+1.5,\y-0.5) {$\cdots$};
}
}
\foreach \y in {-12}
{
\node at (0,\y-0.5) {(i)};
\foreach \x in {1.0,5.5,7.0,14.0}
{\draw[fill=black]  (\x,\y-1) -- (\x+0.5,\y-1) -- (\x+0.5,\y) -- (\x,\y) -- cycle;			
}
\foreach \x in {7.5}
{\draw[fill=gray!40]  (\x,\y-1) -- (\x+0.5,\y-1) -- (\x+0.5,\y) -- (\x,\y) -- cycle;			
}
\foreach \x in {1.5,2.5,3.5,4.5,6.0,10.0,13.0,14.5,15.5,16.5,21.5}
{\draw[fill=white]  (\x,\y-1) -- (\x+1,\y-1) -- (\x+1,\y) -- (\x,\y) -- cycle;				
}
\foreach \x in {8.0,9.0,11.0,12.0,17.5,18.5,19.5,20.5}
{\draw[fill=white]  (\x,\y-1) -- (\x+1,\y-1) -- (\x+1,\y) -- (\x,\y) -- cycle;				
\fill[cyan] (\x+0.5,\y-0.5) circle (0.25);									
}
\foreach \x in {8,9,11,12,17.5,18.5,19.5,20.5}
{\draw[fill=white]  (\x,\y-1) -- (\x+1,\y-1) -- (\x+1,\y) -- (\x,\y) -- cycle;				
\fill[red] (\x+0.5,\y-0.5) circle (0.25);									
}
\foreach \x in {}
{\draw[fill=white,white]  (\x,\y-1) -- (\x+2,\y-1) -- (\x+2,\y) -- (\x,\y) -- cycle;
\draw[-] (\x,\y-1) -- (\x,\y);
\draw[-] (\x,\y-1) -- (\x+2,\y-1);
\draw[-] (\x,\y) -- (\x+2,\y);
\draw[-] (\x+1,\y-1) -- (\x+1,\y);
\node at (\x+1.5,\y-0.5) {$\cdots$};
}
}
}
\caption{A single time-step of the ghost-box-ball evolution, split into the stages that make it up.}
\label{GBBSStepsofevolwithint}
\end{figure}
\n As in Section \ref{bbesubsec}, we employ the blue-red colouring for unmoved-moved balls to aid in the visual tracking of the algorithm. Since the Ghost-Box-Ball Algorithm is more involved than the classical Box-Ball Algorithm, we also provide some supplementary discussion on the first few stages in the above example.

\begin{enumerate}
\item From (a) to (b): we move the left-most blue ball to the nearest empty box to its right. The ball was not neighbouring a ghost initially, so its removal from the initial position does not in itself prompt a materialisation or exorcism of a ghost. However, its new position contains a \textbf{left}-neighbouring filled ghost. By \ref{defnofgbbsevolalg}(4), we must \textbf{exorcise} that ghost.
\item From (b) to (c): the next ball to move does so to a new position not neighboured by a ghost, however its initial position has a \textbf{right}-neighbouring empty ghost. By \ref{defnofgbbsevolalg}(5), an filled ghost must \textbf{materialise}.
\item From (c) to (d): we have ghosts neighbouring both sides of the initial position of the next ball to move. Since we have discussed initial right-neighbouring in the previous step, we focus on the \textbf{left}-neighbouring empty ghost to our moving ball. By \ref{defnofgbbsevolalg}(6), the empty ghost to the \textbf{left} must be \textbf{exorcised}.
\end{enumerate}
\n As a mnemonic device: 
\begin{itemize}
\item \textbf{left}-adjacency leads to \textbf{exorcism}: where a movement would violate a constraint due to a \textbf{left}-adjacent ghost, that ghost must be \textbf{exorcised}.
\item \textbf{right}-adjacency leads to \textbf{materialisation}: where a constraint would be violated due to a \textbf{right}-adjacent ghost, a ghost must \textbf{materialise} to restore order.
\end{itemize}

\n Suppressing the intermediate stages (b)-(h), the single time-step of the ghost box ball evolution corresponding to Example \ref{bbsexamplesteps} is summarised in the following:
\begin{figure}[H]
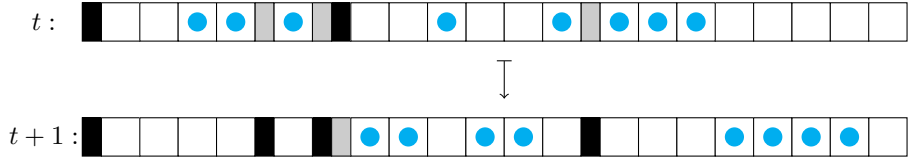

\centering
\tikz[scale=0.51]{
\foreach \y in {0}
{
\node at (0,\y-0.5) {$t:$};
\foreach \x in {1.0,7.5}
{\draw[fill=black]  (\x,\y-1) -- (\x+0.5,\y-1) -- (\x+0.5,\y) -- (\x,\y) -- cycle;			
}
\foreach \x in {5.5,7.0,14.0}
{\draw[fill=gray!40]  (\x,\y-1) -- (\x+0.5,\y-1) -- (\x+0.5,\y) -- (\x,\y) -- cycle;			
}
\foreach \x in {1.5,2.5,8.0,9.0,11.0,12.0,17.5,18.5,19.5,20.5,21.5}
{\draw[fill=white]  (\x,\y-1) -- (\x+1,\y-1) -- (\x+1,\y) -- (\x,\y) -- cycle;				
}
\foreach \x in {3.5,4.5,6.0,10.0,13.0,14.5,15.5,16.5}
{\draw[fill=white]  (\x,\y-1) -- (\x+1,\y-1) -- (\x+1,\y) -- (\x,\y) -- cycle;				
\fill[cyan] (\x+0.5,\y-0.5) circle (0.25);									
}
\foreach \x in {}
{\draw[fill=white]  (\x,\y-1) -- (\x+1,\y-1) -- (\x+1,\y) -- (\x,\y) -- cycle;				
\fill[red] (\x+0.5,\y-0.5) circle (0.25);									
}
\foreach \x in {}
{\draw[fill=white,white]  (\x,\y-1) -- (\x+2,\y-1) -- (\x+2,\y) -- (\x,\y) -- cycle;
\draw[-] (\x,\y-1) -- (\x,\y);
\draw[-] (\x,\y-1) -- (\x+2,\y-1);
\draw[-] (\x,\y) -- (\x+2,\y);
\draw[-] (\x+1,\y-1) -- (\x+1,\y);
\node at (\x+1.5,\y-0.5) {$\cdots$};
}
}
\draw[|->] (12,-1.5) -- (12,-2.5);
\foreach \y in {-3}
{
\node at (0,\y-0.5) {$t+1:$};
\foreach \x in {1.0,5.5,7.0,14.0}
{\draw[fill=black]  (\x,\y-1) -- (\x+0.5,\y-1) -- (\x+0.5,\y) -- (\x,\y) -- cycle;			
}
\foreach \x in {7.5}
{\draw[fill=gray!40]  (\x,\y-1) -- (\x+0.5,\y-1) -- (\x+0.5,\y) -- (\x,\y) -- cycle;			
}
\foreach \x in {1.5,2.5,3.5,4.5,6.0,10.0,13.0,14.5,15.5,16.5,21.5}
{\draw[fill=white]  (\x,\y-1) -- (\x+1,\y-1) -- (\x+1,\y) -- (\x,\y) -- cycle;				
}
\foreach \x in {8.0,9.0,11.0,12.0,17.5,18.5,19.5,20.5}
{\draw[fill=white]  (\x,\y-1) -- (\x+1,\y-1) -- (\x+1,\y) -- (\x,\y) -- cycle;				
\fill[cyan] (\x+0.5,\y-0.5) circle (0.25);									
}
\foreach \x in {}
{\draw[fill=white]  (\x,\y-1) -- (\x+1,\y-1) -- (\x+1,\y) -- (\x,\y) -- cycle;				
\fill[red] (\x+0.5,\y-0.5) circle (0.25);									
}
\foreach \x in {}
{\draw[fill=white,white]  (\x,\y-1) -- (\x+2,\y-1) -- (\x+2,\y) -- (\x,\y) -- cycle;
\draw[-] (\x,\y-1) -- (\x,\y);
\draw[-] (\x,\y-1) -- (\x+2,\y-1);
\draw[-] (\x,\y) -- (\x+2,\y);
\draw[-] (\x+1,\y-1) -- (\x+1,\y);
\node at (\x+1.5,\y-0.5) {$\cdots$};
}
}
}
\caption{A single time-step of the ghost-box-ball evolution (without the intermediate stages).}\label{onetimestepgbbsforref}
\end{figure}
\end{ex}

\begin{lem}
The result of performing this algorithm on a ghost-box-ball system is again a ghost-box-ball system.
\end{lem}
\begin{proof}
This is just a consequence of the construction: (3)-(6) serves the purpose of keeping the constraints of the ghost-box-ball definition (Definition \ref{defnofgbbsstate}) satisfied.
\end{proof}

\begin{defn}\index{$\hat{\varrho}$}
The map $\hat{\varrho}:\text{GBBS}\to\text{GBBS}$ will be defined to be the result of applying this algorithm. 
\end{defn}

\begin{defn}\label{blockdefn}\index{Ghost-Box-Ball Blocks}
In a ghost-box-ball state, a filled block will be either of the following:
\begin{enumerate}
    \item a maximal sequence of adjacent balls, or
    \item a single filled ghost.
\end{enumerate}
Similarly, an empty block will be either of the following:
\begin{enumerate}
    \item a maximal sequence of adjacent empty boxes, or
    \item a single empty ghost.
\end{enumerate}
\end{defn}

\begin{rem}\label{remarkstructuregbbs}
Since filled ghosts cannot neighbour other filled ghosts or filled boxes, and empty ghosts cannot neighbour other empty ghosts or empty boxes, we make the following observations of the general structure of a ghost-box-ball state:
\begin{itemize}
    \item The infinite sequences of empty boxes at the beginning and end constitutes the infinite empty block, with all other blocks (both filled and empty) being finite.
    \item A ghost-box-ball state consists of blocks, alternating between the finite filled blocks and empty blocks, terminating in the infinite empty block.
\end{itemize}
\end{rem}

\n For reference, consider the time $t$ state in Figure \ref{onetimestepgbbsforref}:
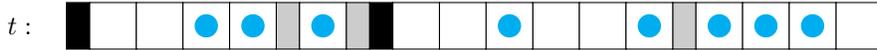
\begin{figure}[H]
\centering
\tikz[scale=0.62]{
\foreach \y in {0}
{
\node at (0,\y-0.5) {$t:$};
\foreach \x in {1.0,7.5}
{\draw[fill=black]  (\x,\y-1) -- (\x+0.5,\y-1) -- (\x+0.5,\y) -- (\x,\y) -- cycle;			
}
\foreach \x in {5.5,7.0,14.0}
{\draw[fill=gray!40]  (\x,\y-1) -- (\x+0.5,\y-1) -- (\x+0.5,\y) -- (\x,\y) -- cycle;			
}
\foreach \x in {1.5,2.5,8.0,9.0,11.0,12.0,17.5}
{\draw[fill=white]  (\x,\y-1) -- (\x+1,\y-1) -- (\x+1,\y) -- (\x,\y) -- cycle;				
}
\foreach \x in {3.5,4.5,6.0,10.0,13.0,14.5,15.5,16.5}
{\draw[fill=white]  (\x,\y-1) -- (\x+1,\y-1) -- (\x+1,\y) -- (\x,\y) -- cycle;				
\fill[cyan] (\x+0.5,\y-0.5) circle (0.25);									
}
\foreach \x in {}
{\draw[fill=white]  (\x,\y-1) -- (\x+1,\y-1) -- (\x+1,\y) -- (\x,\y) -- cycle;				
\fill[red] (\x+0.5,\y-0.5) circle (0.25);									
}
\foreach \x in {}
{\draw[fill=white,white]  (\x,\y-1) -- (\x+2,\y-1) -- (\x+2,\y) -- (\x,\y) -- cycle;
\draw[-] (\x,\y-1) -- (\x,\y);
\draw[-] (\x,\y-1) -- (\x+2,\y-1);
\draw[-] (\x,\y) -- (\x+2,\y);
\draw[-] (\x+1,\y-1) -- (\x+1,\y);
\node at (\x+1.5,\y-0.5) {$\cdots$};
}
}
}
\end{figure}
\n There are fourteen ``blocks'', the initial sequence of which we list in order below:
\begin{enumerate}
    \item Infinitely many empty boxes on the left (suppressed) in the depiction above.
    \item Filled block consisting of one filled ghost.
    \item Empty block consisting of two empty boxes.
    \item Filled block consisting of two balls.
    \item Empty block consisting of one empty ghost.
    \item Filled block consisting of one ball.
    \item Empty block consisting of one empty ghost.
    \item Filled block consisting of one filled ghost.
\end{enumerate}
\textit{etc.}, terminating in the empty block consisting of infinitely many empty boxes.\\[2pt]

\n Between Definition \ref{blockdefn} and Remark \ref{remarkstructuregbbs}, we now have a natural way of extending the earlier coordinate map $C:\text{BBS}\to \mathcal{B}$ as follows:

\begin{defn}\label{coordmapdefn}
We define a coordinatisation $C:\text{GBBS}\to \mathcal{G}^0$ by mapping a ghost-box-ball system to a tuple  $(W_0,Q_1,W_1,\ldots,Q_N,W_N)$, where \vspace{0.35cm}
\begin{align*}
Q_i&=\left\{\begin{array}{cl}
0 & \text{if the $i$-th filled block is a filled ghost}\\
\#\text{length of the $i$-th filled block} & \text{otherwise}
\end{array}\right.\\
W_i&=\left\{\begin{array}{cl}
0 & \text{if the $i$-th empty block is an empty ghost}\\
\#\text{length of the $i$-th empty block} & \text{otherwise}
\end{array}\right.,
\end{align*}
with $W_0=W_N=\infty$.
\end{defn}

\n For each ghost-box-ball system state, its coordinates lie in $\mathcal{G}_n^0$ for some $n\in\N_0$. Therefore, the full set of ghost-box-ball states is identified with $\mathcal{G}^0$.

\subsection{Exorcism, Soliton Behaviour and the Invariant Shape}
In the ghost-box-ball algorithm, ghosts can be exorcised as a result of the movement of balls (either by moving a ball {\it to} the right of an empty ghost or {\it from} the right of an empty ghost). There is a natural map $\maltese:\text{GBBS}\to\text{BBS}$\index{$\maltese$}\index{Global Exorcism} given by exorcising {\it all} ghosts in a ghost-box-ball system, while shifting the remaining balls and boxes to fill in the newly created voids (i.e., create a sequence of empty and filled boxes, scanning the ghost-box-ball system from left to right, ignoring all ghosts). For example, see the following ghost-box-ball state and box-ball state:

\begin{figure}[H]
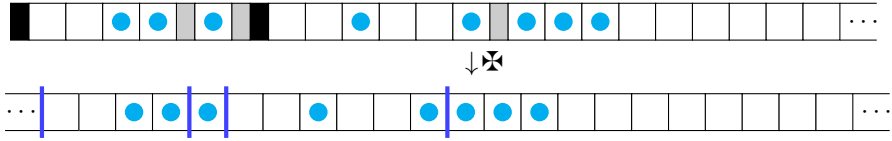

\centering
\tikz[scale=0.49]{
\foreach \y in {0}
{
\foreach \x in {1.0,7.5}
{\draw[fill=black]  (\x,\y-1) -- (\x+0.5,\y-1) -- (\x+0.5,\y) -- (\x,\y) -- cycle;			
}
\foreach \x in {5.5,7.0,14.0}
{\draw[fill=gray!40]  (\x,\y-1) -- (\x+0.5,\y-1) -- (\x+0.5,\y) -- (\x,\y) -- cycle;			
}
\foreach \x in {1.5,2.5,8.0,9.0,11.0,12.0,17.5,18.5,19.5,20.5,21.5,22.5}
{\draw[fill=white]  (\x,\y-1) -- (\x+1,\y-1) -- (\x+1,\y) -- (\x,\y) -- cycle;				
}
\foreach \x in {3.5,4.5,6.0,10.0,13.0,14.5,15.5,16.5}
{\draw[fill=white]  (\x,\y-1) -- (\x+1,\y-1) -- (\x+1,\y) -- (\x,\y) -- cycle;				
\fill[cyan] (\x+0.5,\y-0.5) circle (0.25);									
}
\foreach \x in {}
{\draw[fill=white]  (\x,\y-1) -- (\x+1,\y-1) -- (\x+1,\y) -- (\x,\y) -- cycle;				
\fill[red] (\x+0.5,\y-0.5) circle (0.25);									
}
\foreach \x in {23.5}
{
\draw[-] (\x,\y) -- (\x+1,\y);
\draw[-] (\x,\y-1) -- (\x+1,\y-1);
\node at (\x+0.7,\y-0.5) {$\cdots$};
}
}
}
\tikz[scale=0.49]{
\node at (11,1.81) {$\downarrow\!\maltese$};
\foreach \y in {0}
{
\foreach \x in {0,1,2,3,4,5,6,7,8,9,10,11,12,13,14,15,16,17,18,19,20}
{\draw[fill=white]  (\x,\y) -- (\x+1,\y) -- (\x+1,\y+1) -- (\x,\y+1) -- cycle;			
}
\foreach \x in {1,2,3,6,9,10,11,12}
{
\fill[cyan] (\x+0.5,\y+0.5) circle (0.25);								
}
\foreach \x in {}
{
\fill[red] (\x+0.5,\y+0.5) circle (0.25);									
}
\foreach \x in {21}
{
\draw[-] (\x,\y) -- (\x+1,\y);
\draw[-] (\x,\y+1) -- (\x+1,\y+1);
\node at (\x+0.7,\y+0.5) {$\cdots$};
}
\foreach \x in {0}
{\draw[fill=white,white]  (\x,\y) -- (\x-2,\y) -- (\x-2,\y+1) -- (\x,\y+1) -- cycle;
\draw[-] (\x,\y) -- (\x,\y+1);
\draw[-] (\x,\y) -- (\x-2,\y);
\draw[-] (\x,\y+1) -- (\x-2,\y+1);
\draw[-] (\x-1,\y) -- (\x-1,\y+1);
\node at (\x-1.5,\y+0.5) {$\cdots$};
}
\foreach \x in {-1,3,4,10}
{\draw[-,white] (\x,\y+1) -- (\x,\y);
\draw[-,blue!75,ultra thick] (\x,\y+1.2) -- (\x,\y-0.2);		
}
}
}
\caption{Exorcising ghosts to produce a box-ball state.}
\label{exorcismfirstexample}
\end{figure}

\n The map $\maltese:\text{GBBS}\to\text{BBS}$ shall be referred to as \textit{global exorcism}.
\\

\begin{lem}\label{equivofbbsandgbbs}
The following diagram commutes
\begin{figure}[H]
\centering
\tikz[scale=0.7]{
\node (1) at (0,0) {GBBS};
\node (2) at (3,0) {GBBS};
\node (3) at (0,-3) {BBS};
\node (4) at (3,-3) {BBS};
\draw[->] (1) -- (2) node[above,midway] {$\hat{\varrho}$};
\draw[->] (1) -- (3) node[left,midway] {$\maltese$};
\draw[->] (3) -- (4) node[above,midway] {$\varrho$};
\draw[->] (2) -- (4) node[right,midway] {$\maltese$};
}
\end{figure}
\end{lem}
\begin{proof}
The ghost-box-ball algorithm and the box-ball algorithm only differ in how the ghosts are materialised and exorcised: the new position of a moving ball in the ghost-box-ball algorithm agrees with that of the box-ball algorithm, relative to just the empty boxes and balls. Therefore, the result of globally exorcising the ghosts and then evolving according to the box-ball dynamics coincides with evolving according to the ghost-box-ball dynamics and then globally exorcising.
\end{proof}

\n In light of this result, it also makes sense to ask about how the soliton structure of the box-ball-system translates to the ghost-box-ball system. It is not hard to see that the GBBS algorithm preserves the existence and location of blocks of consecutive ghosts; such a block can grow or shrink, but never disappear altogether. We can view such ghost blocks as single zero length blocks in the soliton structure (see Figure \ref{exorcismfirstexample}). Along with the previous lemma, this observation yileds the following.

\begin{lem}\label{solitonbehavofGBBS}
The ghost-box-ball dynamics exhibits the same soliton behaviour as the box-ball dynamics, subject to the following augmentation of the traditional notion of an $n$-soliton box-ball state: we add the construct of a \textit{ghost soliton}, which is any configuration of ghost blocks. This construct does not move -- it has zero velocity. Asymptotically, all other blocks comprising the soliton state travel with speed equal to their respective lengths. Hence, the ghost-box-ball system still exhibits the sorting property of the box-ball system.
\end{lem}

\n Our final analogue of a classical box-ball system property is its conserved shape, seen in Section \ref{invsofbbs}.

\begin{cor}\label{invshapeofGBBS}
If $G\in \text{GBBS}$, define $B=\maltese(G)$. Representing $B$ as a sequence of $1$'s and $0$'s,
\begin{itemize}
\item let $p_1$ be the number of $10$'s in the sequence.
\item Eliminate all of these $10$'s, and let $p_2$ be the number of $10$'s in the resulting sequence.
\item Repeat this process until no $10$'s remain.
\end{itemize}
We associate the weakly decreasing sequence $(p_1,p_2,\ldots)$, or, equivalently, the Young diagram whose $j$\textsuperscript{th} column has $p_j$ boxes is the shape associated to $G$. This Young diagram is the same for $\hat{\varrho}^k G$ for every $k\in\N$.
\end{cor}
\begin{proof}
This follows from Lemma \ref{equivofbbsandgbbs} and Section \ref{invsofbbs}. It is already known that this Young diagram is conserved for all $\varrho^k(B)=\varrho^k\circ \maltese (G)$. Since $\varrho\circ \maltese=\maltese \circ \hat{\varrho}$, it follows that 
$$\maltese(\hat{\varrho}^k(G))=\varrho^k\circ \maltese(G)=\varrho^k(B).$$
\end{proof}

\begin{ex}
Returning to the ghost-box ball system in Example \ref{bbsexamplesteps}, below are the initial state and the subsequent three evolutions.
\begin{figure}[H]
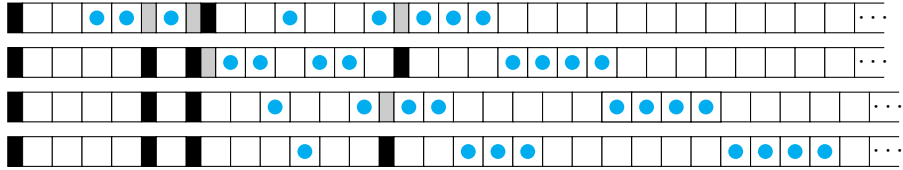

\centering
\tikz[scale=0.395]{
\foreach \y in {0}
{
\foreach \x in {1.0,7.5}
{\draw[fill=black]  (\x,\y-1) -- (\x+0.5,\y-1) -- (\x+0.5,\y) -- (\x,\y) -- cycle;			
}
\foreach \x in {5.5,7.0,14.0}
{\draw[fill=gray!40]  (\x,\y-1) -- (\x+0.5,\y-1) -- (\x+0.5,\y) -- (\x,\y) -- cycle;			
}
\foreach \x in {1.5,2.5,8.0,9.0,11.0,12.0,17.5,18.5,19.5,20.5,21.5,22.5,23.5,24.5,25.5,26.5,27.5,28.5}
{\draw[fill=white]  (\x,\y-1) -- (\x+1,\y-1) -- (\x+1,\y) -- (\x,\y) -- cycle;				
}
\foreach \x in {3.5,4.5,6.0,10.0,13.0,14.5,15.5,16.5}
{\draw[fill=white]  (\x,\y-1) -- (\x+1,\y-1) -- (\x+1,\y) -- (\x,\y) -- cycle;				
\fill[cyan] (\x+0.5,\y-0.5) circle (0.25);									
}
\foreach \x in {}
{\draw[fill=white]  (\x,\y-1) -- (\x+1,\y-1) -- (\x+1,\y) -- (\x,\y) -- cycle;				
\fill[red] (\x+0.5,\y-0.5) circle (0.25);									
}
\foreach \x in {29.5}
{
\draw[-] (\x,\y) -- (\x+1,\y);
\draw[-] (\x,\y-1) -- (\x+1,\y-1);
\node at (\x+0.7,\y-0.5) {$\cdots$};
}
}
\foreach \y in {-1.5}
{
\foreach \x in {1.0,5.5,7.0,14.0}
{\draw[fill=black]  (\x,\y-1) -- (\x+0.5,\y-1) -- (\x+0.5,\y) -- (\x,\y) -- cycle;			
}
\foreach \x in {7.5}
{\draw[fill=gray!40]  (\x,\y-1) -- (\x+0.5,\y-1) -- (\x+0.5,\y) -- (\x,\y) -- cycle;			
}
\foreach \x in {1.5,2.5,3.5,4.5,6.0,10.0,13.0,14.5,15.5,16.5,21.5,22.5,23.5,24.5,25.5,26.5,27.5,28.5}
{\draw[fill=white]  (\x,\y-1) -- (\x+1,\y-1) -- (\x+1,\y) -- (\x,\y) -- cycle;				
}
\foreach \x in {8.0,9.0,11.0,12.0,17.5,18.5,19.5,20.5}
{\draw[fill=white]  (\x,\y-1) -- (\x+1,\y-1) -- (\x+1,\y) -- (\x,\y) -- cycle;				
\fill[cyan] (\x+0.5,\y-0.5) circle (0.25);									
}
\foreach \x in {29.5}
{
\draw[-] (\x,\y) -- (\x+1,\y);
\draw[-] (\x,\y-1) -- (\x+1,\y-1);
\node at (\x+0.7,\y-0.5) {$\cdots$};
}
}
\foreach \y in {-3}
{
\foreach \x in {1.0,5.5,7.0}
{\draw[fill=black]  (\x,\y-1) -- (\x+0.5,\y-1) -- (\x+0.5,\y) -- (\x,\y) -- cycle;			
}
\foreach \x in {13.5}
{\draw[fill=gray!40]  (\x,\y-1) -- (\x+0.5,\y-1) -- (\x+0.5,\y) -- (\x,\y) -- cycle;			
}
\foreach \x in {1.5,2.5,3.5,4.5,6.0,7.5,8.5,10.5,11.5,16.0,17.0,18.0,19.0,20.0,21,22,23,24,25,26,27,28,29}
{\draw[fill=white]  (\x,\y-1) -- (\x+1,\y-1) -- (\x+1,\y) -- (\x,\y) -- cycle;				
}
\foreach \x in {9.5,12.5,14.0,15.0,21.0,22.0,23.0,24.0}
{\draw[fill=white]  (\x,\y-1) -- (\x+1,\y-1) -- (\x+1,\y) -- (\x,\y) -- cycle;				
\fill[cyan] (\x+0.5,\y-0.5) circle (0.25);									
}
\foreach \x in {}
{\draw[fill=white]  (\x,\y-1) -- (\x+1,\y-1) -- (\x+1,\y) -- (\x,\y) -- cycle;				
\fill[red] (\x+0.5,\y-0.5) circle (0.25);									
}
\foreach \x in {30}
{
\draw[-] (\x,\y) -- (\x+1,\y);
\draw[-] (\x,\y-1) -- (\x+1,\y-1);
\node at (\x+0.7,\y-0.5) {$\cdots$};
}
}
\foreach \y in {-4.5}
{
\foreach \x in {1.0,5.5,7.0,13.5}
{\draw[fill=black]  (\x,\y-1) -- (\x+0.5,\y-1) -- (\x+0.5,\y) -- (\x,\y) -- cycle;			
}
\foreach \x in {}
{\draw[fill=gray!40]  (\x,\y-1) -- (\x+0.5,\y-1) -- (\x+0.5,\y) -- (\x,\y) -- cycle;			
}
\foreach \x in {1.5,2.5,3.5,4.5,6.0,7.5,8.5,9.5,11.5,12.5,14.0,15.0,19.0,20.0,21.0,22.0,23.0,24.0,29}
{\draw[fill=white]  (\x,\y-1) -- (\x+1,\y-1) -- (\x+1,\y) -- (\x,\y) -- cycle;				
}
\foreach \x in {10.5,16.0,17.0,18.0,25.0,26.0,27.0,28.0}
{\draw[fill=white]  (\x,\y-1) -- (\x+1,\y-1) -- (\x+1,\y) -- (\x,\y) -- cycle;				
\fill[cyan] (\x+0.5,\y-0.5) circle (0.25);									
}
\foreach \x in {}
{\draw[fill=white]  (\x,\y-1) -- (\x+1,\y-1) -- (\x+1,\y) -- (\x,\y) -- cycle;				
\fill[red] (\x+0.5,\y-0.5) circle (0.25);									
}
\foreach \x in {30}
{
\draw[-] (\x,\y) -- (\x+1,\y);
\draw[-] (\x,\y-1) -- (\x+1,\y-1);
\node at (\x+0.7,\y-0.5) {$\cdots$};
}
}
}
\caption{Three iterations of the ghost-box-ball algorithm.}\label{threelittleghosties}
\end{figure}

\n Applying global exorcism to these four ghost-box-ball states yields

\begin{figure}[H]
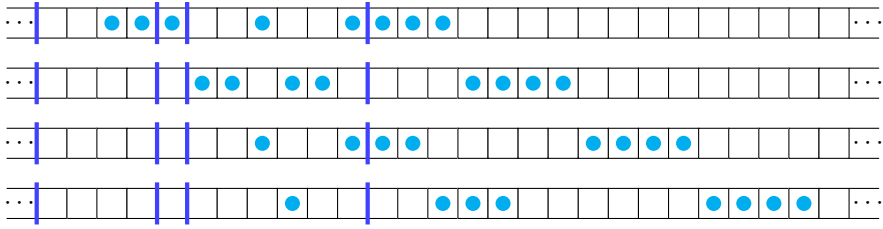

\centering
\tikz[scale=0.4]{
\foreach \y in {0}
{
\foreach \x in {0,1,2,3,4,5,6,7,8,9,10,11,12,13,14,15,16,17,18,19,20,21,22,23,24,25}
{\draw[fill=white]  (\x,\y) -- (\x+1,\y) -- (\x+1,\y+1) -- (\x,\y+1) -- cycle;			
}
\foreach \x in {1,2,3,6,9,10,11,12}
{
\fill[cyan] (\x+0.5,\y+0.5) circle (0.25);								
}
\foreach \x in {}
{
\fill[red] (\x+0.5,\y+0.5) circle (0.25);									
}
\foreach \x in {26}
{
\draw[-] (\x,\y) -- (\x+1,\y);
\draw[-] (\x,\y+1) -- (\x+1,\y+1);
\node at (\x+0.7,\y+0.5) {$\cdots$};
}
\foreach \x in {0}
{\draw[fill=white,white]  (\x,\y) -- (\x-2,\y) -- (\x-2,\y+1) -- (\x,\y+1) -- cycle;
\draw[-] (\x,\y) -- (\x,\y+1);
\draw[-] (\x,\y) -- (\x-2,\y);
\draw[-] (\x,\y+1) -- (\x-2,\y+1);
\draw[-] (\x-1,\y) -- (\x-1,\y+1);
\node at (\x-1.5,\y+0.5) {$\cdots$};
}
\foreach \x in {-1,3,4,10}
{\draw[-,white] (\x,\y+1) -- (\x,\y);
\draw[-,blue!75,ultra thick] (\x,\y+1.2) -- (\x,\y-0.2);		
}
}
\foreach \y in {-2}
{
\foreach \x in {0,1,2,3,4,5,6,7,8,9,10,11,12,13,14,15,16,17,18,19,20,21,22,23,24,25}
{\draw[fill=white]  (\x,\y) -- (\x+1,\y) -- (\x+1,\y+1) -- (\x,\y+1) -- cycle;			
}
\foreach \x in {4,5,7,8,13,14,15,16}
{
\fill[cyan] (\x+0.5,\y+0.5) circle (0.25);								
}
\foreach \x in {}
{
\fill[red] (\x+0.5,\y+0.5) circle (0.25);									
}
\foreach \x in {26}
{
\draw[-] (\x,\y) -- (\x+1,\y);
\draw[-] (\x,\y+1) -- (\x+1,\y+1);
\node at (\x+0.7,\y+0.5) {$\cdots$};
}
\foreach \x in {0}
{\draw[fill=white,white]  (\x,\y) -- (\x-2,\y) -- (\x-2,\y+1) -- (\x,\y+1) -- cycle;
\draw[-] (\x,\y) -- (\x,\y+1);
\draw[-] (\x,\y) -- (\x-2,\y);
\draw[-] (\x,\y+1) -- (\x-2,\y+1);
\draw[-] (\x-1,\y) -- (\x-1,\y+1);
\node at (\x-1.5,\y+0.5) {$\cdots$};
}
\foreach \x in {-1,3,4,10}
{\draw[-,white] (\x,\y+1) -- (\x,\y);
\draw[-,blue!75,ultra thick] (\x,\y+1.2) -- (\x,\y-0.2);		
}
}
\foreach \y in {-4}
{
\foreach \x in {0,1,2,3,4,5,6,7,8,9,10,11,12,13,14,15,16,17,18,19,20,21,22,23,24,25}
{\draw[fill=white]  (\x,\y) -- (\x+1,\y) -- (\x+1,\y+1) -- (\x,\y+1) -- cycle;			
}
\foreach \x in {6,9,10,11,17,18,19,20}
{
\fill[cyan] (\x+0.5,\y+0.5) circle (0.25);								
}
\foreach \x in {}
{
\fill[red] (\x+0.5,\y+0.5) circle (0.25);									
}
\foreach \x in {26}
{
\draw[-] (\x,\y) -- (\x+1,\y);
\draw[-] (\x,\y+1) -- (\x+1,\y+1);
\node at (\x+0.7,\y+0.5) {$\cdots$};
}
\foreach \x in {0}
{\draw[fill=white,white]  (\x,\y) -- (\x-2,\y) -- (\x-2,\y+1) -- (\x,\y+1) -- cycle;
\draw[-] (\x,\y) -- (\x,\y+1);
\draw[-] (\x,\y) -- (\x-2,\y);
\draw[-] (\x,\y+1) -- (\x-2,\y+1);
\draw[-] (\x-1,\y) -- (\x-1,\y+1);
\node at (\x-1.5,\y+0.5) {$\cdots$};
}
\foreach \x in {-1,3,4,10}
{\draw[-,white] (\x,\y+1) -- (\x,\y);
\draw[-,blue!75,ultra thick] (\x,\y+1.2) -- (\x,\y-0.2);		
}
}
\foreach \y in {-6}
{
\foreach \x in {0,1,2,3,4,5,6,7,8,9,10,11,12,13,14,15,16,17,18,19,20,21,22,23,24,25}
{\draw[fill=white]  (\x,\y) -- (\x+1,\y) -- (\x+1,\y+1) -- (\x,\y+1) -- cycle;			
}
\foreach \x in {7,12,13,14,21,22,23,24}
{
\fill[cyan] (\x+0.5,\y+0.5) circle (0.25);								
}
\foreach \x in {}
{
\fill[red] (\x+0.5,\y+0.5) circle (0.25);									
}
\foreach \x in {26}
{
\draw[-] (\x,\y) -- (\x+1,\y);
\draw[-] (\x,\y+1) -- (\x+1,\y+1);
\node at (\x+0.7,\y+0.5) {$\cdots$};
}
\foreach \x in {0}
{\draw[fill=white,white]  (\x,\y) -- (\x-2,\y) -- (\x-2,\y+1) -- (\x,\y+1) -- cycle;
\draw[-] (\x,\y) -- (\x,\y+1);
\draw[-] (\x,\y) -- (\x-2,\y);
\draw[-] (\x,\y+1) -- (\x-2,\y+1);
\draw[-] (\x-1,\y) -- (\x-1,\y+1);
\node at (\x-1.5,\y+0.5) {$\cdots$};
}
\foreach \x in {-1,3,4,10}
{\draw[-,white] (\x,\y+1) -- (\x,\y);
\draw[-,blue!75,ultra thick] (\x,\y+1.2) -- (\x,\y-0.2);		
}
}
}
\caption{Three iterations of the ghost-box-ball evolution (after global exorcism).}
\end{figure}

\n where the blue lines indicate the locations previously occupied by ghosts. The invariant shape (recall Section \ref{invsofbbs}) for this sequence (and all future states) is 
\begin{figure}[H]
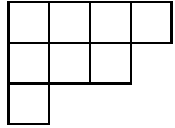

\centering
\ydiagram{4,3,1}
\caption{The invariant shape of the ghost-box-ball system(s) in Figure \ref{threelittleghosties}}
\end{figure}

\end{ex}

\subsection{The Ghost-Box-Ball System and Schensted Insertion}

\n In Definition \ref{phidefnsforref}, we constructed maps $\phi_{\text{RSK}\to\text{BBS}}:\mathcal{R}\to \mathcal{G}^0$ and $\phi_{\text{BBS}\to\text{RSK}}:\mathcal{G}^0\to \mathcal{R}$ to represent Schensted insertion in terms of the natural extension of the coordinatised box-ball evolution to the setting in which some coordinates may vanish. In this section, we demonstrate how to lift the correspondence between $\text{RSK}$ and $\chi^0$ to one between $\text{RSK}$ and the ghost-box-ball evolution, using the coordinate map on $\text{GBBS}$.\\

\n The main tool will be the coordinate map $C:\text{GBBS}\to \mathcal{G}^0$ in Definition \ref{coordmapdefn}. One can encode an RSK pair $(\mb{a},\mb{x})\in\mathcal{R}$ in a ghost-box-ball system by composing the maps $\phi_{\text{RSK}\to\text{BBS}}:\mathcal{R}\to \mathcal{G}^0$ and $C^{-1}:\mathcal{G}^0\to\text{GBBS}$. \\

\n The main result of this section will be to establish the commutativity of the following diagram:

\begin{figure}[H]
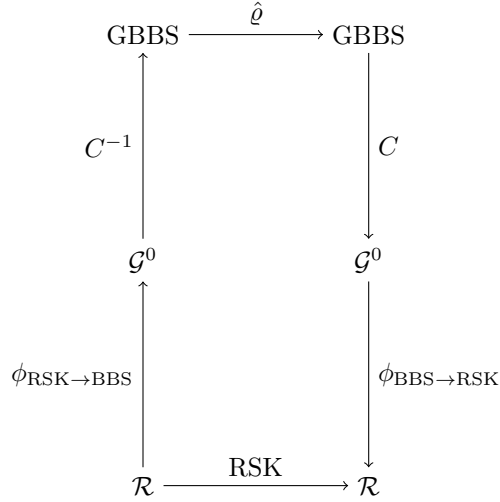

\centering
\tikz[scale=1.5]{
\node (A) at (0,0) {GBBS};
\node (B) at (2,0) {GBBS};
\node (C) at (0,-2) {$\mathcal{G}^0$};
\node (D) at (2,-2) {$\mathcal{G}^0$};
\node (E) at (0,-4) {$\mathcal{R}$};
\node (F) at (2,-4) {$\mathcal{R}$};
\draw[->] (A) -- (B) node[midway,above]  {$\hat{\varrho}$};
\draw[->] (E) -- (F) node[midway,above]  {RSK};
\draw[->] (C) -- (A) node[midway,left]  {$C^{-1}$};
\draw[->] (B) -- (D) node[midway,right]  {$C$};
\draw[->] (E) -- (C) node[midway,left]  {$\phi_{\text{RSK}\to\text{BBS}}$};
\draw[->] (D) -- (F) node[midway,right]  {$\phi_{\text{BBS}\to\text{RSK}}$};
}
\caption{Commutative diagram relating the ghost-box-ball evolution to the RSK dynamics}
\label{commdiagrskgbbs}
\end{figure}

\n or, more succinctly, 

\begin{equation}\text{RSK}=\phi_{\text{BBS}\to \text{RSK}}\circ C\circ \hat{\varrho}\circ C^{-1}\circ \phi_{\text{RSK}\to\text{BBS}}.\end{equation}

\n In establishing this result, we will in fact prove something stronger: if an RSK input pair is encoded in a ghost-box-ball system, the number of stages in the RSK insertion (see, for example, \ref{schenstedexamplewordword}) is equal to the number of stages in the ghost-box-ball evolution (see, for example, Figure \ref{GBBSStepsofevolwithint}), and the data from each stage of the RSK insertion is fully recoverable from the corresponding stage of the ghost-box-ball evolution. \\[3pt]

\n The stages of the ghost-box-ball evolution alluded to here are just the stages given by Definition \ref{defnofgbbsevolalg}. To describe the corresponding stages for the RSK insertion (\textit{cf.} Section \ref{schenstedinsertionsectionrsk}), we introduce the following:

\begin{defn}\label{tripleforrsk}\index{$\mb{r}^i$ (RSK triples)}
For a pair $(\mb{a},\mb{x})\in \mathcal{R}^n$, define a sequence of triples
\begin{equation}\mb{r}^i(\mb{a},\mb{x}):=(\mb{a}^i,\mb{x}^i,\mb{b}^i)\in(\N_0^n)^3\end{equation}
for $i=0,1,\ldots,\sum\limits_{k=1}^n a_k$, where
\begin{enumerate}
\item $\mb{r}^0(\mb{a},\mb{x})=(\mb{a},\mb{x},(0,0,\ldots,0))$.
\item If $j=\min\limits_k\{a^i_k\neq 0\}$, then $\mb{x}^{i+1}$ is the tuple resulting from (Schensted) inserting $j$ into the word with tuple $\mb{x}^i$, $\mb{a}^{i+1}$ is obtained from $\mb{a}^i$ by subtracting 1 from the $j$-th entry, and $\mb{b}^{i+1}$ is the result of adding 1 to the $k$-th entry of $\mb{b}^i$ if a $k$ is bumped from $\mb{x}^i$ to obtain $\mb{x}^{i+1}$, or $\mb{b}^{i+1}=\mb{b}^i$ if nothing is bumped.
\end{enumerate}
\end{defn}

\n We write $\mb{r}^i$ for $\mb{r}^i(\mb{a},\mb{x})$ if it is unambiguous to do so.\\

\n This construction clearly encodes the steps of Schensted insertion. In particular, if $\text{RSK}(\mb{a},\mb{x})=(\mb{b},\mb{y})$ and $m=\sum\limits_{k=1}^n a_k$, then one has $\mb{r}^m=((0,0,\ldots,0),\mb{y},\mb{b})$.\\

\subsubsection{RSK Walls and Conservation Laws in the Ghost-Box-Ball System}
We now introduce a bookkeeping device, referred to as \textit{wall placement}, that will help to establish the stage-by-stage correspondence alluded to at the end of the previous section.\\

\n The key observation underlying this is that there is a conservation law across the stages of RSK insertion: the total number of instances of a given number is conserved, \textit{i.e.} $a_j^i+x_j^i+b_j^i$ is a function of just $j$ (it is constant in $i$). In particular, taking $i=0$, this quantity is
$$a_j^0+x_j^0+b_j^0=a_j+x_j=:w_j$$
which is expressed solely in terms of the input pair. This conserved quantity will be called the $j$-th {\it width}.\\

\n Moreover, this conservation for the RSK stages may be visualised in terms of wall placements in the initial ghost-box-ball configuration using the conserved quantities, the $w_j$'s. Subsequent to the definition, we will then describe how the wall placement evolves during the successive stages of the ghost-box-ball evolution. To represent the wall placement, we will superimpose red zigzags over the ghost-box-ball configurations.\\

\n The following is how we initialise the wall placement on the GBBS associated to an RSK pair:

\begin{defn}\index{Ghost-Box-Ball Walls}
Let $(\mb{a},\mb{x})\in\mathcal{R}^n$ and $G=C^{-1}\circ \phi_{\text{RSK}\to\text{BBS}}(\mb{a},\mb{x})$ be the ghost-box-ball configuration associated to the RSK pair $(\mb{a},\mb{x})$. The walls of $G$, which we will represent by red zigzags, will be placed as follows:
\begin{enumerate}
\item Take the initial filled ghost and the following $w_1$ boxes of $G$, and separate them from the rest of the subsequent boxes by a wall (a red zigzag). Note: If $x_1=0$, include the corresponding empty ghost. Similarly, If $a_1=0$, include the corresponding filled ghost.
\item Take the next $w_2$ boxes of $G$ and place a wall at the end of them. Again, if $a_2=0$, include the corresponding filled ghost before the zigzag.
\item Continue in this manner until the boxes are separated into $n+1$ regions ($n$ finite and one infinite).
\end{enumerate}
\end{defn}

\begin{ex}
Take the RSK input to be $\mathbf{a}=(3,0,2,1,0)$ and $\mathbf{x}=(2,0,0,2,3)$, so that the associated ghost-box-ball system has coordinates
$$\phi_{\text{RSK}\to\text{BBS}}(\mb{a},\mb{x})=(\infty,\mathbf{0},2,\mathbf{3},0,\mathbf{0},0,\mathbf{2},2,\mathbf{1},3,\mathbf{0},\infty).$$
In the above, we embolden the odd positions, which, by definition of $C$, correspond to filled blocks.\\

\n The widths are calculated as follows:
\begin{align*}
    w_1 & = a_1 + x_1 = 3+2 = 5\\
    w_2 & = a_2 + x_2 = 0+0 = 0\\
    w_3 & = a_3 + x_3 = 2+0 = 2\\
    w_4 & = a_4 + x_4 = 1+2 = 3\\
    w_5 & = a_5 + x_5 = 0+3 = 3
\end{align*}
\n We put a wall before the first filled ghost, enclose the two subsequent empty boxes and three filled boxes by another wall (this is now ``Region 1''). In ``Region 2'', we include the filled ghost before the wall because $a_2=0$. Continuing in this manner, we obtain the resulting picture:
\begin{figure}[H]
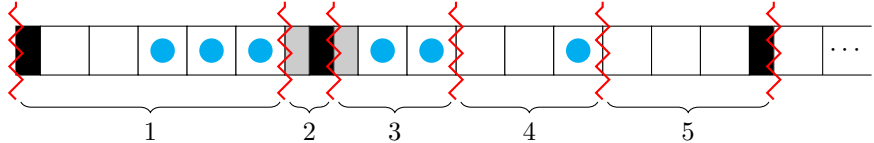

\centering
\tikz[scale=0.65]{
\foreach \x in {0,6,15}
{\draw[fill=black]  (\x,3) -- (\x+0.5,3) -- (\x+0.5,4) -- (\x,4) -- cycle;			
}
\foreach \x in {5.5,6.5}
{\draw[fill=gray!40]  (\x,3) -- (\x+0.5,3) -- (\x+0.5,4) -- (\x,4) -- cycle;			
}
\foreach \x in {0.5,1.5,9,10,12,13,14}
{\draw[fill=white]  (\x,3) -- (\x+1,3) -- (\x+1,4) -- (\x,4) -- cycle;			
}
\foreach \x in {2.5,3.5,4.5,7,8,11}
{\draw[fill=white]  (\x,3) -- (\x+1,3) -- (\x+1,4) -- (\x,4) -- cycle;			
\fill[cyan] (\x+0.5,3.5) circle (0.25);
}
\foreach \x in {15.5}
{\draw[fill=white,white]  (\x,3) -- (\x+2,3) -- (\x+2,4) -- (\x,4) -- cycle;
\draw[-] (\x,3) -- (\x,4);
\draw[-] (\x,3) -- (\x+2,3);
\draw[-] (\x,4) -- (\x+2,4);
\draw[-] (\x+1,3) -- (\x+1,4);
\node at (\x+1.5,3.5) {$\cdots$};
}
\draw[snake=zigzag,thick, red] (0,2.5) -- (0,4.5);
\draw[snake=zigzag,thick, red] (5.5,2.5) -- (5.5,4.5);
\draw[snake=zigzag,thick, red] (6.5,2.5) -- (6.5,4.5);
\draw[snake=zigzag,thick, red] (9,2.5) -- (9,4.5);
\draw[snake=zigzag,thick, red] (12,2.5) -- (12,4.5);
\draw[snake=zigzag,thick, red] (15.5,2.5) -- (15.5,4.5);
\draw [decorate,decoration={brace,amplitude=4pt}] (5.4,2.45) -- (0.1,2.45) node [black,midway,yshift=-0.4cm] {$1$};
\draw [decorate,decoration={brace,amplitude=4pt}] (6.4,2.45) -- (5.6,2.45) node [black,midway,yshift=-0.4cm] {$2$};
\draw [decorate,decoration={brace,amplitude=4pt}] (8.9,2.45) -- (6.6,2.45) node [black,midway,yshift=-0.4cm] {$3$};
\draw [decorate,decoration={brace,amplitude=4pt}] (11.9,2.45) -- (9.1,2.45) node [black,midway,yshift=-0.4cm] {$4$};
\draw [decorate,decoration={brace,amplitude=4pt}] (15.4,2.45) -- (12.1,2.45) node [black,midway,yshift=-0.4cm] {$5$};
}
\caption{The initial ghost-box-ball system with its finite regions labelled.}\label{gbbswithwallsfig}
\end{figure}	
\end{ex}

\n Next, we prescribe how the wall placement evolves at successive stages of the ghost-box-ball evolution:
\begin{enumerate}
\item When a ghost is exorcised: if it is bordered by a wall, the wall then borders the ghost's other neighbour.
\item When a filled ghost materialises by the evacuation of a ball from a box (implying that a wall is present to the right of the evacuated box), the filled ghost that materialises does so to the left of the wall.
\item When an empty ghost materialises by the appearance of a ball, the empty ghost is created in the same region as the ball's new location.
\end{enumerate}

\n The key point is that the position of the walls, relative to the non-ghost boxes, does not change with the steps of the ghost-box-ball evolution. The walls keep a notion of the regions throughout the evolution, and it is precisely the numbers of moved balls, unmoved balls and empty boxes in those regions that we show encodes the corresponding RSK steps. For ease of visualisation, we once again employ a colouring of balls to keep track of balls that have moved and those that have not, colouring unmoved balls blue and moved balls red.\\

\n Before presenting the main theorem of this section, we provide an illustrative example showing the locations of the walls throughout the evolution of the ghost-box-ball system in Figure \ref{gbbswithwallsfig}.

\begin{ex}
Below is the evolution of the ghost-box-ball system in Figure \ref{gbbswithwallsfig}, with the walls included at each step.
\begin{figure}[H]
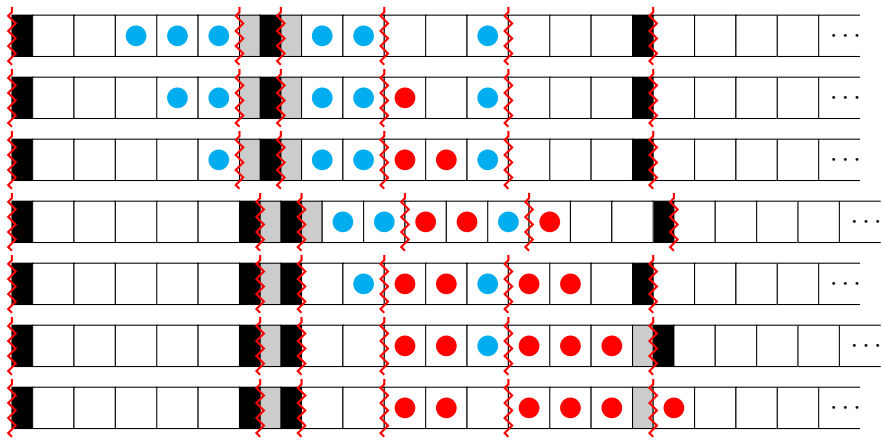

\centering
\tikz[scale=0.55]{
\foreach \y in {0}
{
\foreach \x in {-0.5,5.5,14.5}
{\draw[fill=black]  (\x,\y) -- (\x+0.5,\y) -- (\x+0.5,\y+1) -- (\x,\y+1) -- cycle;			
}
\foreach \x in {5,6}
{\draw[fill=gray!40]  (\x,\y) -- (\x+0.5,\y) -- (\x+0.5,\y+1) -- (\x,\y+1) -- cycle;		
}
\foreach \x in {0,1,2,3,4,6.5,7.5,8.5,9.5,10.5,11.5,12.5,13.5,15,16,17}
{\draw[fill=white]  (\x,\y) -- (\x+1,\y) -- (\x+1,\y+1) -- (\x,\y+1) -- cycle;			
}
\foreach \x in {2,3,4,6.5,7.5,10.5}
{
\fill[cyan] (\x+0.5,\y+0.5) circle (0.25);								
}
\foreach \x in {}
{
\fill[red] (\x+0.5,\y+0.5) circle (0.25);									
}
\foreach \x in {18}
{\draw[fill=white,white]  (\x,\y) -- (\x+2,\y) -- (\x+2,\y+1) -- (\x,\y+1) -- cycle;
\draw[-] (\x,\y) -- (\x,\y+1);
\draw[-] (\x,\y) -- (\x+2,\y);
\draw[-] (\x,\y+1) -- (\x+2,\y+1);
\draw[-] (\x+1,\y) -- (\x+1,\y+1);
\node at (\x+1.7,\y+0.5) {$\cdots$};
}
\foreach \x in {-0.5,5,6,8.5,11.5,15}{
\draw[decorate, decoration={zigzag, segment length=2mm, amplitude=0.5mm},red,thick]  (\x,\y-0.2) -- (\x,\y+1.2);	}
}
\foreach \y in {-1.5}
{
\foreach \x in {-0.5,5.5,14.5}
{\draw[fill=black]  (\x,\y) -- (\x+0.5,\y) -- (\x+0.5,\y+1) -- (\x,\y+1) -- cycle;			
}
\foreach \x in {5,6}
{\draw[fill=gray!40]  (\x,\y) -- (\x+0.5,\y) -- (\x+0.5,\y+1) -- (\x,\y+1) -- cycle;		
}
\foreach \x in {0,1,2,3,4,6.5,7.5,8.5,9.5,10.5,11.5,12.5,13.5,15,16,17}
{\draw[fill=white]  (\x,\y) -- (\x+1,\y) -- (\x+1,\y+1) -- (\x,\y+1) -- cycle;			
}
\foreach \x in {3,4,6.5,7.5,10.5}
{
\fill[cyan] (\x+0.5,\y+0.5) circle (0.25);								
}
\foreach \x in {8.5}
{
\fill[red] (\x+0.5,\y+0.5) circle (0.25);									
}
\foreach \x in {18}
{\draw[fill=white,white]  (\x,\y) -- (\x+2,\y) -- (\x+2,\y+1) -- (\x,\y+1) -- cycle;
\draw[-] (\x,\y) -- (\x,\y+1);
\draw[-] (\x,\y) -- (\x+2,\y);
\draw[-] (\x,\y+1) -- (\x+2,\y+1);
\draw[-] (\x+1,\y) -- (\x+1,\y+1);
\node at (\x+1.7,\y+0.5) {$\cdots$};
}
\foreach \x in {-0.5,5,6,8.5,11.5,15}{
\draw[decorate, decoration={zigzag, segment length=2mm, amplitude=0.5mm},red,thick]  (\x,\y-0.2) -- (\x,\y+1.2);	}
}
\foreach \y in {-3}
{
\foreach \x in {-0.5,5.5,14.5}
{\draw[fill=black]  (\x,\y) -- (\x+0.5,\y) -- (\x+0.5,\y+1) -- (\x,\y+1) -- cycle;			
}
\foreach \x in {5,6}
{\draw[fill=gray!40]  (\x,\y) -- (\x+0.5,\y) -- (\x+0.5,\y+1) -- (\x,\y+1) -- cycle;		
}
\foreach \x in {0,1,2,3,4,6.5,7.5,8.5,9.5,10.5,11.5,12.5,13.5,15,16,17}
{\draw[fill=white]  (\x,\y) -- (\x+1,\y) -- (\x+1,\y+1) -- (\x,\y+1) -- cycle;			
}
\foreach \x in {4,6.5,7.5,10.5}
{
\fill[cyan] (\x+0.5,\y+0.5) circle (0.25);								
}
\foreach \x in {8.5,9.5}
{
\fill[red] (\x+0.5,\y+0.5) circle (0.25);									
}
\foreach \x in {18}
{\draw[fill=white,white]  (\x,\y) -- (\x+2,\y) -- (\x+2,\y+1) -- (\x,\y+1) -- cycle;
\draw[-] (\x,\y) -- (\x,\y+1);
\draw[-] (\x,\y) -- (\x+2,\y);
\draw[-] (\x,\y+1) -- (\x+2,\y+1);
\draw[-] (\x+1,\y) -- (\x+1,\y+1);
\node at (\x+1.7,\y+0.5) {$\cdots$};
}
\foreach \x in {-0.5,5,6,8.5,11.5,15}{
\draw[decorate, decoration={zigzag, segment length=2mm, amplitude=0.5mm},red,thick]  (\x,\y-0.2) -- (\x,\y+1.2);	}
}
\foreach \y in {-4.5}
{
\foreach \x in {-0.5,5,6,15}
{\draw[fill=black]  (\x,\y) -- (\x+0.5,\y) -- (\x+0.5,\y+1) -- (\x,\y+1) -- cycle;			
}
\foreach \x in {5.5,6.5}
{\draw[fill=gray!40]  (\x,\y) -- (\x+0.5,\y) -- (\x+0.5,\y+1) -- (\x,\y+1) -- cycle;		
}
\foreach \x in {0,1,2,3,4,7,8,9,10,11,12,13,14,15.5,16.5,17.5}
{\draw[fill=white]  (\x,\y) -- (\x+1,\y) -- (\x+1,\y+1) -- (\x,\y+1) -- cycle;			
}
\foreach \x in {7,8,11}
{
\fill[cyan] (\x+0.5,\y+0.5) circle (0.25);								
}
\foreach \x in {9,10,12}
{
\fill[red] (\x+0.5,\y+0.5) circle (0.25);									
}
\foreach \x in {18.5}
{\draw[fill=white,white]  (\x,\y) -- (\x+2,\y) -- (\x+2,\y+1) -- (\x,\y+1) -- cycle;
\draw[-] (\x,\y) -- (\x,\y+1);
\draw[-] (\x,\y) -- (\x+2,\y);
\draw[-] (\x,\y+1) -- (\x+2,\y+1);
\draw[-] (\x+1,\y) -- (\x+1,\y+1);
\node at (\x+1.7,\y+0.5) {$\cdots$};
}
\foreach \x in {-0.5,5.5,6.5,9,12,15.5}{
\draw[decorate, decoration={zigzag, segment length=2mm, amplitude=0.5mm},red,thick]  (\x,\y-0.2) -- (\x,\y+1.2);	}
}
\foreach \y in {-6}
{
\foreach \x in {-0.5,5,6,14.5}
{\draw[fill=black]  (\x,\y) -- (\x+0.5,\y) -- (\x+0.5,\y+1) -- (\x,\y+1) -- cycle;			
}
\foreach \x in {5.5}
{\draw[fill=gray!40]  (\x,\y) -- (\x+0.5,\y) -- (\x+0.5,\y+1) -- (\x,\y+1) -- cycle;		
}
\foreach \x in {0,1,2,3,4,6.5,7.5,8.5,9.5,10.5,11.5,12.5,13.5,15,16,17}
{\draw[fill=white]  (\x,\y) -- (\x+1,\y) -- (\x+1,\y+1) -- (\x,\y+1) -- cycle;			
}
\foreach \x in {7.5,10.5}
{
\fill[cyan] (\x+0.5,\y+0.5) circle (0.25);								
}
\foreach \x in {8.5,9.5,11.5,12.5}
{
\fill[red] (\x+0.5,\y+0.5) circle (0.25);									
}
\foreach \x in {18}
{\draw[fill=white,white]  (\x,\y) -- (\x+2,\y) -- (\x+2,\y+1) -- (\x,\y+1) -- cycle;
\draw[-] (\x,\y) -- (\x,\y+1);
\draw[-] (\x,\y) -- (\x+2,\y);
\draw[-] (\x,\y+1) -- (\x+2,\y+1);
\draw[-] (\x+1,\y) -- (\x+1,\y+1);
\node at (\x+1.7,\y+0.5) {$\cdots$};
}
\foreach \x in {-0.5,5.5,6.5,8.5,11.5,15}{
\draw[decorate, decoration={zigzag, segment length=2mm, amplitude=0.5mm},red,thick]  (\x,\y-0.2) -- (\x,\y+1.2);	}
}
\foreach \y in {-7.5}
{
\foreach \x in {-0.5,5,6,15}
{\draw[fill=black]  (\x,\y) -- (\x+0.5,\y) -- (\x+0.5,\y+1) -- (\x,\y+1) -- cycle;			
}
\foreach \x in {5.5,14.5}
{\draw[fill=gray!40]  (\x,\y) -- (\x+0.5,\y) -- (\x+0.5,\y+1) -- (\x,\y+1) -- cycle;		
}
\foreach \x in {0,1,2,3,4,6.5,7.5,8.5,9.5,10.5,11.5,12.5,13.5,15.5,16.5,17.5}
{\draw[fill=white]  (\x,\y) -- (\x+1,\y) -- (\x+1,\y+1) -- (\x,\y+1) -- cycle;			
}
\foreach \x in {10.5}
{
\fill[cyan] (\x+0.5,\y+0.5) circle (0.25);								
}
\foreach \x in {8.5,9.5,11.5,12.5,13.5}
{
\fill[red] (\x+0.5,\y+0.5) circle (0.25);									
}
\foreach \x in {18.5}
{\draw[fill=white,white]  (\x,\y) -- (\x+2,\y) -- (\x+2,\y+1) -- (\x,\y+1) -- cycle;
\draw[-] (\x,\y) -- (\x,\y+1);
\draw[-] (\x,\y) -- (\x+2,\y);
\draw[-] (\x,\y+1) -- (\x+2,\y+1);
\draw[-] (\x+1,\y) -- (\x+1,\y+1);
\node at (\x+1.7,\y+0.5) {$\cdots$};
}
\foreach \x in {-0.5,5.5,6.5,8.5,11.5,15}{
\draw[decorate, decoration={zigzag, segment length=2mm, amplitude=0.5mm},red,thick]  (\x,\y-0.2) -- (\x,\y+1.2);	}
}
\foreach \y in {-9}
{
\foreach \x in {-0.5,5,6}
{\draw[fill=black]  (\x,\y) -- (\x+0.5,\y) -- (\x+0.5,\y+1) -- (\x,\y+1) -- cycle;			
}
\foreach \x in {5.5,14.5}
{\draw[fill=gray!40]  (\x,\y) -- (\x+0.5,\y) -- (\x+0.5,\y+1) -- (\x,\y+1) -- cycle;		
}
\foreach \x in {0,1,2,3,4,6.5,7.5,8.5,9.5,10.5,11.5,12.5,13.5,15,16,17}
{\draw[fill=white]  (\x,\y) -- (\x+1,\y) -- (\x+1,\y+1) -- (\x,\y+1) -- cycle;			
}
\foreach \x in {}
{
\fill[cyan] (\x+0.5,\y+0.5) circle (0.25);								
}
\foreach \x in {8.5,9.5,11.5,12.5,13.5,15}
{
\fill[red] (\x+0.5,\y+0.5) circle (0.25);									
}
\foreach \x in {18}
{\draw[fill=white,white]  (\x,\y) -- (\x+2,\y) -- (\x+2,\y+1) -- (\x,\y+1) -- cycle;
\draw[-] (\x,\y) -- (\x,\y+1);
\draw[-] (\x,\y) -- (\x+2,\y);
\draw[-] (\x,\y+1) -- (\x+2,\y+1);
\draw[-] (\x+1,\y) -- (\x+1,\y+1);
\node at (\x+1.7,\y+0.5) {$\cdots$};
}
\foreach \x in {-0.5,5.5,6.5,8.5,11.5,15}{
\draw[decorate, decoration={zigzag, segment length=2mm, amplitude=0.5mm},red,thick]  (\x,\y-0.2) -- (\x,\y+1.2);	}
}
}
\caption{The evolution of a ghost-box-ball system with walls.}\label{gbbsstepswithwallsfig}
\end{figure}	
\end{ex}
\n The initial ghost-box-ball system came from $(\mb{a},\mb{x})=((3,0,2,1,0),(2,0,0,2,3))$. The result of RSK insertion (see Figure \ref{pictorialrepnofschensted}) is 

\begin{figure}[H]
\centering
\tikz[scale=0.75]{
\node at (-5,0) {$\mathbf{x}=(2,0,0,2,3)$};
\node at (0,2) {$\mathbf{a}=(3,0,2,1,0)$};
\node at (5,0) {$\mathbf{y}=(5,0,2,1,0)$};
\node at (0,-2) {$\mathbf{b}=(0,0,0,2,3)$};
\draw[->,thick] (-2.5,0) -- (2.5,0);
\draw[->,thick] (0,1.25) -- (0,-1.25);
}
\caption{The Schensted evolution encoded by Figure \ref{gbbsstepswithwallsfig}}
\end{figure}	
\n In the final stage of the above ghost-box-ball evolution, we can construct two sequences: the number of empty boxes in each region and the number of (red) balls in each region. These two sequences are $(5,0,2,1,0)$ and $(0,0,0,2,3)$, which are $\mb{y}$ and $\mb{b}$, respectively. We will prove that this holds in general, by proving a stronger result.

\begin{thm}\label{mainresRSKencodedinGBBS}
Let $(\mb{a},\mb{x})\in\mathcal{R}^n$ and let $G=C^{-1}\circ \phi_{\text{RSK}\to\text{BBS}}(\mb{a},\mb{x})$. Let $(\mb{r}^i)_i$ be as defined in Definition \ref{tripleforrsk}. At the $i$-th stage of the ghost-box-ball evolution of $G$, the following holds for each $j\in[n]$:
\begin{enumerate}[(1)]
\item $a_j^i$ is equal to the number of blue (unmoved) balls in the $j$-th region,
\item $b_j^i$ is equal to the number of red (moved) balls in the $j$-th region,
\item $x_j^i$ is equal to the number of empty boxes in the $j$-th region.
\end{enumerate}
\end{thm}

\begin{proof}
We prove this by induction on $i$.\\[4pt]
The base case ($i=0$) is satisfied by construction: the initial ghost-box-ball state and wall structure were built out of the RSK input variables so that (1) and (3) are satisfied, and (2) holds trivially because nothing has been bumped yet (so $b_j^0=0$ for each $j$) and no balls have been moved yet (so all balls, if any, are blue).\\

\n Let us now suppose that, at some stage, say the $k$-th stage, we have for each $j\in[n]$:
\begin{enumerate}[(1)]
\item $a_j^k$ is equal to the number of blue (unmoved) balls in the $j$-th region of the $k$-th step in the GBBS algorithm,
\item $b_j^k$ is equal to the number of red (moved) balls in the $j$-th region of the $k$-th step in the GBBS algorithm,
\item $x_j^k$ is equal to the number of empty boxes in the $j$-th region of the $k$-th step in the GBBS algorithm.
\end{enumerate}

\n We now consider the $(k+1)$-st stage of the ghost-box-ball evolution. We need to show that the equalities hold for the region the ball was in and the region the ball moves to (unless it moves to the infinite region).\\[5pt]
Suppose the ball we moved was in Region $j$. We first make the observation that the ball cannot move within Region $j$: the walls were placed so that each region either has no blue balls or the block of blue balls is precisely the last part of the region (if there are blue balls, the wall comes right after the last of them). Therefore, the ball must move to Region $l$, for some $l>j$ (which may be the infinite region).\\[4pt]
With this observation, we can immediately check the counts for Region $j$ after the ball is moved: the number of red balls has not changed (by the observation), the number of blue balls has decreased by 1, and the number of empty boxes has increased by 1. We therefore need to show:
\begin{align*}
a_j^{k+1}&=a_j^k-1\\
b_j^{k+1}&=b_j^k\\
x_j^{k+1}&=x_j^k+1.
\end{align*}
By the induction hypothesis, for us to be able to move a ball from Region $j$, we must have had $a_j^k\geq 1$ and, since the ball was the left-most unmoved ball, $a_{j'}^k=0$ for all $j'<j$. Thus, in terms of RSK, this means we are inserting a $j$ into the current row for $x^k$. This reduces $a_j^k$ by one (since the $j$ is inserted into $x^k$), increases $x_j^k$ by one (because the $j$ finds a place in the row, and does not alter $b_j^k$, since a number can only bump a number greater than itself). We see, therefore, the validity of the counts (in terms of the GBBS and the RSK insertion) agree for Region $j$.\\

\n Now we split into two cases (based on the destination of the ball):
\begin{itemize}
\item (Case 1): The ball lands in some finite region, say Region $l$, where $n\geq l>k$
\item (Case 2): The ball lands in the infinite region.
\end{itemize}

\n\textbf{(Case 1):} Since the ball moves to the left-most empty box to its right, all spaces between the ball's origin and its new box must be full or ghosts. In particular, at the $k$-th stage, there were no empty boxes in the regions between Region $j$ and Region $l$. By the induction hypothesis, $x_m^k=0$ for all $j<m<l$ and $x_l^k\geq 1$ (i.e. the $j$ to be inserted bumps an $l$). In terms of Region $l$, we lose one empty box and gain a red ball (there is no change to the number of blue balls here). We therefore need to show:
\begin{align*}
a_l^{k+1}&=a_l^k\\
b_l^{k+1}&=b_l^k+1\\
x_l^{k+1}&=x_l^k-1.
\end{align*}
This is clearly the case, since a $j$ is bumping an $l$.

\n\textbf{(Case 2):} By the same reasoning in Case 1, there must be no empty boxes in any of the finite regions beyond the $j$-th, so $x_m^k=0$ for all $j<m\leq n$. By the induction hypothesis, there are no numbers in the row that are strictly greater than $j$. Therefore, in this RSK insertion step, we do not bump anything. Instead, we extend the row by a box and fill it with the $j$. Since the theorem does not contain any conditions relating the current RSK step and the infinite region of the ghost-box-ball system, we have nothing more to check here. Simply by having the only changes be
\begin{align*}
a_j^{k+1}&=a_j^k-1\\
b_j^{k+1}&=b_j^k\\
x_j^{k+1}&=x_j^k+1.
\end{align*}
establishes that nothing has been bumped and that the row has been extended by a $j$; only the variables/counts for Region $j$ are affected in this case.
\end{proof}

\n We now have the following immediate corollary:

\begin{cor}\label{rskisgbbs}
One has $$\text{RSK}=\phi_{\text{BBS}\to \text{RSK}}\circ C\circ \hat{\varrho}\circ C^{-1}\circ \phi_{\text{RSK}\to \text{BBS}}.$$
\end{cor}

\begin{proof} 
Since, at each stage, the RSK triple $\mb{r}^i$ is encoded in the $i$-th stage of the ghost-box-ball evolution, and no blue balls remain at the end of the ghost-box-ball evolution, only empty boxes, empty ghosts, filled ghosts and red balls are left. One can simply apply the coordinate mapping $C:\text{GBBS}\to \mathcal{G}^0$ and read off the RSK variables (using $\phi_{\text{BBS}\to\text{RSK}}$) to find the output pair $(\mb{b},\mb{y})$ for the RSK insertion $\mb{a}\to\mb{x}$.
\end{proof}

\subsection{Ghost-Box-Ball Dynamics Beyond RSK}
Throughout this section so far, we have focused on connections to the RSK algorithm. We showed a precise correspondence between RSK and the ghost-box-ball algorithm (Corollary \ref{rskisgbbs})), which was our original goal. A key construct in doing this was the implementation of a coordinate dynamics for the ghost-box-ball system. These coordinates (\textit{cf.} Definition \ref{coordmapdefn}) are given in terms of lengths of filled and empty blocks of boxes of various types. The dynamics of the ghost-box-ball system (denoted $\hat{\varrho}$) and the coordinate dynamics (denoted $\chi^0$) have a direct relation, independent of RSK, which is of interest in its own right. That is what we will now discuss. More precisely, we show the analogue of Corollary \ref{commdiagbbscoords} holds for the ghost-box-ball system by proving that the following diagram commutes:

\begin{figure}[H]
\centering
\tikz[scale=0.7]{
\node (1) at (0,0) {GBBS};
\node (2) at (3,0) {GBBS};
\node (3) at (0,-3) {$\mathcal{G}^0$};
\node (4) at (3,-3) {$\mathcal{G}^0$};
\draw[->] (1) -- (2) node[above,midway] {$\hat{\varrho}$};
\draw[->] (1) -- (3) node[left,midway] {$C$};
\draw[->] (3) -- (4) node[above,midway] {$\chi^0$};
\draw[->] (2) -- (4) node[right,midway] {$C$};
}.
\end{figure}

\n In this section, we provide a partial answer to this question simply by utilising the results of the previous sections. \\

\n From Corollary \ref{rskisbbscoords}, we have
\begin{equation}\text{RSK}=\phi_{\text{BBS}\to\text{RSK}}\circ \chi^0 \circ \phi_{\text{RSK}\to\text{BBS}}\end{equation}
and Corollary \ref{rskisgbbs} establishes
\begin{equation}\text{RSK}=\phi_{\text{BBS}\to \text{RSK}}\circ C\circ \hat{\varrho}\circ C^{-1}\circ \phi_{\text{RSK}\to \text{BBS}}.\end{equation}

\n Since $\phi_{\text{RSK}\to\text{BBS}}$ is bijective, one obtains from the above 
\begin{equation}\phi_{\text{BBS}\to\text{RSK}}\circ \chi^0=\phi_{\text{BBS}\to \text{RSK}}\circ C\circ \hat{\varrho}\circ C^{-1}.\label{obstructedinvbbsrsk}\end{equation}

\n In what follows, we show that, although $\phi_{\text{BBS}\to \text{RSK}}$ is not injective, we can eliminate this map from both sides of the above equation by a local analysis.

\begin{defn}
For $\mb{z}=(\infty,Q_1=0,W_1,Q_2,W_2,\ldots,Q_{n-1},W_{n-1},Q_{n},\infty)\in\mathcal{G}^0_n$, define $\zeta_n:\mathcal{G}_n^0\to \N_0$ by\vspace{0.1cm}
\begin{equation}
\zeta_n(\mb{z})=\sum_{j=1}^{n} Q_j.
\end{equation}
\end{defn}

\n Since this counts the number of balls in the system with coordinates $\mb{z}$, this should be conserved. To see that this is the case, consider the $Q_{n+1}^{t+1}$:\vspace{0.2cm}
\begin{align}
Q_{n}^{t+1}&=\min\left(W_{n}^t,\sum_{j=1}^{n} Q_j^t-\sum_{j=1}^{n-1}Q_j^{t+1}\right)\\
&=\min\left(\infty,\sum_{j=1}^{n} Q_j^t-\sum_{j=1}^{n-1}Q_j^{t+1}\right)\\
&=\sum_{j=1}^{n} Q_j^t-\sum_{j=1}^{n-1}Q_j^{t+1},
\end{align}

\n Rearranging this shows that $\zeta_n$ is invariant under the coordinate dynamics on $\mathcal{G}^0$.

\begin{thm}\label{thmbbsgbbsequiv}
For each $m\in\N_0$, let $\mathcal{G}_n^{0,m}:=\zeta_n^{-1}(m)$. On this level set of $\zeta_n$, $\phi_{\text{BBS}\to \text{RSK}}$ is injective, and one has
\begin{equation}\chi^0=C\circ \hat{\varrho}\circ C^{-1}\end{equation}
when restricted to $\mathcal{G}_n^{0,m}$.
\end{thm}

\begin{proof}
For $\mb{z}=(\infty,Q_1=0,W_1,Q_2,W_2,\ldots,Q_{n-1},W_{n-1},Q_{n},\infty)$, recall that the explicit form of the mapping $\phi_{\text{BBS}\to \text{RSK}}$ (\textit{cf.} Definition \ref{phidefnsforref}) gives
\begin{equation}
\phi_{\text{BBS}\to\text{RSK}}(\mb{z})=((0,Q_2,\ldots,Q_{n-1}),(W_1,\ldots,W_{n-1})).
\end{equation}
Clearly the obstruction to injectivity is in losing the data of $Q_{n}$. However, on $\mathcal{G}_n^{0,m}$, one has
\begin{equation}
Q_{n}=m-m+Q_{n}=m-\sum_{j=1}^{n} Q_j+Q_{n}=m-\sum_{j=1}^{n-1} Q_j.
\end{equation}
Thus, when restricted to $\mathcal{G}_n^{0,m}$, $\phi_{\text{BBS}\to\text{RSK}}$ is injective and therefore has a left inverse. We post compose Equation \ref{obstructedinvbbsrsk} by this left inverse to complete the proof.
\end{proof}

\begin{rem}
Since $\mathcal{G}_n^0$ is the union of the level sets of $\zeta_n$, Theorem \ref{thmbbsgbbsequiv} gives \ref{chiisvarrhooncoords} on all of $\mathcal{G}_n^0$. Furthermore, since $\mathcal{G}^0$ is itself a disjoint union of the $\mathcal{G}_n^0$ sets, Equation \ref{chiisvarrhooncoords} holds on all of $\mathcal{G}^0$. As a result, we have the following corollary:
\end{rem}

\begin{cor}\label{conjectureaboutgbbsbbs}
The equality
\begin{equation}\label{chiisvarrhooncoords}
\chi^0=C\circ \hat{\varrho}\circ C^{-1}
\end{equation}
holds on all of $\mathcal{G}^0$.
\end{cor}

\n Thus, the ghost-box-ball dynamics on ghost-box-ball systems starting with a filled ghost is in agreement with the coordinate evolution on $\mathcal{G}^0$.


\subsection{Comparison with the Literature}\label{fukudasecfourrems}
There are other works in the literature that describe a direct connection between box-ball systems and the RSK algorithm. We conclude this section with a brief comparison between what we have done and one of the principal and most cited treatments in this regard due to Fukuda \cite{bib:fukuda}. His work introduces an encoding of Schensted insertion in an advanced box-ball system with {\it carrying capacities} and {\it ball colours}. In contrast we emphasise that our work captures Schensted insertion in what is essentially only a slight deviation from the original box-ball system (in the sense that it is governed by the original box-ball coordinate evolution), rather than having to add the complexity of box labels and capacities to the box-ball system. For a more detailed and self-contained description of the advanced box-ball system we  refer the reader to \cite{bib:r}.\\

\n Both the ghost-box-ball and advanced box-ball (with uniform carrying capacity 1) systems reproduce Schensted/RSK insertion, but these two systems are fundamentally different. On the one hand, the ghost-box-ball system exhibits what we have called ghost solitons: particles with velocity zero, whereas no such zero soliton exists in the advanced box-ball system. Conversely, prioritisation of certain balls over others (via colouring/labelling) in the advanced box-ball system is not something currently in the ghost-box-ball system.\\

\n It would be interesting to study the applications of a hybrid of the two systems. As far as Schensted insertion alone is concerned, there is an obvious appeal to the ghost-box-ball system over the advanced box-ball system: the former is simply a manifestation of the original (unlabelled, carrying capacity one) box-ball-system, originally introduced by Takahashi and Satsuma \cite{bib:ts}. The advanced box-ball system is not governed by such simple equations. Therefore, purely from the standpoint of simplicity, the attraction of the ghost-box-ball system seems clear.

\section{Intrinsic Ghost-Box-Ball Dynamics and the Phase Shift}\label{chapterextensions}
We present constructions for generalising the ghost-box-ball system and studying its intrinsic dynamical nature in Section \ref{genGBBS}, and we use these constructions to study the box-ball phase shift phenomenon in Section \ref{sectionphaseshiftbbs}. Key to both sections is the necessity of a backwards time dynamics for the GBBS, and this is what we explore first.

\subsection{Extension to the Full Set of Ghost-Box-Ball Configurations}\label{genGBBS}
For the classical box-ball system, it was important to run the time evolution in both forwards and backwards time: in backwards time, asymptotically, the evolution sorts the blocks in descending order. In making a connection to continuous-time dynamics (see Section \ref{chapterconclusions}), unidirectional dynamics is not enough.

\begin{rem}
A nice property of the classical box-ball dynamics is that one can perform a backwards time step by reflecting the box-ball system horizontally, performing the usual box-ball algorithm, and then reflecting the system again \cite{bib:ts}.\\
By carefully going through (3)-(6) of the ghost-box-ball algorithm (\textit{cf.} Definition \ref{defnofgbbsevolalg}), on can check that the same principle holds for reversing a time-step of the ghost-box-ball algorithm. This is demonstrated below in Example \ref{gbbsinreversetime}.
\end{rem}

\n The restricted set of ghost-box-ball configurations studied in Section \ref{chaptergbbs} (\textit{i.e.} those with a filled ghost on the left) are not closed under this reverse dynamics.
\begin{ex}\label{gbbsinreversetime}
Let us take the following ghost-box-ball state:
\begin{figure}[H]
\centering
\tikz[scale=0.49]{
\foreach \y in {0}
{
\node at (-2.5,\y-0.5) {$G_0 := $};
\foreach \x in {5}
{\draw[fill=black]  (\x,\y-1) -- (\x+0.5,\y-1) -- (\x+0.5,\y) -- (\x,\y) -- cycle;			
}
\foreach \x in {}
{\draw[fill=gray!40]  (\x,\y-1) -- (\x+0.5,\y-1) -- (\x+0.5,\y) -- (\x,\y) -- cycle;			
}
\foreach \x in {1,2,3,4,5.5,7.5,8.5,9.5,10.5}
{\draw[fill=white]  (\x,\y-1) -- (\x+1,\y-1) -- (\x+1,\y) -- (\x,\y) -- cycle;				
}
\foreach \x in {6.5}
{\draw[fill=white]  (\x,\y-1) -- (\x+1,\y-1) -- (\x+1,\y) -- (\x,\y) -- cycle;				
\fill[cyan] (\x+0.5,\y-0.5) circle (0.25);									
}
\foreach \x in {}
{\draw[fill=white]  (\x,\y-1) -- (\x+1,\y-1) -- (\x+1,\y) -- (\x,\y) -- cycle;				
\fill[red] (\x+0.5,\y-0.5) circle (0.25);									
}
\foreach \x in {11.5}
{
\draw[-] (\x,\y) -- (\x+1,\y);
\draw[-] (\x,\y-1) -- (\x+1,\y-1);
\node at (\x+0.7,\y-0.5) {$\cdots$};
}
\foreach \x in {0}
{
\draw[fill=white]  (\x,\y-1) -- (\x+1,\y-1) -- (\x+1,\y) -- (\x,\y) -- cycle;				
\draw[-] (\x,\y) -- (\x-1,\y);
\draw[-] (\x,\y-1) -- (\x-1,\y-1);
\node at (\x-0.6,\y-0.5) {$\cdots$};
}
}
}
\end{figure}
\n Reflecting this system horizontally, we obtain
\begin{figure}[H]
\centering
\tikz[scale=0.49]{
\foreach \y in {0}
{
\node at (-2.5,\y-0.5) {$\bar{G}_0 := $};
\foreach \x in {7}
{\draw[fill=black]  (\x,\y-1) -- (\x+0.5,\y-1) -- (\x+0.5,\y) -- (\x,\y) -- cycle;			
}
\foreach \x in {}
{\draw[fill=gray!40]  (\x,\y-1) -- (\x+0.5,\y-1) -- (\x+0.5,\y) -- (\x,\y) -- cycle;			
}
\foreach \x in {1,2,3,4,6,7.5,8.5,9.5,10.5}
{\draw[fill=white]  (\x,\y-1) -- (\x+1,\y-1) -- (\x+1,\y) -- (\x,\y) -- cycle;				
}
\foreach \x in {5}
{\draw[fill=white]  (\x,\y-1) -- (\x+1,\y-1) -- (\x+1,\y) -- (\x,\y) -- cycle;				
\fill[cyan] (\x+0.5,\y-0.5) circle (0.25);									
}
\foreach \x in {}
{\draw[fill=white]  (\x,\y-1) -- (\x+1,\y-1) -- (\x+1,\y) -- (\x,\y) -- cycle;				
\fill[red] (\x+0.5,\y-0.5) circle (0.25);									
}
\foreach \x in {11.5}
{
\draw[-] (\x,\y) -- (\x+1,\y);
\draw[-] (\x,\y-1) -- (\x+1,\y-1);
\node at (\x+0.7,\y-0.5) {$\cdots$};
}
\foreach \x in {0}
{
\draw[fill=white]  (\x,\y-1) -- (\x+1,\y-1) -- (\x+1,\y) -- (\x,\y) -- cycle;				
\draw[-] (\x,\y) -- (\x-1,\y);
\draw[-] (\x,\y-1) -- (\x-1,\y-1);
\node at (\x-0.6,\y-0.5) {$\cdots$};
}
}
}
\end{figure}
\n We can now apply the ghost-box-ball algorithm to this configuration, which we will do thrice, for good measure:
\begin{figure}[H]
\centering
\tikz[scale=0.49]{
\foreach \y in {0}
{
\node at (-3.15,\y-0.5) {$\hat{\varrho}(\bar{G}_0) := $};
\foreach \x in {7.5}
{\draw[fill=black]  (\x,\y-1) -- (\x+0.5,\y-1) -- (\x+0.5,\y) -- (\x,\y) -- cycle;			
}
\foreach \x in {7}
{\draw[fill=gray!40]  (\x,\y-1) -- (\x+0.5,\y-1) -- (\x+0.5,\y) -- (\x,\y) -- cycle;			
}
\foreach \x in {1,2,3,4,5,8,9,10,11}
{\draw[fill=white]  (\x,\y-1) -- (\x+1,\y-1) -- (\x+1,\y) -- (\x,\y) -- cycle;				
}
\foreach \x in {6}
{\draw[fill=white]  (\x,\y-1) -- (\x+1,\y-1) -- (\x+1,\y) -- (\x,\y) -- cycle;				
\fill[cyan] (\x+0.5,\y-0.5) circle (0.25);									
}
\foreach \x in {}
{\draw[fill=white]  (\x,\y-1) -- (\x+1,\y-1) -- (\x+1,\y) -- (\x,\y) -- cycle;				
\fill[red] (\x+0.5,\y-0.5) circle (0.25);									
}
\foreach \x in {12}
{
\draw[-] (\x,\y) -- (\x+1,\y);
\draw[-] (\x,\y-1) -- (\x+1,\y-1);
\node at (\x+0.7,\y-0.5) {$\cdots$};
}
\foreach \x in {0}
{
\draw[fill=white]  (\x,\y-1) -- (\x+1,\y-1) -- (\x+1,\y) -- (\x,\y) -- cycle;				
\draw[-] (\x,\y) -- (\x-1,\y);
\draw[-] (\x,\y-1) -- (\x-1,\y-1);
\node at (\x-0.6,\y-0.5) {$\cdots$};
}
}
\foreach \y in {-1.75}
{
\node at (-3.3,\y-0.5) {$\hat{\varrho}^2(\bar{G}_0) := $};
\foreach \x in {7}
{\draw[fill=black]  (\x,\y-1) -- (\x+0.5,\y-1) -- (\x+0.5,\y) -- (\x,\y) -- cycle;			
}
\foreach \x in {7.5}
{\draw[fill=gray!40]  (\x,\y-1) -- (\x+0.5,\y-1) -- (\x+0.5,\y) -- (\x,\y) -- cycle;			
}
\foreach \x in {1,2,3,4,5,6,9,10,11}
{\draw[fill=white]  (\x,\y-1) -- (\x+1,\y-1) -- (\x+1,\y) -- (\x,\y) -- cycle;				
}
\foreach \x in {8}
{\draw[fill=white]  (\x,\y-1) -- (\x+1,\y-1) -- (\x+1,\y) -- (\x,\y) -- cycle;				
\fill[cyan] (\x+0.5,\y-0.5) circle (0.25);									
}
\foreach \x in {}
{\draw[fill=white]  (\x,\y-1) -- (\x+1,\y-1) -- (\x+1,\y) -- (\x,\y) -- cycle;				
\fill[red] (\x+0.5,\y-0.5) circle (0.25);									
}
\foreach \x in {12}
{
\draw[-] (\x,\y) -- (\x+1,\y);
\draw[-] (\x,\y-1) -- (\x+1,\y-1);
\node at (\x+0.7,\y-0.5) {$\cdots$};
}
\foreach \x in {0}
{
\draw[fill=white]  (\x,\y-1) -- (\x+1,\y-1) -- (\x+1,\y) -- (\x,\y) -- cycle;				
\draw[-] (\x,\y) -- (\x-1,\y);
\draw[-] (\x,\y-1) -- (\x-1,\y-1);
\node at (\x-0.6,\y-0.5) {$\cdots$};
}
}
\foreach \y in {-3.5}
{
\node at (-3.3,\y-0.5) {$\hat{\varrho}^3(\bar{G}_0) := $};
\foreach \x in {7}
{\draw[fill=black]  (\x,\y-1) -- (\x+0.5,\y-1) -- (\x+0.5,\y) -- (\x,\y) -- cycle;			
}
\foreach \x in {}
{\draw[fill=gray!40]  (\x,\y-1) -- (\x+0.5,\y-1) -- (\x+0.5,\y) -- (\x,\y) -- cycle;			
}
\foreach \x in {1,2,3,4,5,6,7.5,9.5,10.5}
{\draw[fill=white]  (\x,\y-1) -- (\x+1,\y-1) -- (\x+1,\y) -- (\x,\y) -- cycle;				
}
\foreach \x in {8.5}
{\draw[fill=white]  (\x,\y-1) -- (\x+1,\y-1) -- (\x+1,\y) -- (\x,\y) -- cycle;				
\fill[cyan] (\x+0.5,\y-0.5) circle (0.25);									
}
\foreach \x in {}
{\draw[fill=white]  (\x,\y-1) -- (\x+1,\y-1) -- (\x+1,\y) -- (\x,\y) -- cycle;				
\fill[red] (\x+0.5,\y-0.5) circle (0.25);									
}
\foreach \x in {11.5}
{
\draw[-] (\x,\y) -- (\x+1,\y);
\draw[-] (\x,\y-1) -- (\x+1,\y-1);
\node at (\x+0.7,\y-0.5) {$\cdots$};
}
\foreach \x in {0}
{
\draw[fill=white]  (\x,\y-1) -- (\x+1,\y-1) -- (\x+1,\y) -- (\x,\y) -- cycle;				
\draw[-] (\x,\y) -- (\x-1,\y);
\draw[-] (\x,\y-1) -- (\x-1,\y-1);
\node at (\x-0.6,\y-0.5) {$\cdots$};
}
}
}
\end{figure}
\n Finally, reflecting these horizontally, we now obtain the $t=-1,\,t=-2,\,t=-3$ time-steps for the $t=0$ state given by $G_0$:
\begin{figure}[H]
\centering
\tikz[scale=0.49]{
\foreach \y in {0}
{
\node at (-2.5,\y-0.5) {$\bar{G}_0 := $};
\foreach \x in {7}
{\draw[fill=black]  (\x,\y-1) -- (\x+0.5,\y-1) -- (\x+0.5,\y) -- (\x,\y) -- cycle;			
}
\foreach \x in {}
{\draw[fill=gray!40]  (\x,\y-1) -- (\x+0.5,\y-1) -- (\x+0.5,\y) -- (\x,\y) -- cycle;			
}
\foreach \x in {1,2,3,4,6,7.5,8.5,9.5,10.5}
{\draw[fill=white]  (\x,\y-1) -- (\x+1,\y-1) -- (\x+1,\y) -- (\x,\y) -- cycle;				
}
\foreach \x in {5}
{\draw[fill=white]  (\x,\y-1) -- (\x+1,\y-1) -- (\x+1,\y) -- (\x,\y) -- cycle;				
\fill[cyan] (\x+0.5,\y-0.5) circle (0.25);									
}
\foreach \x in {}
{\draw[fill=white]  (\x,\y-1) -- (\x+1,\y-1) -- (\x+1,\y) -- (\x,\y) -- cycle;				
\fill[red] (\x+0.5,\y-0.5) circle (0.25);									
}
\foreach \x in {11.5}
{
\draw[-] (\x,\y) -- (\x+1,\y);
\draw[-] (\x,\y-1) -- (\x+1,\y-1);
\node at (\x+0.7,\y-0.5) {$\cdots$};
}
\foreach \x in {0}
{
\draw[fill=white]  (\x,\y-1) -- (\x+1,\y-1) -- (\x+1,\y) -- (\x,\y) -- cycle;				
\draw[-] (\x,\y) -- (\x-1,\y);
\draw[-] (\x,\y-1) -- (\x-1,\y-1);
\node at (\x-0.6,\y-0.5) {$\cdots$};
}
}
}
\end{figure}
\n We can now apply the ghost-box-ball algorithm to this configuration, which we will do thrice, for good measure:
\begin{figure}[H]
\centering
\tikz[scale=0.49]{
\foreach \y in {0}
{
\node at (-4,\y-0.5) {$t=-3$:};
\foreach \x in {5}
{\draw[fill=black]  (\x,\y-1) -- (\x+0.5,\y-1) -- (\x+0.5,\y) -- (\x,\y) -- cycle;			
}
\foreach \x in {}
{\draw[fill=gray!40]  (\x,\y-1) -- (\x+0.5,\y-1) -- (\x+0.5,\y) -- (\x,\y) -- cycle;			
}
\foreach \x in {1,2,4,5.5,6.5,7.5,8.5,9.5,10.5}
{\draw[fill=white]  (\x,\y-1) -- (\x+1,\y-1) -- (\x+1,\y) -- (\x,\y) -- cycle;				
}
\foreach \x in {3}
{\draw[fill=white]  (\x,\y-1) -- (\x+1,\y-1) -- (\x+1,\y) -- (\x,\y) -- cycle;				
\fill[cyan] (\x+0.5,\y-0.5) circle (0.25);									
}
\foreach \x in {}
{\draw[fill=white]  (\x,\y-1) -- (\x+1,\y-1) -- (\x+1,\y) -- (\x,\y) -- cycle;				
\fill[red] (\x+0.5,\y-0.5) circle (0.25);									
}
\foreach \x in {11.5}
{
\draw[-] (\x,\y) -- (\x+1,\y);
\draw[-] (\x,\y-1) -- (\x+1,\y-1);
\node at (\x+0.7,\y-0.5) {$\cdots$};
}
\foreach \x in {0}
{
\draw[fill=white]  (\x,\y-1) -- (\x+1,\y-1) -- (\x+1,\y) -- (\x,\y) -- cycle;				
\draw[-] (\x,\y) -- (\x-1,\y);
\draw[-] (\x,\y-1) -- (\x-1,\y-1);
\node at (\x-0.6,\y-0.5) {$\cdots$};
}
}
\foreach \y in {-2.5}
{
\node at (-4,\y-0.5) {$t=-2$:};
\foreach \x in {5.5}
{\draw[fill=black]  (\x,\y-1) -- (\x+0.5,\y-1) -- (\x+0.5,\y) -- (\x,\y) -- cycle;			
}
\foreach \x in {5}
{\draw[fill=gray!40]  (\x,\y-1) -- (\x+0.5,\y-1) -- (\x+0.5,\y) -- (\x,\y) -- cycle;			
}
\foreach \x in {1,2,3,6,7,8,9,10,11}
{\draw[fill=white]  (\x,\y-1) -- (\x+1,\y-1) -- (\x+1,\y) -- (\x,\y) -- cycle;				
}
\foreach \x in {4}
{\draw[fill=white]  (\x,\y-1) -- (\x+1,\y-1) -- (\x+1,\y) -- (\x,\y) -- cycle;				
\fill[cyan] (\x+0.5,\y-0.5) circle (0.25);									
}
\foreach \x in {}
{\draw[fill=white]  (\x,\y-1) -- (\x+1,\y-1) -- (\x+1,\y) -- (\x,\y) -- cycle;				
\fill[red] (\x+0.5,\y-0.5) circle (0.25);									
}
\foreach \x in {12}
{
\draw[-] (\x,\y) -- (\x+1,\y);
\draw[-] (\x,\y-1) -- (\x+1,\y-1);
\node at (\x+0.7,\y-0.5) {$\cdots$};
}
\foreach \x in {0}
{
\draw[fill=white]  (\x,\y-1) -- (\x+1,\y-1) -- (\x+1,\y) -- (\x,\y) -- cycle;				
\draw[-] (\x,\y) -- (\x-1,\y);
\draw[-] (\x,\y-1) -- (\x-1,\y-1);
\node at (\x-0.6,\y-0.5) {$\cdots$};
}
}
\foreach \y in {-5}
{
\node at (-4,\y-0.5) {$t=-1$:};
\foreach \x in {5}
{\draw[fill=black]  (\x,\y-1) -- (\x+0.5,\y-1) -- (\x+0.5,\y) -- (\x,\y) -- cycle;			
}
\foreach \x in {5.5}
{\draw[fill=gray!40]  (\x,\y-1) -- (\x+0.5,\y-1) -- (\x+0.5,\y) -- (\x,\y) -- cycle;			
}
\foreach \x in {1,2,3,4,7,8,9,10,11}
{\draw[fill=white]  (\x,\y-1) -- (\x+1,\y-1) -- (\x+1,\y) -- (\x,\y) -- cycle;				
}
\foreach \x in {6}
{\draw[fill=white]  (\x,\y-1) -- (\x+1,\y-1) -- (\x+1,\y) -- (\x,\y) -- cycle;				
\fill[cyan] (\x+0.5,\y-0.5) circle (0.25);									
}
\foreach \x in {}
{\draw[fill=white]  (\x,\y-1) -- (\x+1,\y-1) -- (\x+1,\y) -- (\x,\y) -- cycle;				
\fill[red] (\x+0.5,\y-0.5) circle (0.25);									
}
\foreach \x in {12}
{
\draw[-] (\x,\y) -- (\x+1,\y);
\draw[-] (\x,\y-1) -- (\x+1,\y-1);
\node at (\x+0.7,\y-0.5) {$\cdots$};
}
\foreach \x in {0}
{
\draw[fill=white]  (\x,\y-1) -- (\x+1,\y-1) -- (\x+1,\y) -- (\x,\y) -- cycle;				
\draw[-] (\x,\y) -- (\x-1,\y);
\draw[-] (\x,\y-1) -- (\x-1,\y-1);
\node at (\x-0.6,\y-0.5) {$\cdots$};
}
}
\foreach \y in {-7.5}
{
\node at (-3,\y-0.5) {$G_0 ~~= $};
\foreach \x in {5}
{\draw[fill=black]  (\x,\y-1) -- (\x+0.5,\y-1) -- (\x+0.5,\y) -- (\x,\y) -- cycle;			
}
\foreach \x in {}
{\draw[fill=gray!40]  (\x,\y-1) -- (\x+0.5,\y-1) -- (\x+0.5,\y) -- (\x,\y) -- cycle;			
}
\foreach \x in {1,2,3,4,5.5,7.5,8.5,9.5,10.5}
{\draw[fill=white]  (\x,\y-1) -- (\x+1,\y-1) -- (\x+1,\y) -- (\x,\y) -- cycle;				
}
\foreach \x in {6.5}
{\draw[fill=white]  (\x,\y-1) -- (\x+1,\y-1) -- (\x+1,\y) -- (\x,\y) -- cycle;				
\fill[cyan] (\x+0.5,\y-0.5) circle (0.25);									
}
\foreach \x in {}
{\draw[fill=white]  (\x,\y-1) -- (\x+1,\y-1) -- (\x+1,\y) -- (\x,\y) -- cycle;				
\fill[red] (\x+0.5,\y-0.5) circle (0.25);									
}
\foreach \x in {11.5}
{
\draw[-] (\x,\y) -- (\x+1,\y);
\draw[-] (\x,\y-1) -- (\x+1,\y-1);
\node at (\x+0.7,\y-0.5) {$\cdots$};
}
\foreach \x in {0}
{
\draw[fill=white]  (\x,\y-1) -- (\x+1,\y-1) -- (\x+1,\y) -- (\x,\y) -- cycle;				
\draw[-] (\x,\y) -- (\x-1,\y);
\draw[-] (\x,\y-1) -- (\x-1,\y-1);
\node at (\x-0.6,\y-0.5) {$\cdots$};
}
}
\foreach \z in {1,2,3}{
\node at (6,-2.5*\z+0.7) {$\hat{\varrho}$~\rotatebox[origin=c]{-90}{$\mapsto$}};
}
}
\end{figure}
\end{ex}

\n To close the set under this reverse time evolution, one must include ghost-box-ball systems whose left-most non-empty box contains a ball. On this more general set of ghost-box-ball configurations, Definition \ref{defnofgbbsevolalg} can still be implemented, and it is to this general set that we dedicate this section: extending some of the results of Section \ref{chaptergbbs}.\\

\n At the heart of this extension will be a simple idea: we will define a map, $\vartheta$, which embeds the set of general ghost-box-ball configurations into itself, the image of which will lie in the (restricted) set of ghost-box-ball configurations studied in Section \ref{chaptergbbs} and is invariant under the (forward) time evolution given by the ghost-box-ball algorithm. We show that this map conjugates the dynamics on the general set to that of the restricted set, and hence lift the key results of interest to this general setting. Furthermore, in the next section (Section \ref{sectionphaseshiftbbs}), these constructions play an important role in proving a soliton phase-shift formula for the 2-soliton box-ball system.

\begin{defn}
Let $\text{GBBS}^0$ be the set of ghost-box-ball states for which the left-most non-empty box is a filled ghost. We define the \textbf{augmentation map} $\vartheta: \text{GBBS} \to \text{GBBS}^0$ by taking the empty box that is two spaces to the left of this left-most non-empty box and changing the state of this empty box to a filled ghost.
\end{defn}

\begin{ex}
Below is an example of applying the augmentation map:
\begin{figure}[H]
\centering
\tikz[scale=0.49]{
\foreach \y in {0}
{
\foreach \x in {11.5}
{\draw[fill=black]  (\x,\y-1) -- (\x+0.5,\y-1) -- (\x+0.5,\y) -- (\x,\y) -- cycle;			
}
\foreach \x in {7}
{\draw[fill=gray!40]  (\x,\y-1) -- (\x+0.5,\y-1) -- (\x+0.5,\y) -- (\x,\y) -- cycle;			
}
\foreach \x in {1,2,3,9.5,10.5,12,13}
{\draw[fill=white]  (\x,\y-1) -- (\x+1,\y-1) -- (\x+1,\y) -- (\x,\y) -- cycle;				
}
\foreach \x in {4,5,6,7.5,8.5}
{\draw[fill=white]  (\x,\y-1) -- (\x+1,\y-1) -- (\x+1,\y) -- (\x,\y) -- cycle;				
\fill[cyan] (\x+0.5,\y-0.5) circle (0.25);									
}
\foreach \x in {}
{\draw[fill=white]  (\x,\y-1) -- (\x+1,\y-1) -- (\x+1,\y) -- (\x,\y) -- cycle;				
\fill[red] (\x+0.5,\y-0.5) circle (0.25);									
}
\foreach \x in {14}
{
\draw[-] (\x,\y) -- (\x+1,\y);
\draw[-] (\x,\y-1) -- (\x+1,\y-1);
\node at (\x+0.7,\y-0.5) {$\cdots$};
}
\foreach \x in {0}
{
\draw[fill=white]  (\x,\y-1) -- (\x+1,\y-1) -- (\x+1,\y) -- (\x,\y) -- cycle;				
\draw[-] (\x,\y) -- (\x-1,\y);
\draw[-] (\x,\y-1) -- (\x-1,\y-1);
\node at (\x-0.6,\y-0.5) {$\cdots$};
}
}
\node at (6.5,-2) {$\downarrow \vartheta$};
\foreach \y in {-3}
{
\foreach \x in {2.5,11.5}
{\draw[fill=black]  (\x,\y-1) -- (\x+0.5,\y-1) -- (\x+0.5,\y) -- (\x,\y) -- cycle;			
}
\foreach \x in {7}
{\draw[fill=gray!40]  (\x,\y-1) -- (\x+0.5,\y-1) -- (\x+0.5,\y) -- (\x,\y) -- cycle;			
}
\foreach \x in {0.5,1.5,3,9.5,10.5,12,13}
{\draw[fill=white]  (\x,\y-1) -- (\x+1,\y-1) -- (\x+1,\y) -- (\x,\y) -- cycle;				
}
\foreach \x in {4,5,6,7.5,8.5}
{\draw[fill=white]  (\x,\y-1) -- (\x+1,\y-1) -- (\x+1,\y) -- (\x,\y) -- cycle;				
\fill[cyan] (\x+0.5,\y-0.5) circle (0.25);									
}
\foreach \x in {}
{\draw[fill=white]  (\x,\y-1) -- (\x+1,\y-1) -- (\x+1,\y) -- (\x,\y) -- cycle;				
\fill[red] (\x+0.5,\y-0.5) circle (0.25);									
}
\foreach \x in {14}
{
\draw[-] (\x,\y) -- (\x+1,\y);
\draw[-] (\x,\y-1) -- (\x+1,\y-1);
\node at (\x+0.7,\y-0.5) {$\cdots$};
}
\foreach \x in {-0.5}
{
\draw[fill=white]  (\x,\y-1) -- (\x+1,\y-1) -- (\x+1,\y) -- (\x,\y) -- cycle;				
\draw[-] (\x,\y) -- (\x-1,\y);
\draw[-] (\x,\y-1) -- (\x-1,\y-1);
\node at (\x-0.6,\y-0.5) {$\cdots$};
}
}
}
\end{figure}
\n The augmentation map also applies to ghost-box-ball states that are already in $\text{GBBS}^0$, for example:

\begin{figure}[H]
\centering
\tikz[scale=0.49]{
\foreach \y in {0}
{
\foreach \x in {4}
{\draw[fill=black]  (\x,\y-1) -- (\x+0.5,\y-1) -- (\x+0.5,\y) -- (\x,\y) -- cycle;			
}
\foreach \x in {4.5}
{\draw[fill=gray!40]  (\x,\y-1) -- (\x+0.5,\y-1) -- (\x+0.5,\y) -- (\x,\y) -- cycle;			
}
\foreach \x in {1,2,3,7,9,10,11}
{\draw[fill=white]  (\x,\y-1) -- (\x+1,\y-1) -- (\x+1,\y) -- (\x,\y) -- cycle;				
}
\foreach \x in {5,6,8}
{\draw[fill=white]  (\x,\y-1) -- (\x+1,\y-1) -- (\x+1,\y) -- (\x,\y) -- cycle;				
\fill[cyan] (\x+0.5,\y-0.5) circle (0.25);									
}
\foreach \x in {}
{\draw[fill=white]  (\x,\y-1) -- (\x+1,\y-1) -- (\x+1,\y) -- (\x,\y) -- cycle;				
\fill[red] (\x+0.5,\y-0.5) circle (0.25);									
}
\foreach \x in {12}
{
\draw[-] (\x,\y) -- (\x+1,\y);
\draw[-] (\x,\y-1) -- (\x+1,\y-1);
\node at (\x+0.7,\y-0.5) {$\cdots$};
}
\foreach \x in {0}
{
\draw[fill=white]  (\x,\y-1) -- (\x+1,\y-1) -- (\x+1,\y) -- (\x,\y) -- cycle;				
\draw[-] (\x,\y) -- (\x-1,\y);
\draw[-] (\x,\y-1) -- (\x-1,\y-1);
\node at (\x-0.6,\y-0.5) {$\cdots$};
}
}
\node at (6.5,-2) {$\downarrow \vartheta$};
\foreach \y in {-3}
{
\foreach \x in {2.5,4}
{\draw[fill=black]  (\x,\y-1) -- (\x+0.5,\y-1) -- (\x+0.5,\y) -- (\x,\y) -- cycle;			
}
\foreach \x in {4.5}
{\draw[fill=gray!40]  (\x,\y-1) -- (\x+0.5,\y-1) -- (\x+0.5,\y) -- (\x,\y) -- cycle;			
}
\foreach \x in {1.5,3,7,9,10,11}
{\draw[fill=white]  (\x,\y-1) -- (\x+1,\y-1) -- (\x+1,\y) -- (\x,\y) -- cycle;				
}
\foreach \x in {5,6,8}
{\draw[fill=white]  (\x,\y-1) -- (\x+1,\y-1) -- (\x+1,\y) -- (\x,\y) -- cycle;				
\fill[cyan] (\x+0.5,\y-0.5) circle (0.25);									
}
\foreach \x in {}
{\draw[fill=white]  (\x,\y-1) -- (\x+1,\y-1) -- (\x+1,\y) -- (\x,\y) -- cycle;				
\fill[red] (\x+0.5,\y-0.5) circle (0.25);									
}
\foreach \x in {12}
{
\draw[-] (\x,\y) -- (\x+1,\y);
\draw[-] (\x,\y-1) -- (\x+1,\y-1);
\node at (\x+0.7,\y-0.5) {$\cdots$};
}
\foreach \x in {0.5}
{
\draw[fill=white]  (\x,\y-1) -- (\x+1,\y-1) -- (\x+1,\y) -- (\x,\y) -- cycle;				
\draw[-] (\x,\y) -- (\x-1,\y);
\draw[-] (\x,\y-1) -- (\x-1,\y-1);
\node at (\x-0.6,\y-0.5) {$\cdots$};
}
}
}
\end{figure}
\end{ex}

\begin{rem}
The image of $\vartheta$ is an invariant subset $\hat{\varrho}(\vartheta(\text{GBBS}))\subset \vartheta(\text{GBBS})$. This is just a consequence of the ball dynamics moving to the right, with enough separation of the left-most filled ghost from the ball dynamics. Moreover, since the augmentation map is clearly injective, it is invertible on its image $\vartheta(\text{GBBS})$. Thus, $\vartheta$ embeds $\text{GBBS}$ into $\text{GBBS}^0$ as an invariant subset.
\end{rem}

\begin{thm}\label{thmvarthetaequiv} The following diagram commutes:
\begin{figure}[H]
\centering
\tikz[scale=0.7]{
\node (1) at (0,0) {GBBS};
\node (2) at (5,0) {GBBS};
\node (3) at (0,-5) {$\text{GBBS}^0$};
\node (4) at (5,-5) {$\text{GBBS}^0$};
\draw[->] (1) -- (2) node[above,midway] {$\hat{\varrho}$};
\draw[->] (1) -- (3) node[left,midway] {$\vartheta$};
\draw[->] (3) -- (4) node[above,midway] {$\hat{\varrho}|_{\text{GBBS}^0}$};
\draw[->] (2) -- (4) node[right,midway] {$\vartheta$};
}
\end{figure}
\n Furthermore, since $\vartheta$ is invertible on its image, we have that iterating the ghost-box-ball algorithm on $\text{GBBS}$ is equivalent to augmenting, iterating on the resulting ghost-box-ball state, and then inverting the augmentation. \textit{i.e.},
\begin{equation}
    \hat{\varrho}^k = \vartheta^{-1} \circ \left(\hat{\varrho}|_{\text{GBBS}^0}\right)^k \circ \vartheta.\label{gbbsaugdeaug}
\end{equation}
\end{thm}
\begin{proof}
The commutation of the diagram is equivalent to showing Equation \ref{gbbsaugdeaug} for $k=1$, which is immediate since this just says that if one introduces a filled ghost to the left of the dynamic evolution, applies the ghost-box-ball algorithm, then removes that filled ghost (which was stationary and didn't interact with the dynamics), then that is the same as simply applying the ghost-box-ball algorithm. Thus, we have
\begin{equation}
    \hat{\varrho} = \vartheta^{-1} \circ \left(\hat{\varrho}|_{\text{GBBS}^0}\right) \circ \vartheta.\label{gbbsaugdeaugnisone}
\end{equation}
\n Equation \ref{gbbsaugdeaug} then follows by composition of Equation \ref{gbbsaugdeaugnisone} with itself $k$ times.
\end{proof}

\begin{cor}
It now follows that the soliton structure, sorting property and invariant shape constructions for the subclass $\text{GBBS}^0$ proved in Chapter \ref{chaptergbbs} holds on the entirety of $\text{GBBS}$, the general ghost-box-ball system.
\end{cor}

\n  We present some extensions of earlier definitions to set us up for the next section. The sets we define offer the generalisations of the sets $\mathcal{G}_n^0$ and $\mathcal{G}^0$ for the ghost-box-ball coordinatisations of the ghost-box-balls of Section \ref{chaptergbbs} (those that start with a filled ghost) to the expected coordinate spaces of the more general ghost-box-ball systems.

\begin{defn}
Let $$\mathcal{G}_n=\{\infty\}\x \N_0^{2n-1} \x \{\infty\}$$
and
$$\mathcal{G} = \bigcup_{n\in\N} \mathcal{G}_n.$$
The latter is the domain of the natural extension of the map $C$ in Definition \ref{coordmapdefn} to the general setting of ghost-box-ball settings. We continue to call this extension $C:\text{GBBS}\to \mathcal{G}$.
\end{defn}

\begin{defn}
Let $\Upsilon:\mathcal{G}\to \mathcal{G}^0$ be the map induced by the maps $\Upsilon_n:\mathcal{G}_n\to\mathcal{G}_{n+1}^0$:
$$\Upsilon_n(\infty,z_1,z_2,\ldots,z_{2n-1},\infty) = (\infty,0,1,z_1,z_2,\ldots,z_{2n-1},\infty).$$
On the image $C(\vartheta(\text{GBBS}))$ of $\vartheta(\text{GBBS})$ under $C$, one has enough coordinates to invert $\Upsilon$, by simply omitting the second and third coordinates. Moreover, this map extends to all subsequent $C(\vartheta^k(\text{GBBS}))$ for each $k\in\N$ by still omitting the second and third coordinates. Call this map $\hat{\Upsilon}$.
\end{defn}

\n We can now state relations between $\vartheta$ and $\Upsilon$, as well as their inverses on appropriate sets:

\begin{lem} The following two squares commute:
\begin{figure}[H]
\centering
\tikz[scale=0.7]{
\node (1) at (0,0) {$\text{GBBS}$};
\node (2) at (5,0) {$\text{GBBS}^0$};
\node (3) at (0,-5) {$\mathcal{G}$};
\node (4) at (5,-5) {$\mathcal{G}^0$};
\draw[->] (1) -- (2) node[above,midway] {$\vartheta$};
\draw[->] (1) -- (3) node[left,midway] {$C$};
\draw[->] (3) -- (4) node[above,midway] {$\Upsilon$};
\draw[->] (2) -- (4) node[right,midway] {$C$};
}
~~~~~
\tikz[scale=0.7]{
\node (1) at (0,0) {$\vartheta(\text{GBBS})$};
\node (2) at (5,0) {$\text{GBBS}$};
\node (3) at (0,-5) {$C(\vartheta(\text{GBBS}))$};
\node (4) at (5,-5) {$\mathcal{G}$};
\draw[->] (1) -- (2) node[above,midway] {$\vartheta^{-1}$};
\draw[->] (1) -- (3) node[left,midway] {$C$};
\draw[->] (3) -- (4) node[above,midway] {$\hat{\Upsilon}$};
\draw[->] (2) -- (4) node[right,midway] {$C$};
}.
\end{figure}
\end{lem}

\n Ultimately, in studying the general set of ghost-box-ball systems, we would like to extend Corollary \ref{conjectureaboutgbbsbbs} to this more general ghost-box-ball setting. For now, with these definitions at our disposal, we demonstrate an application of these maps in the next section on the box-ball phase shift.

\subsection{Phase-Shift Formula}\label{sectionphaseshiftbbs}
\n In this section, we make use of these maps $\vartheta$ and $\Upsilon$, as well as their inverses on appropriate sets, to prove a phase-shift formula for the classical box-ball system, which comes by means of passing over to the ghost-box-ball setting. The subsequent utility of the map $\vartheta$ is in its introduction of a stationary object (an initial filled ghost), relative to which motion is measured. Essentially, where the classical box-ball system's coordinates exhibit left- and right-shift invariance (due to the padding by infinity on each end), the introduced filled ghost plays the role of an origin point, against which the dynamics is measured.\\[2pt]

\n We begin by introducing the notion of the box-ball phase shift phenomenon.

\subsubsection{The Box-Ball Phase Shift}\label{bbsphsshftsecmn}
We have seen how, as $t\to +\infty$, the blocks sort themselves by increasing length. The same holds in reverse time: as $t\to-\infty$, the blocks are ordered by decreasing lengths. The asymptotic sequence of lengths are revealed at any finite time using the invariant shape construction (Section \ref{invsofbbs}). The following example demonstrates why one cannot simply just count block lengths (the third state does not show the soliton structure of blocks of length 1 and 3):

\begin{figure}[H]
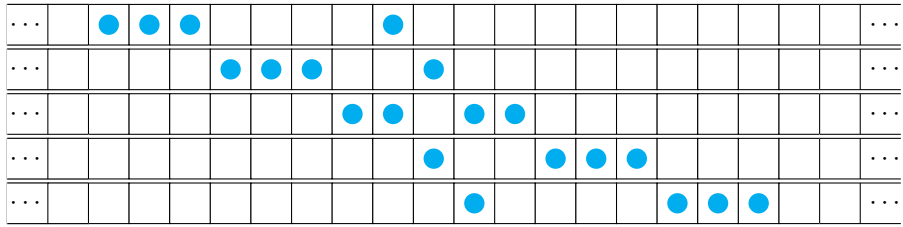

\centering
\tikz[scale=0.54]{
\foreach \y in {3}{
\foreach \x in {-1,0,1,2,3,4,5,6,7,8,9,10,11,12,13,14,15,16,17,18}
{\draw[fill=white]  (\x,\y) -- (\x+1,\y) -- (\x+1,\y+1) -- (\x,\y+1) -- cycle;			
}
\foreach \x in {1,2,3,8}
{
\fill[cyan] (\x+0.5,\y+0.5) circle (0.25);
}
\foreach \x in {19}
{\draw[fill=white,white]  (\x,\y) -- (\x+2,\y) -- (\x+2,\y+1) -- (\x,\y+1) -- cycle;
\draw[-] (\x,\y) -- (\x,\y+1);
\draw[-] (\x,\y) -- (\x+2,\y);
\draw[-] (\x,\y+1) -- (\x+2,\y+1);
\draw[-] (\x+1,\y) -- (\x+1,\y+1);
\node at (\x+1.5,\y+0.5) {$~\cdots$};
}
\foreach \x in {1}
{\draw[fill=white,white]  (\x,\y) -- (\x-2,\y) -- (\x-2,\y+1) -- (\x,\y+1) -- cycle;
\draw[-] (\x,\y) -- (\x,\y+1);
\draw[-] (\x,\y) -- (\x-2,\y);
\draw[-] (\x,\y+1) -- (\x-2,\y+1);
\draw[-] (\x-1,\y+1) -- (\x-1,\y+1);
\draw[-] (\x-1,\y+1) -- (\x-1,\y);
\node at (\x-1.5,\y+0.5) {$\cdots$};
}}
\foreach \y in {1.9}{
\foreach \x in {-1,0,1,2,3,4,5,6,7,8,9,10,11,12,13,14,15,16,17,18}
{\draw[fill=white]  (\x,\y) -- (\x+1,\y) -- (\x+1,\y+1) -- (\x,\y+1) -- cycle;			
}
\foreach \x in {4,5,6,9}
{
\fill[cyan] (\x+0.5,\y+0.5) circle (0.25);
}
\foreach \x in {19}
{\draw[fill=white,white]  (\x,\y) -- (\x+2,\y) -- (\x+2,\y+1) -- (\x,\y+1) -- cycle;
\draw[-] (\x,\y) -- (\x,\y+1);
\draw[-] (\x,\y) -- (\x+2,\y);
\draw[-] (\x,\y+1) -- (\x+2,\y+1);
\draw[-] (\x+1,\y) -- (\x+1,\y+1);
\node at (\x+1.5,\y+0.5) {$~\cdots$};
}
\foreach \x in {1}
{\draw[fill=white,white]  (\x,\y) -- (\x-2,\y) -- (\x-2,\y+1) -- (\x,\y+1) -- cycle;
\draw[-] (\x,\y) -- (\x,\y+1);
\draw[-] (\x,\y) -- (\x-2,\y);
\draw[-] (\x,\y+1) -- (\x-2,\y+1);
\draw[-] (\x-1,\y+1) -- (\x-1,\y+1);
\draw[-] (\x-1,\y+1) -- (\x-1,\y);
\node at (\x-1.5,\y+0.5) {$\cdots$};
}}
\foreach \y in {0.8}{
\foreach \x in {-1,0,1,2,3,4,5,6,7,8,9,10,11,12,13,14,15,16,17,18}
{\draw[fill=white]  (\x,\y) -- (\x+1,\y) -- (\x+1,\y+1) -- (\x,\y+1) -- cycle;			
}
\foreach \x in {7,8,10,11}
{
\fill[cyan] (\x+0.5,\y+0.5) circle (0.25);
}
\foreach \x in {19}
{\draw[fill=white,white]  (\x,\y) -- (\x+2,\y) -- (\x+2,\y+1) -- (\x,\y+1) -- cycle;
\draw[-] (\x,\y) -- (\x,\y+1);
\draw[-] (\x,\y) -- (\x+2,\y);
\draw[-] (\x,\y+1) -- (\x+2,\y+1);
\draw[-] (\x+1,\y) -- (\x+1,\y+1);
\node at (\x+1.5,\y+0.5) {$~\cdots$};
}
\foreach \x in {1}
{\draw[fill=white,white]  (\x,\y) -- (\x-2,\y) -- (\x-2,\y+1) -- (\x,\y+1) -- cycle;
\draw[-] (\x,\y) -- (\x,\y+1);
\draw[-] (\x,\y) -- (\x-2,\y);
\draw[-] (\x,\y+1) -- (\x-2,\y+1);
\draw[-] (\x-1,\y+1) -- (\x-1,\y+1);
\draw[-] (\x-1,\y+1) -- (\x-1,\y);
\node at (\x-1.5,\y+0.5) {$\cdots$};
}}
\foreach \y in {-0.3}{
\foreach \x in {-1,0,1,2,3,4,5,6,7,8,9,10,11,12,13,14,15,16,17,18}
{\draw[fill=white]  (\x,\y) -- (\x+1,\y) -- (\x+1,\y+1) -- (\x,\y+1) -- cycle;			
}
\foreach \x in {9,12,13,14}
{
\fill[cyan] (\x+0.5,\y+0.5) circle (0.25);
}
\foreach \x in {19}
{\draw[fill=white,white]  (\x,\y) -- (\x+2,\y) -- (\x+2,\y+1) -- (\x,\y+1) -- cycle;
\draw[-] (\x,\y) -- (\x,\y+1);
\draw[-] (\x,\y) -- (\x+2,\y);
\draw[-] (\x,\y+1) -- (\x+2,\y+1);
\draw[-] (\x+1,\y) -- (\x+1,\y+1);
\node at (\x+1.5,\y+0.5) {$~\cdots$};
}
\foreach \x in {1}
{\draw[fill=white,white]  (\x,\y) -- (\x-2,\y) -- (\x-2,\y+1) -- (\x,\y+1) -- cycle;
\draw[-] (\x,\y) -- (\x,\y+1);
\draw[-] (\x,\y) -- (\x-2,\y);
\draw[-] (\x,\y+1) -- (\x-2,\y+1);
\draw[-] (\x-1,\y+1) -- (\x-1,\y+1);
\draw[-] (\x-1,\y+1) -- (\x-1,\y);
\node at (\x-1.5,\y+0.5) {$\cdots$};
}}
\foreach \y in {-1.4}{
\foreach \x in {-1,0,1,2,3,4,5,6,7,8,9,10,11,12,13,14,15,16,17,18}
{\draw[fill=white]  (\x,\y) -- (\x+1,\y) -- (\x+1,\y+1) -- (\x,\y+1) -- cycle;			
}
\foreach \x in {10,15,16,17}
{
\fill[cyan] (\x+0.5,\y+0.5) circle (0.25);
}
\foreach \x in {19}
{\draw[fill=white,white]  (\x,\y) -- (\x+2,\y) -- (\x+2,\y+1) -- (\x,\y+1) -- cycle;
\draw[-] (\x,\y) -- (\x,\y+1);
\draw[-] (\x,\y) -- (\x+2,\y);
\draw[-] (\x,\y+1) -- (\x+2,\y+1);
\draw[-] (\x+1,\y) -- (\x+1,\y+1);
\node at (\x+1.5,\y+0.5) {$~\cdots$};
}
\foreach \x in {1}
{\draw[fill=white,white]  (\x,\y) -- (\x-2,\y) -- (\x-2,\y+1) -- (\x,\y+1) -- cycle;
\draw[-] (\x,\y) -- (\x,\y+1);
\draw[-] (\x,\y) -- (\x-2,\y);
\draw[-] (\x,\y+1) -- (\x-2,\y+1);
\draw[-] (\x-1,\y+1) -- (\x-1,\y+1);
\draw[-] (\x-1,\y+1) -- (\x-1,\y);
\node at (\x-1.5,\y+0.5) {$\cdots$};
}}
}
\label{figcolsdhufhdf}
\caption{A phase shift interaction between two colliding blocks.}
\end{figure}	

\begin{rem}\label{remarkaboutspacingforsoliton}
In the above example, we can discern the asymptotic ordering in the first, second, fourth and fifth rows, simply by counting the numbers of balls in each block of adjacent balls. The middle row (the third) could be misleading, since it reveals a $(2,2)$ structure for the blocks (although, the invariant shape construction would reveal the correct soliton structure here). If two blocks are spaced far enough apart, then no such obfuscation occurs. 
\end{rem} 

\n Barring this intricacy (\textit{i.e.} when there is enough space between consecutive blocks), one can take two blocks, evolve sufficiently many times according to the box-ball evolution, and compare the position of the blocks to where they would have been if it had not have been for the collision.\\

\n In the figure below, we replicate Figure \ref{figcolsdhufhdf}. However, we use green balls to keep track of where the block of three balls would have been without the collision, and magenta balls to keep track of where the block of one ball would have been.
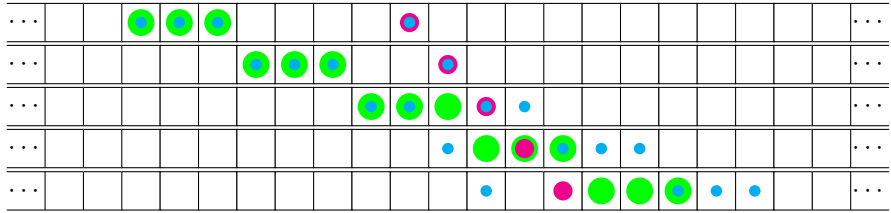
\vspace{0.4cm}

\begin{figure}[H]
\centering
\tikz[scale=0.51]{
\foreach \x in {0,1,2,3,4,5,6,7,8,9,10,11,12,13,14,15,16,17,18}
{\draw[fill=white]  (\x,3) -- (\x+1,3) -- (\x+1,4) -- (\x,4) -- cycle;			
}
\foreach \x in {1,2,3}
{			
\fill[green] (\x+0.5,3.5) circle (0.35);
}
\foreach \x in {8}
{			
\fill[magenta] (\x+0.5,3.5) circle (0.25);
}
\foreach \x in {1,2,3,8}
{			
\fill[cyan] (\x+0.5,3.5) circle (0.15);
}
\foreach \x in {19}
{\draw[fill=white,white]  (\x,3) -- (\x+2,3) -- (\x+2,4) -- (\x,4) -- cycle;
\draw[-] (\x,3) -- (\x,4);
\draw[-] (\x,3) -- (\x+2,3);
\draw[-] (\x,4) -- (\x+2,4);
\draw[-] (\x+1,3) -- (\x+1,4);
\node at (\x+1.5,3.5) {$\cdots$};
}
\foreach \x in {0}
{\draw[fill=white,white]  (\x,3) -- (\x-2,3) -- (\x-2,4) -- (\x,4) -- cycle;
\draw[-] (\x,3) -- (\x,4);
\draw[-] (\x,3) -- (\x-2,3);
\draw[-] (\x,4) -- (\x-2,4);
\draw[-] (\x-1,3) -- (\x-1,4);
\node at (\x-1.5,3.5) {$\cdots$};
}
}
\tikz[scale=0.51]{
\foreach \x in {0,1,2,3,4,5,6,7,8,9,10,11,12,13,14,15,16,17,18}
{\draw[fill=white]  (\x,3) -- (\x+1,3) -- (\x+1,4) -- (\x,4) -- cycle;			
}
\foreach \x in {4,5,6}
{			
\fill[green] (\x+0.5,3.5) circle (0.35);
}
\foreach \x in {9}
{			
\fill[magenta] (\x+0.5,3.5) circle (0.25);
}
\foreach \x in {4,5,6,9}
{			
\fill[cyan] (\x+0.5,3.5) circle (0.15);
}
\foreach \x in {19}
{\draw[fill=white,white]  (\x,3) -- (\x+2,3) -- (\x+2,4) -- (\x,4) -- cycle;
\draw[-] (\x,3) -- (\x,4);
\draw[-] (\x,3) -- (\x+2,3);
\draw[-] (\x,4) -- (\x+2,4);
\draw[-] (\x+1,3) -- (\x+1,4);
\node at (\x+1.5,3.5) {$\cdots$};
}
\foreach \x in {0}
{\draw[fill=white,white]  (\x,3) -- (\x-2,3) -- (\x-2,4) -- (\x,4) -- cycle;
\draw[-] (\x,3) -- (\x,4);
\draw[-] (\x,3) -- (\x-2,3);
\draw[-] (\x,4) -- (\x-2,4);
\draw[-] (\x-1,3) -- (\x-1,4);
\node at (\x-1.5,3.5) {$\cdots$};
}
}
\tikz[scale=0.51]{
\foreach \x in {0,1,2,3,4,5,6,7,8,9,10,11,12,13,14,15,16,17,18}
{\draw[fill=white]  (\x,3) -- (\x+1,3) -- (\x+1,4) -- (\x,4) -- cycle;			
}
\foreach \x in {7,8,9}
{			
\fill[green] (\x+0.5,3.5) circle (0.35);
}
\foreach \x in {10}
{			
\fill[magenta] (\x+0.5,3.5) circle (0.25);
}
\foreach \x in {7,8,10,11}
{			
\fill[cyan] (\x+0.5,3.5) circle (0.15);
}
\foreach \x in {19}
{\draw[fill=white,white]  (\x,3) -- (\x+2,3) -- (\x+2,4) -- (\x,4) -- cycle;
\draw[-] (\x,3) -- (\x,4);
\draw[-] (\x,3) -- (\x+2,3);
\draw[-] (\x,4) -- (\x+2,4);
\draw[-] (\x+1,3) -- (\x+1,4);
\node at (\x+1.5,3.5) {$\cdots$};
}
\foreach \x in {0}
{\draw[fill=white,white]  (\x,3) -- (\x-2,3) -- (\x-2,4) -- (\x,4) -- cycle;
\draw[-] (\x,3) -- (\x,4);
\draw[-] (\x,3) -- (\x-2,3);
\draw[-] (\x,4) -- (\x-2,4);
\draw[-] (\x-1,3) -- (\x-1,4);
\node at (\x-1.5,3.5) {$\cdots$};
}
}
\tikz[scale=0.51]{
\foreach \x in {0,1,2,3,4,5,6,7,8,9,10,11,12,13,14,15,16,17,18}
{\draw[fill=white]  (\x,3) -- (\x+1,3) -- (\x+1,4) -- (\x,4) -- cycle;			
}
\foreach \x in {10,11,12}
{			
\fill[green] (\x+0.5,3.5) circle (0.35);
}
\foreach \x in {11}
{			
\fill[magenta] (\x+0.5,3.5) circle (0.25);
}
\foreach \x in {9,12,13,14}
{			
\fill[cyan] (\x+0.5,3.5) circle (0.15);
}
\foreach \x in {19}
{\draw[fill=white,white]  (\x,3) -- (\x+2,3) -- (\x+2,4) -- (\x,4) -- cycle;
\draw[-] (\x,3) -- (\x,4);
\draw[-] (\x,3) -- (\x+2,3);
\draw[-] (\x,4) -- (\x+2,4);
\draw[-] (\x+1,3) -- (\x+1,4);
\node at (\x+1.5,3.5) {$\cdots$};
}
\foreach \x in {0}
{\draw[fill=white,white]  (\x,3) -- (\x-2,3) -- (\x-2,4) -- (\x,4) -- cycle;
\draw[-] (\x,3) -- (\x,4);
\draw[-] (\x,3) -- (\x-2,3);
\draw[-] (\x,4) -- (\x-2,4);
\draw[-] (\x-1,3) -- (\x-1,4);
\node at (\x-1.5,3.5) {$\cdots$};
}
}
\tikz[scale=0.51]{
\foreach \x in {0,1,2,3,4,5,6,7,8,9,10,11,12,13,14,15,16,17,18}
{\draw[fill=white]  (\x,3) -- (\x+1,3) -- (\x+1,4) -- (\x,4) -- cycle;			
}
\foreach \x in {13,14,15}
{			
\fill[green] (\x+0.5,3.5) circle (0.35);
}
\foreach \x in {12}
{			
\fill[magenta] (\x+0.5,3.5) circle (0.25);
}
\foreach \x in {10,15,16,17}
{			
\fill[cyan] (\x+0.5,3.5) circle (0.15);
}
\foreach \x in {19}
{\draw[fill=white,white]  (\x,3) -- (\x+2,3) -- (\x+2,4) -- (\x,4) -- cycle;
\draw[-] (\x,3) -- (\x,4);
\draw[-] (\x,3) -- (\x+2,3);
\draw[-] (\x,4) -- (\x+2,4);
\draw[-] (\x+1,3) -- (\x+1,4);
\node at (\x+1.5,3.5) {$\cdots$};
}
\foreach \x in {0}
{\draw[fill=white,white]  (\x,3) -- (\x-2,3) -- (\x-2,4) -- (\x,4) -- cycle;
\draw[-] (\x,3) -- (\x,4);
\draw[-] (\x,3) -- (\x-2,3);
\draw[-] (\x,4) -- (\x-2,4);
\draw[-] (\x-1,3) -- (\x-1,4);
\node at (\x-1.5,3.5) {$\cdots$};
}
}
\end{figure}	

\n When the collision has concluded, we see that the three-block is two positions ahead of where it would have been, and the one-block is two positions behind where it would have been. Therefore, we say that the three-block experiences a $+2$ phase shift, and the one-block experiences a $-2$ phase shift.\\

\subsubsection{The 2-Soliton Phase Shift Formula}
\n We want to study the phase shift for 2-soliton box-ball collision. For a collision to occur, we must have a larger block of balls to the left of a smaller block of balls. By definition of blocks, the two must be separated by a sequence of empty boxes, which we refer to as the \textit{gap}. With this in mind, we shall initialise with a configuration of a block of $k$ balls on the left, a gap of $g$ empty boxes, followed by a block of $l<k$ balls:

\begin{figure}[H]
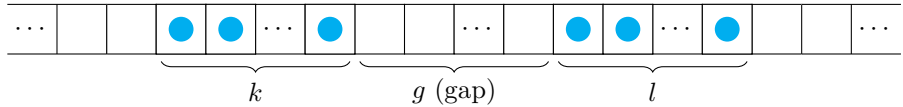

\centering
\tikz[scale=0.66]{
\foreach \x in {0,1,2,3,4,5,6,7,8,9,10,11,12,13}
{\draw[fill=white]  (\x,3) -- (\x+1,3) -- (\x+1,4) -- (\x,4) -- cycle;			
}
\foreach \x in {1,2,4,9,10,12}
{\draw[fill=white]  (\x,3) -- (\x+1,3) -- (\x+1,4) -- (\x,4) -- cycle;			
\fill[cyan] (\x+0.5,3.5) circle (0.25);
}
\foreach \x in {}
{\draw[fill=white]  (\x,3) -- (\x+1,3) -- (\x+1,4) -- (\x,4) -- cycle;			
\fill[red] (\x+0.5,3.5) circle (0.25);
}
\foreach \x in {14}
{\draw[fill=white,white]  (\x,3) -- (\x+2,3) -- (\x+2,4) -- (\x,4) -- cycle;
\draw[-] (\x,3) -- (\x,4);
\draw[-] (\x,3) -- (\x+2,3);
\draw[-] (\x,4) -- (\x+2,4);
\draw[-] (\x+1,3) -- (\x+1,4);
\node at (\x+1.5,3.5) {$\cdots$};
}
\foreach \x in {0}
{\draw[fill=white,white]  (\x,3) -- (\x-2,3) -- (\x-2,4) -- (\x,4) -- cycle;
\draw[-] (\x,3) -- (\x,4);
\draw[-] (\x,3) -- (\x-2,3);
\draw[-] (\x,4) -- (\x-2,4);
\draw[-] (\x-1,3) -- (\x-1,4);
\node at (\x-1.5,3.5) {$\cdots$};
}
\foreach \x in {3,11,7}
{
\node at (\x+0.5,3.5) {$\cdots$};
}
\foreach \x in {4}{
\draw [decorate,decoration={brace,amplitude=4pt}] (\x+0.9,2.85) -- (\x-2.9,2.85) node [black,midway,yshift=-0.4cm] {$k$};
}
\foreach \x in {8}{
\draw [decorate,decoration={brace,amplitude=4pt}] (\x+0.9,2.85) -- (\x-2.9,2.85) node [black,midway,yshift=-0.4cm] {$g\text{ (gap)}$};
}
\foreach \x in {12}{
\draw [decorate,decoration={brace,amplitude=4pt}] (\x+0.9,2.85) -- (\x-2.9,2.85) node [black,midway,yshift=-0.4cm] {$l$};
}
}
\caption{BBS with a $(k,l)$ structure, prior to the collision}\label{bbspriorcoll}
\end{figure}

\n We assume that $g$ is large enough so that $(k,l)$ is representative of the solitonic structure of this box-ball configuration (as discussed in Remark \ref{remarkaboutspacingforsoliton}). We also motivate some terminology in the following:

\begin{defn}
For a 2-soliton system, if one cannot simply see the soliton structure (the asymptotic sizes of blocks) in a box-ball state by counting the block sizes at a given time, that state will be said to be in the \textit{collision phase}. We say that a state is \textit{pre-collision} (or that we are a time \textit{prior to collision}) if the blocks are ordered from largest to smallest and not in the collision phase.
\end{defn}

\n It is natural to ask how one can say in general whether a box-ball state of the type in Figure \ref{bbspriorcoll} is in its collision phase or is pre-collision. We answer this using the invariant shape construction of Section \ref{invsofbbs}.

\begin{lem}
A box-ball configuration of the type shown in Figure \ref{bbspriorcoll} is pre-collision if and only if $g\geq l$.
\end{lem}
\begin{proof}
Using the ``10'' construction of Section \ref{invsofbbs}, and viewing the configuration as a sequence of $1$'s and $0$'s, we recall that we must count all instances of $10$'s, remove them, and repeat the process to get a sequence of counts. \\[2pt]
If $g\geq l$, then one has enough $0$'s between the two blocks of $1$'s to get a sequence $(p_i)_{i=1}^k$:
$$p_i = \left\{ \begin{array}{cl} 2 & i\leq l\\ 1 & l<i\leq k \end{array}\right.$$
The resulting Young diagram has $l$ boxes in the bottom row and $(k-l)+l=k$ boxes in the top row. Therefore, if $g\geq l$, then the box-ball state in Figure \ref{bbspriorcoll} is pre-collision.\\[2pt]
Conversely, if $g<l$, one would exhaust the $g$ zeroes in between the two blocks before finishing counting off the $1$'s in the $l$-block. At that point, the remaining $k-g$ balls on the left would form a single block with the $l-g$ balls on the right. The resulting sequence would be  $(p_i)_{i=1}^k$, where
$$p_i = \left\{ \begin{array}{cl} 2 & i\leq g\\ 1 & g<i\leq k+l-g \end{array}\right.$$
The Young diagram here would have $g$ boxes in its bottom row and $k+l-g$ boxes in its top row. Since $g<l<k$, $g$ is not equal to $k$ or $l$, so the block sizes in such a state would not be indicative of the soliton structure. Therefore, such a configuration is in its collision phase.
\end{proof}

\n Before proving the 2-soliton result, we characterise the inception of a collision phase.

\begin{lem}
Taking the box-ball configuration in Figure \ref{bbspriorcoll} to be the time $t=0$ state, assuming $g\geq l$, the state at time
$$t_\text{max}:=\left\lfloor \dfrac{l-g}{l-k}\right\rfloor$$
is the last time prior to the collision, any time beyond this is either part of the collision phase or a time for which the sorting has concluded. Additionally, this is the unique time for which the gap lies in the interval $[l,k)$.
\end{lem}
\begin{proof}
Prior to the collision, the $k$-block moves forwards (reducing the gap by) $k$ units, and the $l$-block moves forwards (increasing the gap by) $l$ units. Therefore, for each time-step prior to the collision, the gap becomes
$$g+t(l-k).$$
To then be pre-collision at time $t$, we require
$$g+t(l-k)\geq l.$$
The maximal such $t$ is $t_\text{max}$ given in the theorem.\\[2pt]

\n At this time, the gap is given by
\begin{align*}
g+t_\text{max}(l-k) &= g+(l-k)\left\lfloor \dfrac{l-g}{l-k}\right\rfloor\\
&<g+(l-k)\left(\dfrac{l-g}{l-k}-1\right)\\
&=k.
\end{align*}
The next gap would then be strictly less than $k+(l-k)=l$, hence it would not lie in the correct interval.
\end{proof}

\n In proving the main theorem of this section, we assume that we are beginning with the configuration in Figure \ref{bbspriorcoll}, with $l\leq g<k$, so that we are at the last pre-collision stage.

\begin{thm}
Take a box-ball system consisting of just two blocks of adjacent balls, subject to the following:
\begin{enumerate}
\item The left-most block has $k$ balls.
\item The right-most block has $l$ balls.
\item $k>l$
\item The two blocks are separated by at least $l$ empty boxes.
\end{enumerate}
After sufficiently many time steps of the box-ball evolution, after the blocks have collided and ordered themselves, the $k$-block will have experienced a phase shift of $+2\min(k,l)=2l$, and the $l$-block will have experienced a phase shift of $-2\min(k,l)=-2l$.
\end{thm}

\begin{proof}
To begin the proof, we apply the augmentation map to the box-ball system to obtain the following ghost-box-ball configuration in $\mathcal{G}^0$:
\begin{figure}[H]
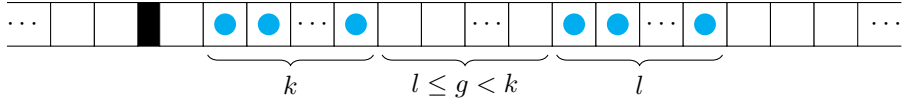

\centering
\tikz[scale=0.58]{
\foreach \y in {0}
{
\foreach \x in {4}
{\draw[fill=black]  (\x,\y-1) -- (\x+0.5,\y-1) -- (\x+0.5,\y) -- (\x,\y) -- cycle;			
}
\foreach \x in {}
{\draw[fill=gray!40]  (\x,\y-1) -- (\x+0.5,\y-1) -- (\x+0.5,\y) -- (\x,\y) -- cycle;			
}
\foreach \x in {3,4.5,9.5,10.5,11.5,12.5,17.5,18.5,19.5,7.5,15.5}
{\draw[fill=white]  (\x,\y-1) -- (\x+1,\y-1) -- (\x+1,\y) -- (\x,\y) -- cycle;				
}
\foreach \x in {5.5,6.5,8.5,13.5,14.5,16.5}
{\draw[fill=white]  (\x,\y-1) -- (\x+1,\y-1) -- (\x+1,\y) -- (\x,\y) -- cycle;				
\fill[cyan] (\x+0.5,\y-0.5) circle (0.25);									
}
\foreach \x in {}
{\draw[fill=white]  (\x,\y-1) -- (\x+1,\y-1) -- (\x+1,\y) -- (\x,\y) -- cycle;				
\fill[red] (\x+0.5,\y-0.5) circle (0.25);									
}
\foreach \x in {20.5}
{
\draw[-] (\x,\y) -- (\x+1,\y);
\draw[-] (\x,\y-1) -- (\x+1,\y-1);
\node at (\x+0.7,\y-0.5) {$\cdots$};
}
\foreach \x in {2}
{
\draw[fill=white]  (\x,\y-1) -- (\x+1,\y-1) -- (\x+1,\y) -- (\x,\y) -- cycle;		
\draw[-] (\x,\y) -- (\x-1,\y);
\draw[-] (\x,\y-1) -- (\x-1,\y-1);
\node at (\x-0.6,\y-0.5) {$\cdots$};
}
\foreach \x in {7.5,11.5,15.5}
{
\node at (\x+0.55,\y-0.5) {$\cdots$};
}
\foreach \y in {-1.2}{
\foreach \x in {8.5}{
\draw [decorate,decoration={brace,amplitude=4pt}] (\x+0.9,\y) -- (\x-2.9,\y) node [black,midway,yshift=-0.4cm] {$k$};
}
\foreach \x in {12.5}{
\draw [decorate,decoration={brace,amplitude=4pt}] (\x+0.9,\y) -- (\x-2.9,\y) node [black,midway,yshift=-0.4cm] {$l\leq g<k$};
}
\foreach \x in {16.5}{
\draw [decorate,decoration={brace,amplitude=4pt}] (\x+0.9,\y) -- (\x-2.9,\y) node [black,midway,yshift=-0.4cm] {$l$};
}
}
}
}
\caption{The augmentation of the canonical representative of a $(k,l)$ 2-soliton box-ball system.}\label{augmentedtwosolitonphasepic}
\end{figure}

\n By Theorem \ref{thmvarthetaequiv}, studying this augmented ghost-box-ball system for subsequent time-steps is equivalent to the corresponding study for the original box-ball system.\\[2pt]

\n Coordinatising this ghost-box-ball system yields the coordinates:
$$(\infty,0,1,k,g,l,\infty).$$
Since this is in $\mathcal{G}^0$, we may apply Corollary \ref{conjectureaboutgbbsbbs}: evolving these coordinates by any number of iterations of $\chi^0$, and then producing the corresponding ghost-box-ball system will yield the same result as iterating the ghost-box-ball algorithm on the GBBS in Figure \ref{augmentedtwosolitonphasepic}.\\

\n Lining these up with the ghost-box-ball coordinates for the evolution equations, we have
$$
W_0^0=\infty,~~~
Q_1^0=0,~~~
W_1^0=1,~~~
Q_2^0=k,~~~
W_2^0=g,~~~
Q_3^0=l,~~~
W_3^0=\infty.
$$

\n We recall the evolution rules below:
\begin{align*}
W_0^{t+1}&=W_3^{t+1}=\infty\\
W_n^{t+1}&=Q_{n+1}^t+W_n^t-Q_n^{t+1},~~~~~~~~~~~~~~~~~~~~~n=1,\ldots,2\\
Q_n^{t+1}&=\min\left(W_n^{t},\sum_{j=1}^n Q_j^t-\sum_{j=1}^{n-1} Q_j^{t+1}\right),~~~~~n=1,\ldots,3,
\end{align*}
Studying the above equations, we make the following observation/simplifications to the above:
\begin{itemize}
    \item We see that $Q_1^{t}=0$ for all $t$ because $Q_1^{t+1}= \min(\infty,Q_1^t)=Q_1^t$ and $Q_1^0=0$.
    \item Since $Q_1^{t+1}=0$, it also follows that $W_1^{t+1}=Q_2^t+W_1^t$.
    \item Since $Q_1^t=Q_1^{t+1}=0$, we have $Q_2^{t+1}=\min(W_2^t,Q_2^t)$.
    \item Since $W_3^t=\infty$, we have $Q_3^{t+1}=Q_2^t+Q_3^t-Q_2^{t+1}$.
\end{itemize} 
Thus, at time $t=1$, we have:
$$
W_0^1=\infty,~~~
Q_1^1=0,~~~
W_1^1=k+1,~~~
Q_2^1=g,~~~
W_2^1=l,~~~
Q_3^1=k+l-g,~~~
W_3^1=\infty.
$$
At time $t=2$, we have:
$$
W_0^2=\infty,~~~
Q_1^2=0,~~~
W_1^2=k+g+1,~~~
Q_2^2=l,~~~
W_2^2=k+l-g,~~~
Q_3^2=k,~~~
W_3^2=\infty.
$$
At this point, the gap between the two blocks is $k+l-g>\min(k,l)=l$ and the shorter block is behind, so we know that the collision and sorting phenomenon has resolved.\\[2pt]

\n In the absence of collisions, a block travels with velocity equal to its length. From this, we deduce that

\begin{equation}
    W_1^t=k+g+1+l(t-2),~~
Q_2^t=l,~~
W_2^t=k+l-g+(t-2)(k-l),~~
Q_3^t=k\label{eqnbbstgeqtwo}
\end{equation}

\n for $t\geq 2$.\\

\n To capture the phase shift, we make use of the filled ghost we introduced at the start. We will label the box immediately to the right of this ghost as the $0^\text{th}$ box, with the next box to the right labelled as the $1^\text{st}$ box, and so on. Below we demonstrate this by showing the label for the first box in each block after the initial filled ghost in Figure \ref{augmentedtwosolitonphasepic}:

\begin{figure}[H]
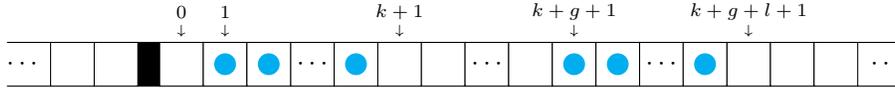

\centering
\tikz[scale=0.58]{
\foreach \y in {0}
{
\foreach \x in {4}
{\draw[fill=black]  (\x,\y-1) -- (\x+0.5,\y-1) -- (\x+0.5,\y) -- (\x,\y) -- cycle;			
}
\foreach \x in {}
{\draw[fill=gray!40]  (\x,\y-1) -- (\x+0.5,\y-1) -- (\x+0.5,\y) -- (\x,\y) -- cycle;			
}
\foreach \x in {3,4.5,9.5,10.5,11.5,12.5,17.5,18.5,19.5,7.5,15.5}
{\draw[fill=white]  (\x,\y-1) -- (\x+1,\y-1) -- (\x+1,\y) -- (\x,\y) -- cycle;				
}
\foreach \x in {5.5,6.5,8.5,13.5,14.5,16.5}
{\draw[fill=white]  (\x,\y-1) -- (\x+1,\y-1) -- (\x+1,\y) -- (\x,\y) -- cycle;				
\fill[cyan] (\x+0.5,\y-0.5) circle (0.25);									
}
\foreach \x in {}
{\draw[fill=white]  (\x,\y-1) -- (\x+1,\y-1) -- (\x+1,\y) -- (\x,\y) -- cycle;				
\fill[red] (\x+0.5,\y-0.5) circle (0.25);									
}
\foreach \x in {20.5}
{
\draw[-] (\x,\y) -- (\x+1,\y);
\draw[-] (\x,\y-1) -- (\x+1,\y-1);
\node at (\x+0.7,\y-0.5) {$\cdots$};
}
\foreach \x in {2}
{
\draw[fill=white]  (\x,\y-1) -- (\x+1,\y-1) -- (\x+1,\y) -- (\x,\y) -- cycle;		
\draw[-] (\x,\y) -- (\x-1,\y);
\draw[-] (\x,\y-1) -- (\x-1,\y-1);
\node at (\x-0.6,\y-0.5) {$\cdots$};
}
\foreach \x in {7.5,11.5,15.5}
{
\node at (\x+0.55,\y-0.5) {$\cdots$};
}
\node at (5,0.7) {\scriptsize{$0$}};
\node at (6,0.7) {\scriptsize{$1$}};
\node at (10,0.7) {\scriptsize{$k+1$}};
\node at (14,0.7) {\scriptsize{$k+g+1$}};
\node at (18,0.7) {\scriptsize{$k+g+l+1$}};
\foreach \x in {5,6,10,14,18}{
\node at (\x,0.27) {\tiny$\downarrow$};
}
}
}
\caption{The initial ghost-box-ball state with numbering relative to the filled ghost.}
\end{figure}
\n In the above, we see that the position of the first ball in the first block is
$$W_1^0 = 1$$
and the position of the first ball in the second block is 
$$W_1^0+Q_2^0+W_2^0 = 1+k+g.$$
This holds in general since $W_1^0$ represents the number of spaces between the filled ghost and the first ball, and the position of the first ball of the second block is the number of initial empty boxes ($W_1^0$) plus the number of balls in the first block ($Q_2^0$) plus the number of empty spaces between the first and second block ($W_2^0$).\\

\n Since the phase shift pertains to the positions of the blocks, the quantities of interest will be $W_1^t$ and $W_1^t+Q_2^t+W_2^t$. From Equation \ref{eqnbbstgeqtwo} we have for $t\geq 2$:
$$W_1^t = k+g+1+l(t-2),~~~W_1^t+Q_2^t+W_2^t = 2l+kt+1.$$
Now, to see the phase-shift, we suppose the blocks in Figure \ref{augmentedtwosolitonphasepic} could travel freely, passing through each other, at their respective velocities. After sufficiently many time steps, this dynamics would see the $k$-block overtake the $l$-block and one would have:
$$\tilde{W}_1^t = 1 + k + g + lt$$
because the $l$-block was initially $1+k+g$ boxes from the filled ghost. Here, we use $\tilde{W}_1^t$ to distinguish between the ``would-be'' value and $W_1^t$ (the actual value). \\

\n If $t$ is sufficiently large, the $k$-block would be $1+kt$ spaces away from the filled ghost. Therefore,
$$\tilde{W}_1^t+\tilde{Q}_2^t+\tilde{W}_2^t= 1+kt.$$

\n We are finally able to reveal the phase shifts. For the $k$-block, we look at
$$(W_1^t+Q_2^t+W_2^t)-(\tilde{W}_1^t+\tilde{Q}_2^t+\tilde{W}_2^t)
= (2l+kt+1) - (1+kt) = 2l. $$
For the $l$-block, we look at 
$$W_1^t - \tilde{W}_1^t = (k+g+1+l(t-2)) - (1 + k + g + lt) = -2l.$$
\end{proof}

\n This theorem extends to the following conjectured formula for the phase shifts for box-ball systems with any number of blocks.

\begin{con}
If one has a box-ball system with a total of $n$ blocks, with $Q_k$-many balls in the $k$-th block, and with blocks separated sufficiently so as to be able to identify the asymptotic soliton structure by simply ordering $(Q_k)_{k=1}^n$. After sufficiently many time-steps have passed (i.e. after the blocks have finished all collisions), the $k$-th block will have experienced a total phase shift of 
\begin{equation}2\sum_{\substack{j>k\\Q_j<Q_k}} \min(Q_j,Q_k) - 2\sum_{\substack{j<k\\Q_j>Q_k}} \min(Q_j,Q_k).\end{equation}
As a special case of this, if the initial configuration is such that 
$$Q_1>Q_2>\cdots>Q_n,$$
then the phase shift experienced by the $k$-th block is given by
\begin{equation}2\sum_{j>k} \min(Q_j,Q_k) - 2\sum_{j<k} \min(Q_j,Q_k).\label{bbsphsshftfrmconj}\end{equation}
\end{con}

\n To interpret Formula \ref{bbsphsshftfrmconj}: the first sum is the total (positive) phase shift experienced by the $Q_k$-block as a result of colliding with slower blocks in front, and the second sum is the total (negative) phase shift experienced by the $Q_k$-block as a result of colliding with faster blocks initially behind it.\\

\n This reflects the analogous phase shift behaviour of classical soliton theory \cite{bib:moser}: phase shifts propagate collision-by-collision.

\section{Conclusions}\label{chapterconclusions}
Our main results, detailed in Theorem \ref{mainresRSKencodedinGBBS}, provide a complete, rigorous and comparatively simple correspondence between Schensted insertion and a particle system of box-ball type. This opens the door for further connections to algorithms, dynamics and integrable systems theory. To be sure some of these connections have been noticed in the literature before; however, we believe the simplicity of our correspondence provides avenues for deeper insights and broader connections. The results in this paper concerned what may be characterized as discrete time - discrete space systems. 

The connections we refer to here have to do with passing to continuous versions, going both forwards and back. Some of that is already evident in the passage from RSK to gRSK (which is a discrete time - continuous space system) described in Section \ref{kirillovgrsksection}.
gRSK may in fact be reformulated as a matrix factorization dynamics as described in Section \ref{sectmatrixformgrsk}.  It is also known \cite{bib:h} that gRSK is related to a discrete-time form of the Toda lattice, a well-known continuous time - continuous space integrable system. In \cite{bib:r} it was shown how to relate gRSK in terms of matrix factorization to a standard lower-upper matrix factorization representation for the discrete-time Toda Lattice. 
The latter such representation was developed and studied by Symes \cite{bib:symes} and Deift-Nanda-Tomei \cite{bib:dnt}. Our results here make it possible to make all these connections precise. Moreover they provide a means to relate all the solitonic properties of our ghost box-ball system to those of the integrable Toda lattice.  A particular instance of this will be to deduce the phase shift formulae established in Section \ref{sectionphaseshiftbbs} directly from the classical continuous time - continuous space phase shift formulae \cite{bib:moser}. 

Going in the other direction, from gRSK to the continuous time Toda lattice, we can compare our approach with that of O’Connell and collaborators \cite{bib:osurvey} \cite{bib:o} \cite{bib:cosz}. That work used this direction as a means to push forward random algorithmic structures to the setting of random directed polymers in a way that remarkably connects to semiclassical limits of the quantum Toda lattice.  We are able to look more deeply into these constructions from a strictly integrable systems perspective to find  a number of simplifications of these constructions that replace some of the more technical aspects of quantization by more classical constructions based on Painlev\'{e} balances and dressing transformations \cite{bib:efh}. This is more than just a reformulation since it may lead to novel approaches for randomization as well as extensions to general Lie theoretic settings of the Toda lattice.

\section*{Acknowledgments}
The authors wish to thank Joceline C. Lega and Sergey Cherkis for their helpful comments and suggestions on this project.


\end{document}